# Spectra of CO$_2$-Rg$_2$ and CO$_2$-Rg-He trimers (Rg = Ne, Ar, Kr, and Xe): intermolecular CO$_2$ rock, vibrational shifts and three-body effects

A.J. Barclay,[1] A.R.W. McKellar,[2] and N. Moazzen-Ahmadi[1]

[1] *Department of Physics and Astronomy, University of Calgary, 2500 University Drive North West, Calgary, Alberta T2N 1N4, Canada*

[2] *National Research Council of Canada, Ottawa, Ontario K1A 0R6, Canada*

## Abstract

Weakly-bound CO$_2$-Rg$_2$ trimers are studied by high resolution (0.002 cm$^{-1}$) infrared spectroscopy in the region of the CO$_2$ $\nu_3$ fundamental band ($\approx$2350 cm$^{-1}$), using a tunable OPO to probe a pulsed supersonic slit jet expansion with an effective rotational temperature of about 2 K. CO$_2$-Ar$_2$ spectra have been reported previously, but are extended here to include Rg = Ne, Kr, and Xe as well as new combination and hot bands. For Kr and Xe, a unified scaled parameter scheme is used to account for the many possible isotopic species. Vibrational shifts of CO$_2$-Rg$_2$ trimers are compared to those of CO$_2$-Rg dimers, and in all cases the trimer shifts are slightly more positive (blue-shifted) than expected on the basis of linear extrapolation from the dimer. Combination bands directly measure an intermolecular vibrational mode (the CO$_2$ rock), and gives values of about 32.2, 33.8, and 34.7 cm$^{-1}$ for CO$_2$-Ar$_2$, -Kr$_2$, and -Xe$_2$. Structural parameters derived for CO$_2$-Rg$_2$ trimers are compared with those of CO$_2$-Rg and Rg$_2$ dimers. Spectra of the mixed trimers CO$_2$-Rg-He are also reported.

## 1. Introduction

Thanks to their fundamental nature and experimental accessibility, $CO_2$-rare gas (Rg) clusters serve as useful probes of intermolecular force effects to help improve theoretical modeling. The $CO_2$-Rg dimers have been extensively studied by high resolution spectroscopy, beginning with the 1979 work of Steed et al.[1] on pure rotational transitions of $CO_2$-Ar. Some key experimental spectroscopic results in the microwave and infrared regions are as follows: for $CO_2$-He,[2-4] for $CO_2$-Ne,[5-9] for $CO_2$-Ar,[1,5,6,10-15] for $CO_2$-Kr,[5,6, 16-18] and for $CO_2$-Xe.[5,8,16,19] The equilibrium dimer structures were found to be T-shaped, with the Rg atom located to the 'side' of the $CO_2$, adjacent to the C atom. Effective intermolecular distances (C-Rg) ranged from 3.3 Å for $CO_2$-Ne to 3.8 Å for $CO_2$-Xe.

In a notable 1993 paper, Xu et al.[20] observed microwave transitions of the trimer $CO_2$-$Ar_2$, and this was followed up with infrared observations of $CO_2$-$Ar_2$ by Sperhac et al.[21] in 1996. As illustrated in Fig. 1, the trimer structure places the second Ar atom in a 'side' position equivalent to that of the first, with an Ar-Ar distance of about 3.8 Å, similar to that in the isolated $Ar_2$ dimer. The trimer is thus an asymmetric rotor with $C_{2v}$ point group symmetry, as is $CO_2$-Ar. Analogous trimer structures and infrared spectra have also been observed for $CO_2$-$He_2$,[22] $N_2O$-$Ar_2$, and $N_2O$-$Ne_2$.[23] Other weakly bound trimers containing a linear molecule plus two rare gas atoms have also been studied, including OCS-$Ar_2$,[24] OCS-$Ne_2$,[25] HCN-$Ar_2$,[26] and HBr-$Ar_2$.[27]

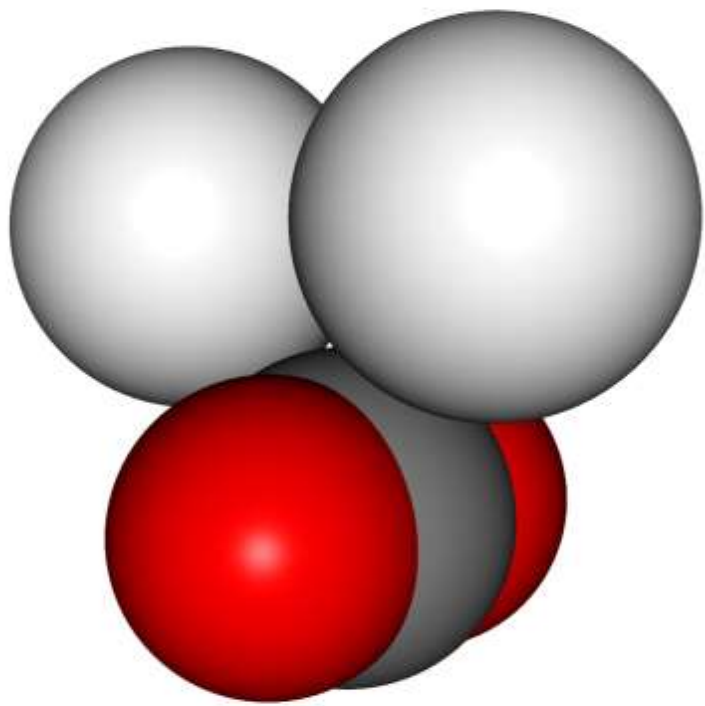

Fig. 1. Illustration of the $CO_2$-$Rg_2$ trimer structure. The two Rg atoms at the top in this picture occupy equivalent positions on the equatorial plane of the $CO_2$ molecule. The 2-fold rotational symmetry axis (vertical in this picture) is the *a*-inertial axis for Rg = He and Ne, and the *b* axis for Rg = Ar, Kr, and Xe.

In the present paper, we extend the study of $CO_2$-$Rg_2$ to include Ne, Kr, and Xe by observing infrared spectra in the region of the $\nu_3$ fundamental band of $CO_2$ ($\approx$2350 cm⁻¹), the same as for $CO_2$-$Ar_2$ in Ref. 21. We also observe combination bands for Ar, Kr, and Xe in addition to the fundamental band. These combinations involve low frequency ($\approx$30 cm⁻¹) intermolecular modes and thus give new and direct information on trimer vibrational dynamics.

As well, we observe (limited) trimer spectra corresponding to the $CO_2$ $(v_1, v_2^{l2}, v_3) = (01^11) - (01^10)$ hot band transition ($\approx 2337$ cm$^{-1}$). All the spectra are recorded using supersonic expansion gas mixtures with helium as the main component, and we are able to observe spectra of the mixed trimers $CO_2$-Rg-He (Rg = Ne, Ar, Kr, and Xe). As mentioned, $CO_2$-Rg dimers and $CO_2$-Rg$_2$ trimers have $C_{2v}$ symmetry. For the dimers, where the $a$-inertial axis is the $C_2$ symmetry axis, nuclear spin statistics dictate that only levels with $(K_a) =$ (even) are allowed in the ground vibrational state (for $^{12}C^{16}O_2$). The same is true for $CO_2$-Ne$_2$, but for $CO_2$-Ar$_2$, it turns out that $b$ is the symmetry axis, so only levels with $(K_a, K_c) =$ (even, even) or (odd, odd) are allowed. For $CO_2$-Kr$_2$ and $CO_2$-Xe$_2$ trimers, $b$ is again the symmetry axis, but the multiplicity of atomic isotopes means that spin statistics have less influence on the spectra.

## 2. Results

The spectra were recorded as described previously,[9,19,28] using a rapid-scan optical parametric oscillator source to probe a pulsed supersonic slit jet expansion. The typical gas expansion mixture contained about 0.04% carbon dioxide plus 0.8% neon, argon, krypton, or xenon in helium carrier gas with a jet backing pressure of about 13 atmospheres. Wavenumber calibration was carried out by simultaneously recording signals from a fixed etalon and a $CO_2$ reference gas cell. Simulation and fitting were carried out using PGOPHER software,[29] using the Mergeblend option to fit blended lines to an intensity weighted average of their components.

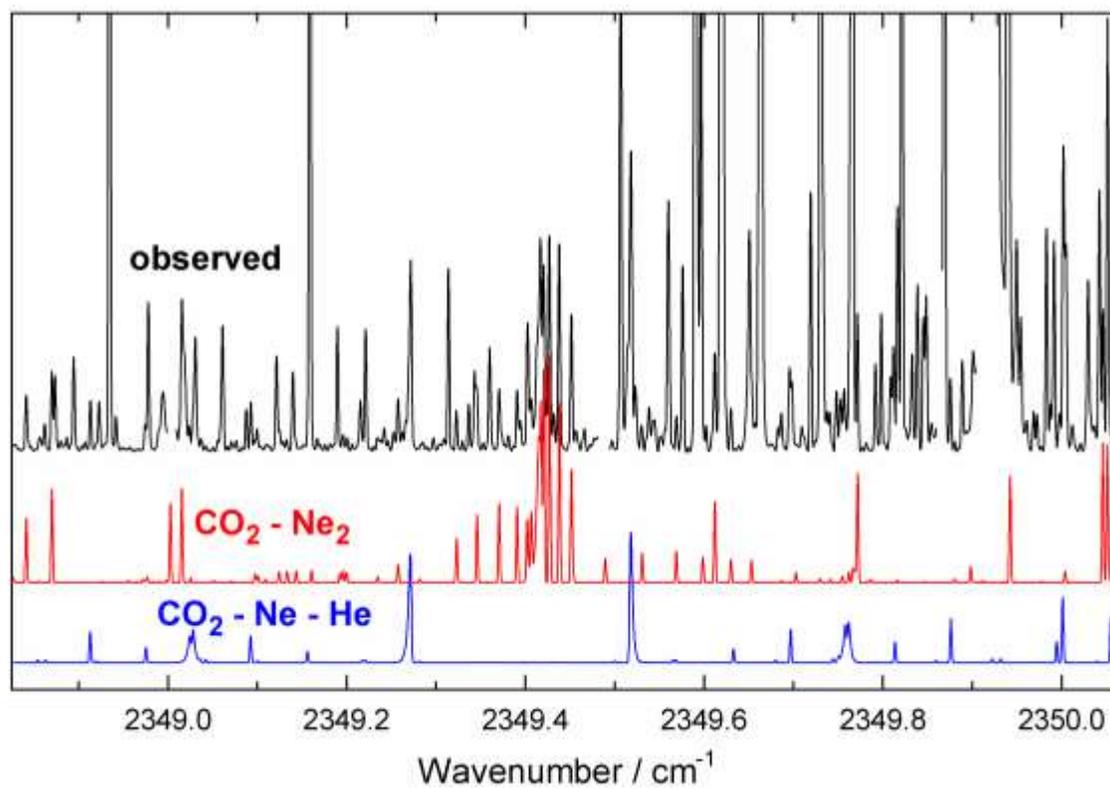

Fig. 2. Observed and simulated spectra showing $CO_2$-$Ne_2$ and $CO_2$-Ne-He in the region of the $CO_2$ $\nu_3$ fundamental band. The simulation shows only the $CO_2$-$^{20}Ne_2$ contribution, but transitions of $CO_2$-$^{20}Ne$-$^{22}Ne$ were also assigned.

Table 1. Molecular parameters for $CO_2$-$Ne_2$ (in $cm^{-1}$) [a]

| | $CO_2$-$^{20}Ne_2$ ground State | $CO_2$-$^{20}Ne_2$ fundamental | $CO_2$-$^{20}Ne$-$^{22}Ne$ ground State | $CO_2$-$^{20}Ne$-$^{22}Ne$ fundamental |
|---|---|---|---|---|
| $\nu_0$ | | 2349.4201(1) | | 2349.4215(3) |
| $A$ | 0.112244(37) | 0.111971(28) | 0.108665(70) | 0.108197(99) |
| $B$ | 0.080417(28) | 0.080118(26) | 0.078629(29) | 0.078321(43) |
| $C$ | 0.060634(64) | 0.060548(63) | 0.05854(18) | [0.05845] |
| $10^5 \times \Delta_K$ | 2.51(31) | [2.51] | [2.51] | [2.51] |
| $10^5 \times \Delta_{JK}$ | -1.66(39) | [-1.66] | [-1.66] | [-1.66] |
| $10^6 \times \Delta_J$ | 5.4(12) | [5.4] | [5.4] | [5.4] |
| $10^6 \times \delta_J$ | -1.05(62) | [-1.05] | [-1.05] | [-1.05] |

[a] Quantities in parentheses are 1σ from the least-squares fit, in units of the last quoted digit. Centrifugal distortion parameters were fixed to $CO_2$-$^{20}Ne_2$ ground state values as indicated by square brackets. For $CO_2$-$^{20}Ne$-$^{22}Ne$, ($C'$ - $C''$) was fixed to the $CO_2$-$^{20}Ne_2$ value.

### 2.1. $CO_2$-$Ne_2$

Part of the observed spectrum showing $CO_2$-$Ne_2$ is illustrated in Fig. 2 (a broader view of the same spectrum is available in Fig. 1 of Ref. 9). Here the strongest lines (and many of the weaker ones) belong to $CO_2$-Ne, $CO_2$-He, and $(CO_2)_2$. In spite of these many strong interfering lines, much of the relatively weak $Q$-branch of $CO_2$-$Ne_2$ (2349.3 – 2349.45 $cm^{-1}$) is fortunately clear of interference from other species, and many $P$- and $R$-branch transitions were also well resolved. This is a $c$-type perpendicular band ($\Delta K_a = \pm 1$, $\Delta K_c = 0$) with $K_a$ = even levels in the ground state and $K_a$ = odd in the excited state. The central $Q$-branch transitions noted in Fig. 2 have $K_a = 1 \leftarrow 0$, and we also assigned transitions with $K_a = 5 \leftarrow 6$, $3 \leftarrow 4$, $1 \leftarrow 2$, and $3 \leftarrow 2$.

Ultimately 31 lines were assigned in terms of 34 transitions (there were a few blends) and fitted to obtain the parameters listed in Table 1. The root mean square (rms) deviation was 0.00021 $cm^{-1}$, approximately equal to our estimated experimental precision. Detailed line positions and assignments are given as Supplementary Information. It was also possible to assign about 15 transitions to the mixed trimer $CO_2$-$^{20}$Ne-$^{22}$Ne (the natural abundance of $^{22}$Ne is about 9%), with results as shown in Table 1. For this species, there are of course no restrictions on the values of $K_a$. The rms deviation was 0.00019 $cm^{-1}$.

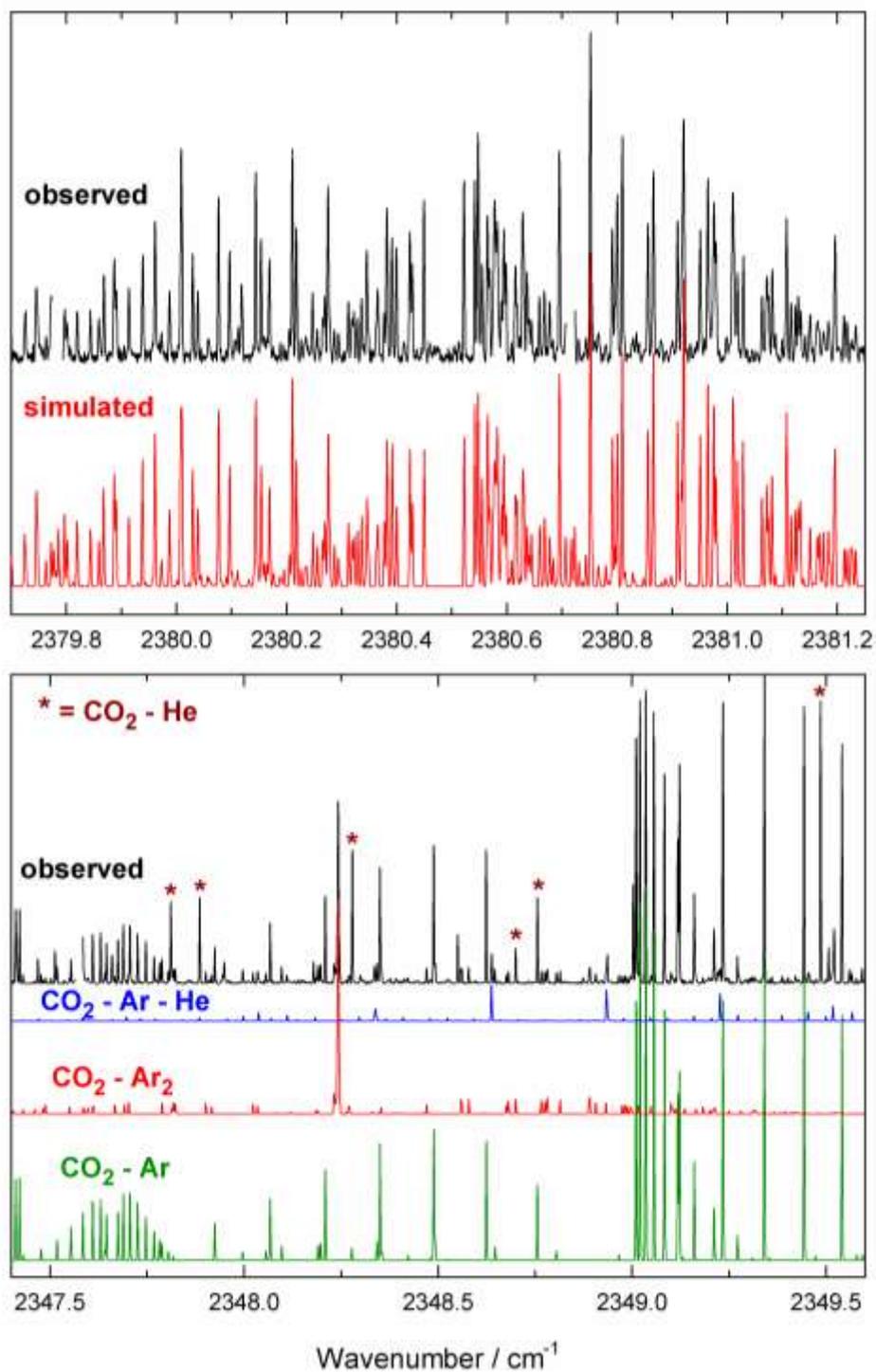

Fig. 3. Observed and simulated spectra showing the fundamental bands of $CO_2$-$Ar_2$ and $CO_2$-Ar-He (lower panel) and the combination band of $CO_2$-$Ar_2$ (upper panel).

## 2.2. CO$_2$-Ar$_2$

The lower panel of Fig. 3 shows the fundamental band of CO$_2$-Ar$_2$, previously studied by Sperhac et al.[21] This is a *c*-type band like that of CO$_2$-Ne$_2$, but now the symmetry axis is *b*, so the allowed levels in the lower state have $(K_a, K_c)$ = (even, even) and (odd, odd). Note the relatively prominent *Q*-branch at 2348.24 cm$^{-1}$. The upper panel of Fig. 3 shows the new CO$_2$-Ar$_2$ combination band, which is a *b*-type band ($\Delta K_a = \pm 1$, $\Delta K_c = \pm 1, \pm 3$) with a noticeable gap in the center. This band looks complicated, but was fairly easy to assign since we already had good ground state rotational parameters.[20,21] It has the advantage, compared to the fundamental, of being free from interference from stronger transitions due to other molecular species.

In order to analyze the spectra, we made a combined fit to the two present infrared bands and the appropriately weighted microwave data of Xu et al.,[20] with results as shown in Table 2. A total of 52 lines were assigned in the fundamental and 105 in the combination band, and fitted with rms errors of 0.00025 and 0.00027 cm$^{-1}$, respectively. The 21 microwave transitions had an rms error of 1.3 kHz.

The earlier CO$_2$-Ar$_2$ studies[20,21] included sextic centrifugal distortion terms in their analyses, but we felt that the slight improvement of the fit was not worth the complication. Also, the earlier studies used a III$^r$ representation, but CO$_2$-Ar$_2$ is not especially close to the oblate limit, and we found that a I$^r$ representation (as used here) actually improves the fit (for example by a factor of about 2.3 for the sum of squares in both 8 and 11 parameter fits to the microwave data[20]). This I$^r$ vs. III$^r$ difference is probably not deeply meaningful, but the result encouraged us to stay with the more common I$^r$ scheme. The switch to I$^r$ from III$^r$ explains the differences between our distortion parameters and the previous ones (e.g., the sign of $\delta_K$).

Table 2. Molecular parameters for $CO_2$-$Ar_2$ (in $cm^{-1}$) [a]

|  | Ground State | Fundamental | Combination |
|---|---|---|---|
| $\nu_0$ |  | 2348.2445(1) | 2380.4904(1) |
| $A$ | 0.059000938(21) | 0.0588947(20) | 0.0567837(67) |
| $B$ | 0.050120164(16) | 0.0500721(26) | 0.0504864(27) |
| $C$ | 0.031241461(12) | 0.0312255(57) | 0.0307406(12) |
| $10^7 \times \Delta_K$ | 7.769(11) | [7.769] |  |
| $10^7 \times \Delta_{JK}$ | -4.726(11) | [-4.726] | 5.76(81) |
| $10^7 \times \Delta_J$ | 3.4716(28) | [3.4716] |  |
| $10^7 \times \delta_K$ | 0.2838(87) | [0.2838] |  |
| $10^7 \times \delta_J$ | 1.3460(11) | [1.3460] |  |

[a] Quantities in parentheses correspond to 1σ from the least-squares fit, in units of the last quoted digit. Fit includes ground state microwave data of Ref. 20. Ground and excited fundamental state centrifugal distortion parameters were constrained to be equal. This analysis uses a $I^r$ representation, while those in Refs. 20, 21 use a $III^r$ representation.

Table 3. Molecular parameters for the $(01^11) \leftarrow (01^10)$ hot band of $CO_2 - Ar_2$ (in $cm^{-1}$).[a]

|  | $(01^10)$ | $(01^11)$ |
|---|---|---|
| $\sigma_0$ (i-p) | X [b] | 2335.7499(1) + X |
| $\sigma_0$ (o-p) | 0.581(8) + $\sigma_0$ (i-p) | 0.580(8) + $\sigma_0$ (i-p) |
| $A$ | 0.058920(16) | [0.058814] |
| $B$ | 0.050142(20) | [0.050094] |
| $C$ | [0.031241] | [0.031225] |
| $\xi_c$ | [0.062] | [0.062] |

[a] Quantities in parentheses correspond to 1σ from the least-squares fit, in units of the last quoted digit. Quantities in square brackets were constrained at the indicated values (see text).

[b] X is equal to the free $CO_2$ $\nu_2$ frequency (667.380 $cm^{-1}$) plus or minus a (small) unknown vibrational shift.

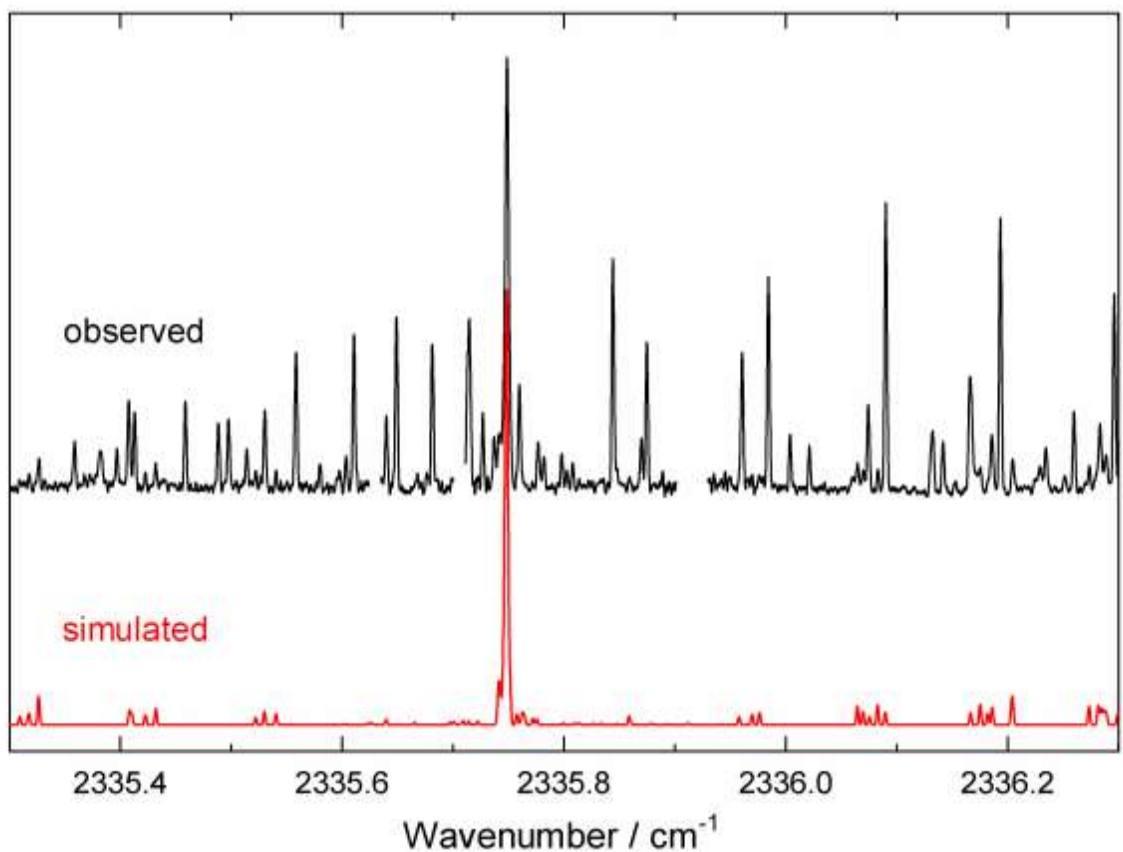

Fig. 4. Observed and simulated spectra of $CO_2$-$Ar_2$ in the region of the $CO_2$ $(01^11) \leftarrow (01^10)$ hot band. Gaps in the observed spectrum correspond to regions of $CO_2$ monomer absorption. The simulated spectrum includes both the i-p and o-p modes. The many observed lines not in the simulation are due to $CO_2$-Ar.[15]

Part of the observed spectrum of $CO_2$-$Ar_2$ in the region of the $CO_2$ $(01^11) \leftarrow (01^10)$ hot band is shown in Fig. 4. Its existence depends on the small fraction of $CO_2$ molecules which remain in the $(01^10)$ vibrational state following the supersonic expansion. As explained previously for the case of the dimer $CO_2$-Ar,[15] the presence of the nearby Ar atom(s) break the symmetry of the degenerate $CO_2$ $(01^10)$ bending mode into in-plane (i-p) and out-of-plane (o-p) components. In the present case the plane referred to is the symmetry plane containing $CO_2$ and bisecting the angle between the two Ar atoms. These i-p and o-p modes have $A_1$ and $B_2$ symmetry, respectively, for the lower $(01^10)$ vibrational state, and $B_1$ and $A_2$ symmetry for the upper $(01^11)$ state. The hot band is $c$-type, the same as the fundamental, with $(K_a, K_c)$ = (even, even) and (odd, odd) in the lower $(01^10)$ vibrational state for the i-p mode and $(K_a, K_c)$ = (even, odd) and (odd, even) for the o-p mode. From previous studies,[9,15,18,19] we expect strong Coriolis mixing ($c$-type in this case) between the i-p and o-p modes. The unresolved $Q$-branch of the $CO_2$-$Ar_2$ hot band at 2335.75 cm[-1] is fairly strong, but the individual resolved $P$- and $R$-branch transitions are weak, as shown in Fig. 4. We assigned a total of 28 lines and fitted them with an rms error of 0.00027 cm[-1] to obtain the parameters listed in Table 3. Due to the weakness of the spectrum and the relatively few assigned lines, it was prudent to highly constrain the fit. Thus rotational constants were assumed to be the same for the i-p and o-p modes, changes in rotational constants between $(01^10)$ and $(01^11)$ were assumed to be the same as between $(00^00)$ and $(00^01)$ (i.e. the fundamental band, Table 2), $C$ was fixed at fundamental band values, and the Coriolis parameter, $\xi_c$, was assumed to equal $2C$. These are all reasonable approximations, based on the case of $CO_2$-Ar.[15] With these constraints, we obtained a value of 0.581(8) cm[-1] for the separation between the i-p and o-p modes, with o-p lying above i-p. However we feel that the real uncertainty in this value is significantly larger, based on our experience of trying different constraints in the fit. In the case of $CO_2$-Ar, the separation of 0.8773(2) cm[-1] was much better determined, and had the same sign.[15]

Is it possible to relate the splitting in the trimer to that in the dimer? If we denote the angle between the $CO_2$ bending plane and the $CO_2$-Ar dimer plane as $\phi$, then for $\phi = 0°$ the splitting of the $CO_2$ bending mode is the measured value of 0.877 cm[-1], for $\phi = 45°$ the splitting must be zero, and for $\phi = 90°$ it is -0.877 cm[-1]. This dependence can be simply modeled as sinusoidal: (splitting) = $0.877 \times \cos(2\phi)$. For $CO_2$-$Ar_2$ trimer, the angle between each C-Ar bond and the symmetry plane bisecting the two bonds is $\phi = 33.2°$,[20] and the model would predict: (splitting) = $2 \times 0.877 \times \cos (66.4°) = 0.702$ cm$^{-1}$. This does not agree particularly well with the measured value of 0.59 cm$^{-1}$, but the splitting dependence is not necessarily sinusoidal, and, as mentioned, the experimental uncertainty is large.

### 2.3. CO$_2$-Kr$_2$

The fundamental and combination bands of CO$_2$-Kr$_2$ are illustrated in Fig. 5. The spectrum in the fundamental region is distinguished by a strong unresolved $Q$-branch feature at 2347.45 cm$^{-1}$ surrounded by much weaker $P$- and $R$-branches, similar to CO$_2$-Ar$_2$ in Fig. 2 but with more structure. The added structure comes from the greater mass of CO$_2$-Kr$_2$ and also from the presence of more allowed rotational transitions due to the many trimers which contain unlike Kr isotopes and thus have all values allowed for ($K_a$, $K_c$). The weakness and complexity made assignment of the $c$-type fundamental band spectrum somewhat challenging, but fortunately it was possible to predict good preliminary rotational parameters in advance by assuming that the trimer structure was analogous to that of CO$_2$-Ar$_2$.

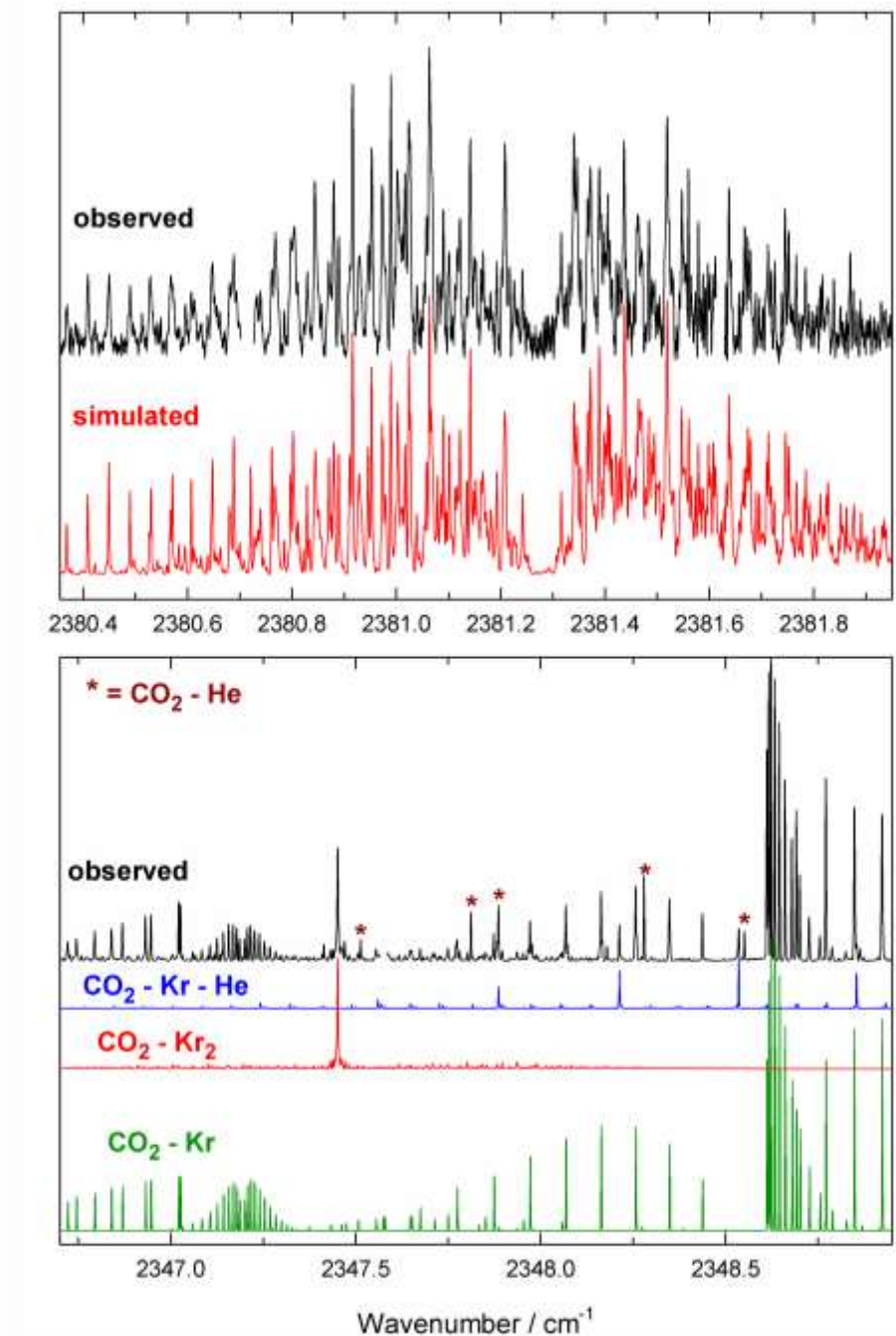

Fig. 5. Observed and simulated spectra showing the fundamental bands of $CO_2$-$Kr_2$ and $CO_2$-Kr-He (lower panel) and the combination band of $CO_2$-$Kr_2$ (upper panel). The simulations for $CO_2$-$Kr_2$ represent the sum of 15 isotopologues with scaled rotational and (for the combination band) vibrational parameters. Lines marked with an asterisk are due to $CO_2$-He.

Interpretation of the combination band (upper panel of Fig. 5) posed further difficulties, in part because there is more Kr isotopic splitting here. To deal with this we used a unified approach with scaled vibrational and rotational parameters to include all Kr isotopes, similar to the scheme developed previously for $CO_2$-Kr and $CO_2$-Xe.[18,19] The isotopic dependence of the rotational constants $A$, $B$, and $C$ was first calculated for all combinations of the 5 most abundant Kr isotopes using a simple rigid model for the trimer structure. There are fifteen such combinations with abundances ranging from 0.32 for $CO_2$-$^{84}Kr_2$ to 0.0005 for $CO_2$-$^{80}Kr_2$. Ten of these are unlike combinations with no spin statistics and a total abundance of 0.61, four are like combinations with statistical weights of 1:0 for levels with $(K_a, K_c)$ = (ee, oo) : (eo, oe) and a total abundance of 0.37, and one ($CO_2$-$^{83}Kr_2$) is a like combination with statistical weight 9:11 and an abundance of 0.013. The fifteen isotopologues, their statistical abundances, and their rotational scaling factors are analogous to those described[18] for $CO_2$-Kr, and details are given here as Supplementary Information. The same rotational scaling factors were used for all vibrational states. Previously for the dimers $CO_2$-Kr and -Xe, we introduced additional empirical fitting parameters to further refine the rotational isotope dependence, but this could not be done here because there are no precise microwave results available for $CO_2$-$Kr_2$ (unlike $CO_2$-Rg dimers and $CO_2$-$Ar_2$).

The isotope dependence of the combination band origin was modeled as

$$\nu_0(N_a, N_b) = \nu_0(N_0) + \textit{Offset} \times ((N_a + N_b) - 2N_0),$$

where *Offset* is an adjustable parameter, $N_a$ and $N_b$ are the Kr atomic mass numbers of a particular isotopologue, and $N_0$ is the "standard" mass number, taken to be 84, the most abundant. Atomic mass *numbers* were used for convenience rather than the actual atomic masses. Of course we know that vibrational frequencies do not scale linearly with Kr mass, but rather (for example) as the square root of a reduced mass, but the difference is negligible here. Possible isotope dependence of the fundamental band origin seemed to be negligible here (it was previously found to be very small for $CO_2$-Kr and -Xe). It must be acknowledged that including 15 $CO_2$-$Kr_2$ isotopologues with abundances as low as 0.0005 in our fitting procedure amounted to overkill! But it was actually fairly easy to do thanks to our previous experience with $CO_2$-Kr and -Xe, and to the convenient features of the PGOPHER software, as highlighted in Ref. 19.

Table 4. Molecular parameters for $CO_2$-$^{84}Kr_2$ (in cm$^{-1}$). [a]

|  | Ground State | Fundamental | Combination |
| --- | --- | --- | --- |
| $\nu_0$ |  | 2347.4522(2) | 2381.2807(1) |
| *A* | 0.047285(10) | 0.047228(12) | 0.045208(13) |
| *B* | 0.022930(11) | 0.022916(11) | 0.022927(11) |
| *C* | 0.0166835(65) | 0.0166796(93) | 0.0165239(66) |
| *Offset* |  |  | -0.00114(15) |

[a] Quantities in parentheses correspond to 1σ from the least-squares fit, in units of the last quoted digit. The parameter *Offset* expresses the Kr isotope dependence of the combination band origin, and has effective units of cm$^{-1}$/Dalton.

The unified fitting procedure just described greatly improved the fit and simulation of the combination band, as compared to fitting to a single "average" isotopologue, and it introduced only one additional variable (the *Offset* parameter). We also used the unified procedure for the fundamental (though it was less necessary), and analyzed both bands simultaneously. The results of this fit are listed in Table 4, and shown by the simulations in Fig. 5. These parameters apply specifically to the "standard" species $CO_2$-$^{84}Kr_2$, and the scaled parameter values for the 14 other isotope combinations are given as Supplementary Information. For the fundamental, 42 lines were fit with an rms error of 0.00062 cm$^{-1}$, and for the combination band, 81 lines were fit with an error of 0.00048 cm$^{-1}$. The *Offset* parameter in Table 4 describes how much the combination band origin shifts with Kr atomic mass. Its value of -0.001 cm$^{-1}$/Dalton implies a total shift of the origin of only about 0.01 cm$^{-1}$, from 2381.290 cm$^{-1}$ for $CO_2$-$^{80}Kr_2$, to 2381.276 cm$^{-1}$ for $CO_2$-$^{86}Kr_2$.

In the $CO_2$ $(01^10) - (01^10)$ hot band region, we observed a $Q$-branch due to $CO_2$-$Kr_2$ at 2334.948 cm$^{-1}$ (it is visible in Fig. 3 of Ref. 18), analogous to that of $CO_2$-$Ar_2$ at 2335.750 cm$^{-1}$ as described above. But it was not possible to reliably assign enough $P$- and $R$-branch transitions to make a meaningful rotational analysis of the $CO_2$-$Kr_2$ hot band.

### 2.4. $CO_2$-$Xe_2$

The fundamental and combination bands of $CO_2$-$Xe_2$ are illustrated in Fig. 6. The fundamental spectrum has a strong unresolved $Q$-branch feature at 2346.34 cm$^{-1}$. Line assignments in the weaker $P$- and $R$-branches were, like $CO_2$-$Kr_2$, challenging but still possible. The $b$-type combination band (upper panel of Fig. 6) has a central gap (like $CO_2$-$Ar_2$ and -$Kr_2$), considerable rotational structure in the $P$-branch region, and more blended structure in the $R$-branch.

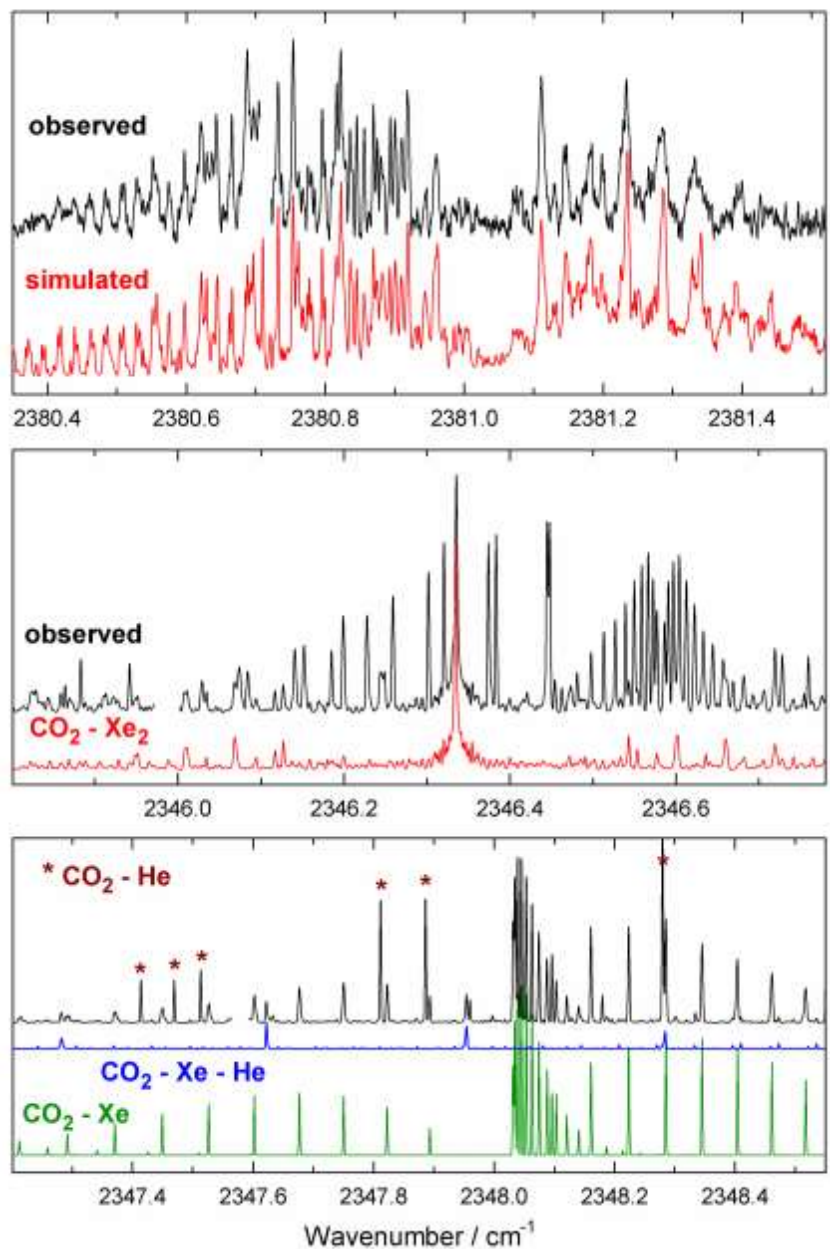

Fig. 6. Observed and simulated spectra showing the fundamental bands of $CO_2$-$Xe_2$ and $CO_2$-Xe-He (lower and middle panels) and the combination band of $CO_2$-$Xe_2$ (upper panel). The simulations for $CO_2$-$Xe_2$ represent the sum of 21 isotopologues with scaled rotational and (for the combination band) vibrational parameters. Observed lines in the $CO_2$-$Xe_2$ fundamental region (middle panel) not present in the simulation are due to $CO_2$-Xe.

To deal with Xe isotope effects we used the unified scaled approach as described above for $CO_2$-$Kr_2$. This included the six most abundant Xe isotopes, giving a total of 21 isotopic combinations ranging in mass from $CO_2$-$^{129}Xe_2$ to $CO_2$-$^{136}Xe_2$. Fifteen of these are unlike combinations with no spin statistics and a total abundance of 0.75, five are like combinations with statistical weights of 1:0 for levels with $(K_a, K_c)$ = (ee,oo) : (eo,oe) and a total abundance of 0.16, and one ($CO_2$-$^{131}Xe_2$) is a like combination with statistical weight 3:5 and an abundance of 0.05. (Abundances do not quite total to 1 because $^{128}Xe$ was not included). Further details are given as Supplementary Information.

Table 5. Molecular parameters for $CO_2$-$^{131}Xe_2$ (in $cm^{-1}$) [a]

|  | Ground State | Fundamental | Combination |
|---|---|---|---|
| $\nu_0$ |  | 2346.3355(2) | 2381.0385(5) |
| $A$ | 0.041232(12) | 0.041205(12) | 0.040601(24) |
| $B$ | 0.012773(10) | 0.012779(11) | 0.012777(12) |
| $C$ | 0.010244(12) | 0.010242(12) | 0.010183(14) |
| *Offset* |  |  | -0.00078(20) |

[a] Quantities in parentheses correspond to $1\sigma$ from the least-squares fit, in units of the last quoted digit. The parameter *Offset* expresses the Xe isotope dependence of the combination band origin, and has effective units of $cm^{-1}$/Dalton.

The unified fit was especially useful for the $CO_2$-$Xe_2$ combination band (top of Fig. 6), where it enabled us to achieve a rather good simulation which reproduces the dramatic difference between the *P*- and *R*-branch regions. The results of the fit are given in Table 5. These parameters apply to $CO_2$-$^{131}Xe_2$ which was the "standard" species. Appropriately scaled values for the other isotopic species combinations are given as Supplementary Information. For the fundamental, 44 lines were fit with an rms error of 0.00043 cm$^{-1}$, and for the combination band, 31 lines were fit with an error of 0.00048 cm$^{-1}$. The *Offset* parameter describing the shift of the combination band origin with Xe atomic mass has a value similar to that determined for $CO_2$-$Kr_2$. It represents a shift of the origin from 2381.042 cm$^{-1}$ for $CO_2$-$^{129}Xe$ to 2381.031 cm$^{-1}$ for $CO_2$-$^{136}Xe$.

As in the case of $CO_2$-$Kr_2$, we observed a $CO_2$-$Xe_2$ $Q$-branch in the $CO_2$ $(01^11) - (01^10)$ hot band region but could not assign further transitions sufficient for an analysis. The $CO_2$-$Xe_2$ $Q$-branch is located at 2333.811 cm$^{-1}$, and it may be seen in Fig. 4 of Ref. 19, located in among $K = 1 \leftarrow 2$ subband transitions of $CO_2$-Xe.

### 2.5. $CO_2$-Ne-He, $CO_2$-Ar-He, $CO_2$-Kr-He, and $CO_2$-Xe-He

Additional unexplained lines were observed in the fundamental band region that we eventually realized must be due to trimers containing $CO_2$ and the Rg atom being studied plus He. Their possible presence in the spectrum was not surprising since the supersonic expansion mixtures were predominantly composed of the helium carrier gas. Perhaps the most obvious of these new trimers was $CO_2$-Ne-He, which is the source of two noticeable lines located close to the $Q$-branch of $CO_2$-$Ne_2$. These lines, at 2349.272 and 2349.518 cm$^{-1}$ (see Fig. 2) turn out to be unresolved $Q$-branches with $K_a = 0 \leftarrow 1$ and $1 \leftarrow 0$, respectively. Parts of the spectra due to $CO_2$-Ar-He, $CO_2$-Kr-He, and $CO_2$-Xe-He can be seen in Figs. 3, 5, and 6, respectively.

Continuing with $CO_2$-Ne-He as the example, we were able to assign about 18 lines, though some were very weak and perhaps a bit uncertain. In trying to assign and fit the lines, we faced a challenge: the light helium atom and its very weak interactions with $CO_2$ and Ne mean that we expect large amplitude motions and significant centrifugal distortion effects. But allowing many distortion parameters to vary in both the ground and excited states might be dangerous, considering the limited number of assigned lines.

There is a further complication, namely the possibility of He atom tunneling. In the case of $CO_2$-$He_2$,[22] there is significant tunneling involving motion of the He atoms around the 'equator' of the $CO_2$ through a transition state with a linear He-C-He configuration. The tunneling barrier is not large since the He-He attraction is weak and the He mass is small. The result is a relatively large tunneling splitting, calculated to be about 0.50 cm$^{-1}$ for $CO_2$-$He_2$.[22] Rotational levels of the ground tunneling state ($v_t = 0$) have even values of $K_a''$, while those of the first excited tunneling state ($v_t = 1$, at 0.5 cm$^{-1}$) have odd values of $K_a''$. The two tunneling states have significantly different rotational constants (since their vibrational wave functions are different), for example $A \approx 0.30$ cm$^{-1}$ for $v_t = 0$ and $A \approx 0.21$ cm$^{-1}$ for $v_t = 1$, so that different rotational constants are needed for even and odd $K_a''$ transitions.[22] Of course for the present case of $CO_2$-Ne-He, the tunneling barrier will certainly be larger. But the possibility of needing different rotational constants for $K_a'' =$ even and odd transitions further challenged our assignment and fitting of the spectrum.

Table 6. Molecular parameters for $CO_2$-Rg-He trimers (in $cm^{-1}$).[a]

|  | $CO_2$-Ne-He $c$-type | $CO_2$-Ar-He ($c$-type) | $CO_2$-Kr-He $c$-type | $CO_2$-Xe-He $c$-type |
|---|---|---|---|---|
| $\nu_0$ | 2349.3975(4) | 2348.7877(4) | 2348.3750(5) | 2347.7889(2) |
| $A'$ | 0.21051(16) | 0.20392(10) | 0.20044(10) | 0.197344(61) |
| $A''$ | 0.21490(18) | 0.20564(12) | 0.20226(14) | 0.199258(74) |
| $(B + C)/2$ | 0.089301(61) | 0.056473(45) | 0.03968(11) | 0.031440(25) |
| $(B - C)$ | 0.00042(9) | 0.00001(6) | 0.00005(12) | 0.000111(31) |
| n | 18 | 33 | 13 | 32 |
| rmsd | 0.0010 | 0.0013 | 0.0008 | 0.0008 |

[a] Quantities in parentheses are $1\sigma$ from the least-squares fit, in units of the last quoted digit. n is the number of assigned lines (some of which are blends of multiple transitions), and rmsd is the root mean square error of the fit.

We tried including various centrifugal distortion parameters as well as separately fitting $K_a''$ = even and odd transitions for $CO_2$-Ne-He and the other mixed trimers. Of course the fits improved as more parameters were varied, but at the same time our confidence in the results diminished! So in the end we decided to use highly constrained fits with few parameters, and these are the results given in Table 6. The values of $(B + C)/2$ and $(B - C)$ were constrained to be equal in the ground and upper vibrational states, while $A$ was allowed to vary separately in the two states. The constraint is not unreasonable, as demonstrated in Tables 1, 3, 4, and 5. Interestingly, all the $CO_2$-Rg-He trimers turn out to be (accidental) near-symmetric rotors. In all cases, the observed transitions appeared to be $c$-type. But since the values of $(B - C)$ were very small (insignificant for $CO_2$-Ar-He and $CO_2$-Kr-He) the distinction between $b$- and $c$-type transitions is not very meaningful.

## 3. Discussion

### 3.1. Combination bands

The combination bands observed here result in intermolecular mode frequency values for $CO_2\text{-}Ar_2$, $\text{-}Kr_2$, and $\text{-}Xe_2$ as given in Table 7, where they are compared to the previously measured intermolecular bending frequencies of the corresponding $CO_2\text{-}Rg$ dimers. (When we say 'intermolecular', each Rg atom is considered to be a 'molecule'.) Note that the trimer frequencies are consistently about 15% larger than the dimer ones. The trimer combination bands were observed to have $b$-type rotational selection rules, which tells us that the combination mode has $A_1$ symmetry (in the $C_{2v}$ point group appropriate for trimers with indistinguishable Rg atoms). Since the intramolecular fundamental ($CO_2$ $\nu_3$) is $B_1$, this means that the intermolecular mode must also be $B_1$ in order to get the $A_1$ combination band ($B_1 \otimes B_1 = A_1$).

Table 7. Comparison of observed intermolecular bending frequencies for $CO_2\text{-}Rg_2$ trimers and $CO_2\text{-}Rg$ dimers (in $cm^{-1}$).

| Rg | $CO_2\text{-}Rg_2$ | $CO_2\text{-}Rg$ [a] |
|---|---|---|
| Ne | | 17.716 |
| Ar | 32.246 | 27.818 |
| Kr | 33.829 | 29.429 |
| Xe | 34.703 | 30.574 |

[a] From Refs. 9, 12, 18, and 19, respectively.

$CO_2\text{-}Rg_2$ trimers have five intermolecular vibrational modes, described as follows:

1. $Rg_2\text{-}CO_2$ van der Waals stretch ($A_1$ symmetry);

2. Rg-Rg van der Waals stretch ($A_1$);

3. torsion, or asymmetric Rg-C-O bend ($A_2$);

4. $CO_2$ rock, or "$Rg_2$ flap",[20] or symmetric Rg-C-O bend ($B_1$);

5. $Rg_2$ rock, or asymmetric Rg-C stretch ($B_2$).

In Ref. 20 these modes correspond to coordinates $S_3$, $S_4$, $S_5$, $S_7$, and $S_9$, and in Ref. 21 they correspond to $R$, $\rho$, $\theta_x$, $\theta_y$, and $\chi$, respectively. There is only one possible $B_1$ mode, so it is evident that our observed combination bands must involve the $CO_2$ rocking motion. This is not surprising since the trimer rocking mode is analogous to the $CO_2$-Rg dimer bending mode, and they have similar values as noted above. Arguments based on $C_{2v}$ symmetry do not strictly apply to trimers with inequivalent Rg atoms, in which case the torsional mode could in principle also cause the observed combination bands. However there can be no doubt that that the $CO_2$ rock remains the real assignment. In the case of $CO_2$-$Ne_2$, a combination band has not yet been observed. Since the $CO_2$-$Ne_2$ $a$- and $b$-inertial axes are interchanged relative to the other trimers, we expect its analogous combination band to be $a$-type.

### 3.2. Vibrational shifts

Vibrational shifts observed for $CO_2$-$Rg_2$ trimers relate directly to the interesting question of non-additive intermolecular effects, as was carefully analyzed in the $CO_2$-$Ar_2$ paper of Sperhac et al. [21] The present results are listed in Table 8, where the first column gives the Rg atom, the second column gives (previously known) shifts for $CO_2$-Rg dimers relative to the free $CO_2$ molecule, and the third column gives the additional shift induced by adding the second Rg atom to form a trimer.

Table 8. Vibrational frequency shifts in $CO_2$-Rg dimers and $CO_2$-$Rg_2$ trimers (in $cm^{-1}$).

| Rg | $CO_2$-Rg minus $CO_2$ | $CO_2$-$Rg_2$ minus $CO_2$-Rg | Difference | $CO_2$-Rg-He minus $CO_2$-Rg |
|-----|-----|-----|-----|-----|
| Ne | +0.1364 | +0.1405 | +0.0041 | +0.1179 |
| Ar | -0.4706 | -0.4281 | +0.0425 | +0.1151 |
| Kr | -0.8847 | -0.8063 | +0.0784 | +0.1165 |
| Xe | -1.4719 | -1.3358 | +0.1361 | +0.1176 |

The shifts in columns two and three are similar, as expected since each Rg atom in a trimer occupies a position equivalent (relative to $CO_2$) to the position of the Rg atom in the dimer. The notable trend in Table 8 is that the incremental shifts in column three are always more positive (i.e. blue-shifted) compared to column two. This deviation from linearity (column three minus column two) is shown in column four of Table 8, and it is especially significant since it is precisely ($\approx$0.0002 $cm^{-1}$) determined and has a clear meaning. Sperhac et al.[21] argued that this quantity (in their case, the difference between the incremental shifts in $CO_2$-Ar and $CO_2$-$Ar_2$) is an essentially exact measure of differential three body effects (i.e. the *change* in three body effects between the $CO_2$ ground state and excited $\nu_3$ state). Their model calculation suggested that most of this effect was due to competing induced dipole-induced dipole and exchange quadrupole terms. But their quantitative agreement with experiment was not very good, and the need to integrate over the $CO_2$ $\nu_3$ motion was emphasized. When this was done in a subsequent *ab initio* calculation by Rak et al.,[30] much better agreement with experiment was indeed achieved. However, the authors pointed out that the agreement might have been fortuitous "in view of the fact that the effects of intermolecular vibrations were not included". In addition to these effects of zero-point intermolecular motion, the effects of the other intramolecular ($CO_2$) modes could also be significant, as shown for $\nu_1$ in the case of $CO_2$-He. [31,32] In any case, the new results reported here for $CO_2$-Ne$_2$, -Kr$_2$, and -Xe$_2$ provide further direct and straightforward tests for these calculations.

The final column of Table 8 shows the incremental vibrational shift resulting from addition of a helium atom to a $CO_2$-Rg dimer to form a $CO_2$-Rg-He trimer, and we see that the shift is almost the same ($\approx$+0.117 cm$^{-1}$) for all the mixed trimers. This value is somewhat greater than the shift of +0.095 cm$^{-1}$ reported by Weida et al. for $CO_2$-He relative to $CO_2$ itself. But in a reanalysis (unpublished), we obtain a more similar shift of +0.116 cm$^{-1}$ for $^{12}C^{16}O_2$-He (see also Ref. 4 for other $CO_2$ isotopologues of $CO_2$-He).

### 3.3. Structures

Determining precise geometrical structures for weakly-bound molecular complexes can be a frustrating or even meaningless task because of their large-amplitude intermolecular motions, but of course we still like to try! Given the $C_{2v}$ symmetry of the $CO_2$-Rg$_2$ trimers, only two parameters are required to specify an equilibrium structure: the Rg-Rg distance, $r$(Rg-Rg), and the C-Rg distance, $r$(C-Rg). Alternately, one can use the $CO_2$ to Rg$_2$ center of mass distance, $R_{c.m.}$, since $r(C\text{-}Rg)^2 = R_{c.m.}^2 + (r(\text{Rg-Rg})/2)^2$. The trimer structures are thus over-determined since it is not possible to exactly match three rotational parameters by varying two structural parameters. In Table 9, we show the results of two-parameter least-squares fits to obtain $r$(Rg-Rg) and $r$(C-Rg), made using the simple and effective strfit and pmifst programs of Z. Kisiel.[33] As expected, the present structural parameters for $CO_2$-Ar$_2$ are very similar to those of Refs. 20 and 21.

Table 9. Comparison of structural parameters in $CO_2$-$Rg_2$, $CO_2$-Rg, and $Rg_2$ (lengths in Å, angles in degrees).

| Rg | $CO_2$-$Rg_2$ $r_0$(C-Rg) | $CO_2$-$Rg_2$ $r_0$(Rg-Rg) | $CO_2$-$Rg_2$ $r_0$(C-Rg) | $CO_2$-$Rg_2$ $r_0$(Rg-Rg) | $CO_2$-$Rg_2$ $\theta_y$ [a] | $CO_2$-Rg $r_0$(C-Rg)[b] | $Rg_2$ $r_0$(Rg-Rg) | $Rg_2$ $r_e$(Rg-Rg)[c] |
|---|---|---|---|---|---|---|---|---|
| | 2 parameter | | 3 parameter | | | | | |
| Ne | 3.276 | 3.295 | 3.268 | 3.307 | 11.5 | 3.290 | 3.337 [d] | 3.084 |
| Ar | 3.504 | 3.839 | 3.499 | 3.843 | 11.2 | 3.504 | 3.827 [e] | 3.755 |
| Kr | 3.629 | 4.066 | 3.623 | 4.069 | 13.2 | 3.624 | 4.059 [f] | 4.012 |
| Xe | 3.828 | 4.436 | 3.823 | 4.437 | 12.3 | 3.815 | | 4.375 |

[a] Deviation from 90° of the angle ($Rg_2$ c.m.-C-O); $Rg_2$ c.m. = $Rg_2$ center of mass.

[b] Experimental $r_0$ values from Randall, Walsh, and Howard.[5]

[c] Theoretical $r_e$ from Deiters and Sadus.[34]

[d] Experimental $r_0$ for $Ne_2$ from Wüest and Merkt.[35]

[e] Experimental $r_0$ for $Ar_2$ from Mizuse, et al.[36]

[f] Experimental $r_0$ for $Kr_2$ from LaRocque et al.[37]

To extend their two-parameter $CO_2$-$Ar_2$ fit, Sperhac et al.[21] proceeded to make an effective three-parameter fit by varying the parameter $\chi$ in addition to $r$(Rg-Rg) and $r$(C-Rg). $\chi$ is the angle describing the intermolecular $Rg_2$ rocking mode, or, more precisely, the deviation from 90° of the angle subtended by an Ar atom, the $Ar_2$ center of mass, and the C atom. But we think the description[21] of their three-parameter fit must be incorrect. It is clear from their Table III that varying $\chi$ affects only the $P_x$ and $P_y$ planar moments. $P_x$ is also sensitive to $r$(Rg-Rg) and $P_y$ to $r$(C-Rg). Thus $\chi$ cannot improve the result already achieved in a two-parameter fit, since $P_z$ is not influenced by $r$(Rg-Rg), $r$(C-Rg) or $\chi$. Rather, it is necessary to vary one of the remaining structural parameters: $\theta_x$ (which we call torsion above), or $\theta_y$ (which we call $CO_2$ rock). It appears that Sperhac et al. actually varied $\theta_y$ in their three-parameter fit, and this is also what we have varied for our three-parameter fits (Table 9). Our $\theta_y$ value of 11.2° for $CO_2$-$Ar_2$ is almost the same as the 11.4° value of Sperhac et al., which they labeled as $\chi$.

The angle $\theta_y$ is a logical choice for the third parameter, since it is similar to the angle $\theta$ which is varied in two-parameter fits of $CO_2$-Rg dimers.[5] We see in Table 9 that $\theta_y$ has fairly similar values (11.2° to 13.2°) for all four $CO_2$-$Rg_2$ trimers, and that $r$(Rg-Rg) and $r$(C-Rg) do not change very much between the two- and three-parameter fits. In the case of the $CO_2$-Rg dimers, $\theta$ has values of 6.6° to 8.6°.[5] But in a trimer, $\theta$ (deviation from 90° of angle Rg-C-O) and $\theta_y$ (deviation from 90° of angle $Rg_2$ c.m.-C-O) are not quite the same geometrically, with $\theta$ necessarily being smaller than $\theta_y$. So the actual trimer $\theta$-values range from about 9° to 11°, which is really quite similar to the dimer values. Of course these effective angles do not correspond to equilibrium structures (where $\theta$ and $\theta_y$ equal zero). Rather they provide measures of the rms value of the particular coordinate averaged over its zero-point wavefunction. For $CO_2$-Rg dimers, with just two intermolecular coordinates, the effective values of $R$ and $\theta$ are relatively meaningful and easy to interpret. For $CO_2$-$Rg_2$ trimers with five intermolecular coordinates, the meaning of a three-parameter fit is not so clear since the effects of the remaining two coordinates have necessarily been ignored.

The $r$(C-Rg) bond length values for $CO_2$-$Rg_2$ trimers in Table 9 are very similar to the corresponding $CO_2$-Rg dimer values, as expected. For a comparison of Rg-Rg bond lengths, we have experimental $r_0$(Rg-Rg) (zero-point average) values from rotationally resolve electronic spectra of $Ne_2$, $Ar_2$ and $Kr_2$, and these compare quite well with our $CO_2$-$Rg_2$ trimer results. Rotationally resolved $Xe_2$ spectra are not available, so we only have theoretical $r_e$(Rg-Rg) (equilibrium) values for comparison, and these of course are systematically smaller than the $CO_2$-$Rg_2$ trimer $r_0$(Rg-Rg) results.

## 4. Conclusions

We have studied high-resolution spectra of $CO_2$-$Rg_2$ trimers (Rg = Ne, Ar, Kr, Xe,) as observed in the $CO_2$ $\nu_3$ fundamental band region ($\approx$2350 cm$^{-1}$), thus extending previous work on $CO_2$-$Ar_2$.[20,21] The precisely measured vibrational frequency shifts and intermolecular $CO_2$ rocking mode frequencies provide sensitive new tests for intermolecular potential energy calculations. When compared to the vibrational shifts in the $CO_2$-Rg, the measured vibrational shifts in the $CO_2$-$Rg_2$ provide precise and straightforward measure of non-additive three body effects. As well, the reported rotational constants should aid the possible detection of pure rotational microwave spectra for Rg = Ne, Kr, and Xe, though the latter two will be complicated by myriad isotopic species. We are continuing to analyze spectra of larger $CO_2$-$Rg_n$ clusters and have recently reported the observation of highly symmetric structures for $CO_2$-$Ar_{15}$ and $CO_2$-$Ar_{17}$, with the latter marking the completion of the first solvation shell for $CO_2$ in Ar.[38]

**Supplementary Information**

Supplementary Information includes tables giving observed and fitted line positions for $CO_2$-$Rg_2$ trimers and isotope specific parameters for $CO_2$-$Kr_2$ and $CO_2$-$Xe_2$.

**Acknowledgements**

The financial support of the Natural Sciences and Engineering Research Council of Canada is gratefully acknowledged.

Appendix to:

Table A-1. Krypton isotope atomic mass and abundance

| atomic mass number N | atomic mass (Dalton) | abundance |
|---|---|---|
| 80 | 79.916 | 0.0229 |
| 82 | 81.913 | 0.1159 |
| 83 | 82.914 | 0.1150 |
| 84 | 83.912 | 0.5699 |
| 86 | 85.911 | 0.1728 |

Table A-2. Krypton isotope atomic masses, abundance, and rigid model scaling factors for $CO_2$-$Kr_2$

| atomic mass numbers N | abundance | $F_A$ | $F_B$ | $F_C$ |
|---|---|---|---|---|
| 80-80 | 0.00052 | 1.009116 | 1.046920 | 1.037303 |
| 80-82 | 0.00530 | 1.006809 | 1.034873 | 1.027780 |
| 80-83 | 0.00526 | 1.005713 | 1.029007 | 1.023142 |
| 80-84 | 0.02605 | 1.004656 | 1.023269 | 1.018605 |
| 80-86 | 0.00790 | 1.002643 | 1.012085 | 1.009757 |
| 82-82 | 0.01344 | 1.004451 | 1.022927 | 1.018282 |
| 82-83 | 0.02666 | 1.003329 | 1.017110 | 1.013656 |
| 82-84 | 0.13213 | 1.002249 | 1.011418 | 1.009130 |
| 82-86 | 0.04006 | 1.000188 | 1.000325 | 1.000304 |
| 83-83 | 0.01323 | 1.002196 | 1.011315 | 1.009036 |
| 83-84 | 0.13107 | 1.001103 | 1.005647 | 1.004515 |
| 83-86 | 0.03974 | 0.999021 | 0.994597 | 0.995700 |
| 84-84 | 0.32475 | 1.000000 | 1.000000 | 1.000000 |
| 84-86 | 0.19694 | 0.997895 | 0.988992 | 0.991195 |
| 86-86 | 0.02986 | 0.995749 | 0.978066 | 0.982409 |

Table A-3. Molecular parameters for $CO_2$-$Kr_2$ (in $cm^{-1}$). See Table 4 of the paper for 84-84.
These are the more abundant species. Others can easily be calculated using the scaling factors in Table A-2.

|  | 80-84 | 82-83 | 82-84 | 82-86 | 83-84 | 83-86 | 84-86 | 86-86 |
|---|---|---|---|---|---|---|---|---|
| Ab | 0.02605 | 0.02666 | 0.13213 | 0.04006 | 0.13107 | 0.03974 | 0.19694 | 0.02986 |
| $\nu_{0C}$ | 2381.2853 | 2381.2842 | 2381.2831 | 2381.2809 | 2381.2820 | 2381.2798 | 2381.2787 | 2381.2765 |
| $A'$ | 0.045369 | 0.045309 | 0.045260 | 0.045167 | 0.045208 | 0.045114 | 0.045063 | 0.044967 |
| $B'$ | 0.023464 | 0.023323 | 0.023192 | 0.022938 | 0.023060 | 0.022807 | 0.022678 | 0.022428 |
| $C'$ | 0.016839 | 0.016758 | 0.016683 | 0.016537 | 0.016606 | 0.016461 | 0.016386 | 0.016241 |
| $\nu_{0F}$ | 2347.4519 | 2347.4519 | 2347.4519 | 2347.4519 | 2347.4519 | 2347.4519 | 2347.4519 | 2347.4519 |
| $A'$ | 0.047434 | 0.047372 | 0.047321 | 0.047223 | 0.047266 | 0.047168 | 0.047115 | 0.047014 |
| $B'$ | 0.023453 | 0.023312 | 0.023181 | 0.022927 | 0.023049 | 0.022796 | 0.022667 | 0.022417 |
| $C'$ | 0.017006 | 0.016923 | 0.016848 | 0.016700 | 0.016771 | 0.016623 | 0.016548 | 0.016402 |
| $A''$ | 0.047475 | 0.047413 | 0.047362 | 0.047264 | 0.047308 | 0.047209 | 0.047156 | 0.047055 |
| $B''$ | 0.023474 | 0.023333 | 0.023203 | 0.022948 | 0.023070 | 0.022817 | 0.022688 | 0.022437 |
| $C''$ | 0.017002 | 0.016919 | 0.016844 | 0.016696 | 0.016767 | 0.016620 | 0.016544 | 0.016398 |

Table A-4. Xenon isotope atomic mass and abundance

| atomic mass number N | atomic mass (Dalton) | abundance |
|---|---|---|
| 129 | 128.905 | 0.264 |
| 130 | 129.904 | 0.041 |
| 131 | 130.905 | 0.212 |
| 132 | 131.904 | 0.269 |
| 134 | 133.905 | 0.104 |
| 136 | 135.907 | 0.089 |

Table A-5. Xenon isotope atomic masses, abundance, and rigid model scaling factors for $CO_2-Xe_2$

| atomic mass numbers N | abundance | $F_A$ | $F_B$ | $F_C$ |
|---|---|---|---|---|
| 129-129 | 0.06970 | 1.001996 | 1.014999 | 1.012514 |
| 129-130 | 0.02150 | 1.001495 | 1.011224 | 1.009370 |
| 129-131 | 0.11211 | 1.001001 | 1.007489 | 1.006256 |
| 129-132 | 0.14208 | 1.000518 | 1.003814 | 1.003190 |
| 129-134 | 0.05510 | 0.999576 | 0.996599 | 0.997162 |
| 129-136 | 0.04677 | 0.998666 | 0.989570 | 0.991282 |
| 130-130 | 0.00166 | 1.000992 | 1.007454 | 1.006226 |
| 130-131 | 0.01729 | 1.000497 | 1.003725 | 1.003113 |
| 130-132 | 0.02191 | 1.000012 | 1.000054 | 1.000046 |
| 130-134 | 0.00850 | 0.999066 | 0.992848 | 0.994020 |
| 130-136 | 0.00721 | 0.998154 | 0.985829 | 0.988141 |
| 131-131 | 0.04508 | 1.000000 | 1.000000 | 1.000000 |
| 131-132 | 0.11427 | 0.999514 | 0.996334 | 0.996934 |
| 131-134 | 0.04431 | 0.998565 | 0.989138 | 0.990909 |
| 131-136 | 0.03761 | 0.997650 | 0.982127 | 0.985030 |
| 132-132 | 0.07241 | 0.999025 | 0.992673 | 0.993869 |
| 132-134 | 0.05616 | 0.998074 | 0.985486 | 0.987845 |
| 132-136 | 0.04767 | 0.997156 | 0.978484 | 0.981967 |
| 134-134 | 0.01089 | 0.997117 | 0.978316 | 0.981822 |
| 134-136 | 0.01849 | 0.996192 | 0.971332 | 0.975946 |
| 136-136 | 0.00785 | 0.995263 | 0.964364 | 0.970071 |

Table A-6. Molecular parameters for $CO_2$-$Xe_2$ (in cm$^{-1}$). See Table 5 of the paper for 131-131. These are the 7 most abundant species. Others can easily be calculated using the scaling factors in Table A-5.

|          | 129-129   | 129-131   | 129-132   | 129-134   | 131-132   | 132-132   | 132-134   |
|----------|-----------|-----------|-----------|-----------|-----------|-----------|-----------|
| Ab       | 0.06970   | 0.11211   | 0.14208   | 0.05510   | 0.11427   | 0.07241   | 0.05616   |
| $\nu_{0C}$ | 2381.0416 | 2381.0401 | 2381.0393 | 2381.0378 | 2381.0377 | 2381.0369 | 2381.0354 |
| $A'$     | 0.040778  | 0.040737  | 0.040718  | 0.040679  | 0.040677  | 0.040657  | 0.040618  |
| $B'$     | 0.012968  | 0.012873  | 0.012826  | 0.012733  | 0.012730  | 0.012683  | 0.012591  |
| $C'$     | 0.010311  | 0.010247  | 0.010216  | 0.010514  | 0.010152  | 0.010121  | 0.010060  |
| $\nu_{0F}$ | 2346.3355 | 2346.3355 | 2346.3355 | 2346.3355 | 2346.3355 | 2346.3355 | 2346.3355 |
| $A'$     | 0.041384  | 0.041343  | 0.041323  | 0.041284  | 0.041282  | 0.041262  | 0.041222  |
| $B'$     | 0.012971  | 0.012875  | 0.012828  | 0.012735  | 0.012732  | 0.012685  | 0.012593  |
| $C'$     | 0.010370  | 0.010306  | 0.010274  | 0.010213  | 0.010210  | 0.010179  | 0.010117  |
| $A''$    | 0.041411  | 0.041370  | 0.041350  | 0.041311  | 0.041309  | 0.041289  | 0.041249  |
| $B''$    | 0.012965  | 0.012869  | 0.012822  | 0.012730  | 0.012727  | 0.012680  | 0.012588  |
| $C''$    | 0.010372  | 0.010308  | 0.010276  | 0.010215  | 0.010212  | 0.010181  | 0.010119  |

Table A-7. Observed and calculated line positions for $CO_2$-$^{20}Ne_2$ and $CO_2$-$^{20}Ne$-$^{22}Ne$ (in cm$^{-1}$).

```
***********************************************************
  J'   Ka'  Kc'   J"   Ka"  Kc"    Obs         Calc        O-C
***********************************************************
CO2-20Ne2
   5    5    1     6    6    1   2348.1146    2348.1145
   5    5    0     6    6    0   2348.1146    2348.1146
                              Blend          2348.1145     0.0001
   5    3    3     6    4    3   2348.2615    2348.2618   -0.0004
   5    1    4     6    2    4   2348.3609    2348.3605    0.0004
   4    1    4     5    2    4   2348.4960    2348.4958    0.0002
   3    3    1     4    4    1   2348.5620    2348.5613
   3    3    0     4    4    0   2348.5620    2348.5625
                              Blend          2348.5619     0.0001
   3    1    3     4    2    3   2348.6718    2348.6720   -0.0002
   3    1    2     4    2    2   2348.7140    2348.7140    0.0000
   2    1    2     3    2    2   2348.8414    2348.8415    0.0000
   2    1    1     3    2    1   2348.8703    2348.8703    0.0000
   1    1    0     2    2    0   2349.0160    2349.0156    0.0004
   2    1    2     2    2    0   2349.2578    2349.2577    0.0001
   2    1    1     2    2    1   2349.3230    2349.3232   -0.0002
   3    1    2     3    2    2   2349.3457    2349.3461   -0.0003
   4    1    3     4    2    3   2349.3709    2349.3707    0.0002
   5    1    4     5    2    4   2349.3909    2349.3906    0.0003
   7    1    6     7    2    6   2349.4065    2349.4065   -0.0001
   4    1    4     4    0    4   2349.4206    2349.4208   -0.0002
   3    1    3     3    0    3   2349.4270    2349.4271   -0.0001
   2    1    2     2    0    2   2349.4380    2349.4381   -0.0001
   1    1    1     1    0    1   2349.4517    2349.4516    0.0001
   5    3    3     5    2    3   2349.5303    2349.5308   -0.0005
   4    3    2     4    2    2   2349.5693    2349.5692    0.0001
   1    1    0     0    0    0   2349.6121    2349.6122   -0.0001
   2    1    1     1    0    1   2349.7719    2349.7720   -0.0001
   3    3    0     2    2    0   2350.0468    2350.0464    0.0004
   4    1    3     3    0    3   2350.1259    2350.1258    0.0001
   4    3    1     3    2    1   2350.1813    2350.1811    0.0001
   5    1    4     4    0    4   2350.3162    2350.3156
   5    3    2     4    2    2   2350.3162    2350.3166
                              Blend          2350.3162     0.0000
   5    3    3     4    2    3   2350.3568    2350.3570   -0.0002
   6    1    5     5    0    5   2350.5046    2350.5046    0.0000
   7    3    5     6    2    5   2350.6893    2350.6891    0.0002
***********************************************************
CO2-20Ne-22Ne
   4    4    1     5    5    1   2348.3700    2348.3699
   4    4    0     5    5    0   2348.3700    2348.3702
                              Blend          2348.3700     0.0000
   4    3    1     5    4    1   2348.4517    2348.4514    0.0004
   4    2    2     5    3    2   2348.5375    2348.5373
   4    0    4     5    1    4   2348.5375    2348.5374
                              Blend          2348.5374     0.0001
   3    3    1     4    4    1   2348.5882    2348.5877
   3    3    0     4    4    0   2348.5882    2348.5891
                              Blend          2348.5884    -0.0002
   3    2    2     4    3    2   2348.6550    2348.6551   -0.0001
   3    2    1     4    3    1   2348.6747    2348.6749   -0.0002
   3    1    3     4    2    3   2348.6927    2348.6927   -0.0001
   3    0    3     4    1    3   2348.7252    2348.7253   -0.0001
   2    1    1     3    2    1   2348.8868    2348.8867    0.0001
   2    0    2     3    1    2   2348.9073    2348.9073    0.0000
   1    0    1     2    1    1   2349.0767    2349.0766    0.0001
   2    1    1     2    2    1   2349.3298    2349.3297    0.0001
```

```
    3    1    2         3    2    2    2349.3527    2349.3525     0.0002
    5    1    4         5    2    4    2349.3944    2349.3947    -0.0003
    4    2    3         4    1    3    2349.4567    2349.4566     0.0001
************************************************************
```

Table A-8. Observed and calculated line positions for $CO_2$-$Ar_2$ fundamental band (in cm$^{-1}$).

```
************************************************************
   J'   Ka'  Kc'    J"   Ka"  Kc"    Obs          Calc         O-C
************************************************************
   10    7    4        11    8    4    2347.0459    2347.0456     0.0003
    9    9    0        10   10    0    2347.0733    2347.0736    -0.0003
    9    8    1        10    9    1    2347.1064    2347.1064    -0.0001
    8    8    1         9    9    1    2347.1930    2347.1928     0.0002
    9    5    4        10    6    4    2347.1960    2347.1960     0.0000
    8    7    2         9    8    2    2347.2244    2347.2248    -0.0004
    8    6    3         9    7    3    2347.2532    2347.2535    -0.0003
    7    6    1         8    7    1    2347.3462    2347.3460     0.0003
    7    5    2         8    6    2    2347.3845    2347.3850
    7    2    5         8    3    5    2347.3845    2347.3834
                              Blend              2347.3842     0.0002
    7    3    4         8    4    4    2347.3950    2347.3953    -0.0003
    7    4    3         8    5    3    2347.4059    2347.4054     0.0005
    6    5    2         7    6    2    2347.4610    2347.4609     0.0001
    6    4    3         7    5    3    2347.4821    2347.4819     0.0002
    6    3    4         7    4    4    2347.4888    2347.4890
    6    2    5         7    3    5    2347.4888    2347.4884
                              Blend              2347.4887     0.0001
    5    5    0         6    6    0    2347.5504    2347.5503     0.0002
    5    0    5         6    1    5    2347.5959    2347.5957     0.0002
    5    1    4         6    2    4    2347.5984    2347.5988    -0.0003
    4    4    1         5    5    1    2347.6677    2347.6678    -0.0001
    4    2    3         5    3    3    2347.7034    2347.7030
    4    1    4         5    2    4    2347.7034    2347.7039
                              Blend              2347.7033     0.0001
    3    2    1         4    3    1    2347.8184    2347.8187    -0.0003
    3    1    2         4    2    2    2347.8235    2347.8235     0.0000
    2    2    1         3    3    1    2347.9020    2347.9020     0.0001
    2    1    2         3    2    2    2347.9171    2347.9169     0.0002
    1    1    0         2    2    0    2348.0238    2348.0236     0.0002
    5    2    3         5    3    3    2348.2323    2348.2320
    7    3    4         7    4    4    2348.2323    2348.2322
    3    1    2         3    2    2    2348.2323    2348.2326
                              Blend              2348.2323     0.0000
    1    0    1         1    1    1    2348.2356    2348.2355     0.0001
    5    0    5         5    1    5    2348.2432    2348.2437
    3    0    3         3    1    3    2348.2432    2348.2437
    6    1    6         6    0    6    2348.2432    2348.2435
    4    1    4         4    0    4    2348.2432    2348.2440
    8    1    8         8    0    8    2348.2432    2348.2429
    5    1    4         5    2    4    2348.2432    2348.2426
    6    2    5         6    1    5    2348.2432    2348.2429
    7    1    6         7    2    6    2348.2432    2348.2424
    7    0    7         7    1    7    2348.2432    2348.2432
                              Blend              2348.2433    -0.0001
    2    2    1         2    1    1    2348.2708    2348.2706
    6    4    3         6    3    3    2348.2708    2348.2695
    4    3    2         4    2    2    2348.2708    2348.2724
                              Blend              2348.2710    -0.0002
    2    2    1         1    1    1    2348.4707    2348.4711    -0.0004
    3    2    1         2    1    1    2348.5593    2348.5596    -0.0003
```

```
    3   1   2       2   0   2   2348.5619   2348.5624   -0.0005
    3   3   0       2   2   0   2348.5792   2348.5796   -0.0004
    4   2   3       3   1   3   2348.6763   2348.6760    0.0002
    4   3   2       3   2   2   2348.6817   2348.6815    0.0002
    5   3   2       4   2   2   2348.7665   2348.7664    0.0001
    5   2   3       4   1   3   2348.7757   2348.7756    0.0001
    5   1   4       4   0   4   2348.7815   2348.7827
    5   4   1       4   3   1   2348.7815   2348.7807
                            Blend   2348.7814    0.0002
    5   5   0       4   4   0   2348.8149   2348.8147    0.0002
    6   4   3       5   3   3   2348.8911   2348.8914
    6   2   5       5   1   5   2348.8911   2348.8909
                            Blend   2348.8912   -0.0001
    6   5   2       5   4   2   2348.9063   2348.9062    0.0001
    6   6   1       5   5   1   2348.9331   2348.9326    0.0005
    7   4   3       6   3   3   2348.9744   2348.9744   -0.0001
    7   5   2       6   4   2   2348.9813   2348.9815   -0.0002
    7   3   4       6   2   4   2348.9884   2348.9883    0.0000
    7   2   5       6   1   5   2348.9959   2348.9958    0.0001
    7   6   1       6   5   1   2349.0148   2349.0150   -0.0003
    7   7   0       6   6   0   2349.0490   2349.0484    0.0006
    8   5   4       7   4   4   2349.1008   2349.1009
    8   4   5       7   3   5   2349.1008   2349.1010
                            Blend   2349.1009   -0.0001
    8   6   3       7   5   3   2349.1117   2349.1117    0.0000
    8   7   2       7   6   2   2349.1353   2349.1354   -0.0001
    9   6   3       8   5   3   2349.1831   2349.1825
    9   5   4       8   4   4   2349.1831   2349.1836
                            Blend   2349.1830    0.0000
    9   8   1       8   7   1   2349.2491   2349.2495   -0.0004
********************************************************************
```

Table A-9. Observed and calculated line positions for $CO_2$-$Ar_2$ combination band (in cm$^{-1}$).

```
********************************************************************
  J'  Ka' Kc'    J"  Ka" Kc"    Obs         Calc        O-C
********************************************************************
    7   7   1       8   8   0   2379.4597   2379.4597   -0.0001
    7   6   2       8   7   1   2379.5255   2379.5255    0.0000
    7   5   3       8   6   2   2379.5533   2379.5542   -0.0008
    6   6   0       7   7   1   2379.6060   2379.6064   -0.0004
   10   2   8      11   3   9   2379.6270   2379.6276   -0.0006
    6   5   1       7   6   2   2379.6832   2379.6833   -0.0001
    5   5   1       6   6   0   2379.7461   2379.7457    0.0005
    5   3   3       6   4   2   2379.7634   2379.7633    0.0001
    7   4   4       8   3   5   2379.8015   2379.8017
    4   1   3       5   4   2   2379.8015   2379.8024
                            Blend   2379.8018   -0.0004
    8   1   7       9   2   8   2379.8197   2379.8199   -0.0002
    7   3   5       8   2   6   2379.8440   2379.8443   -0.0003
    6   3   3       7   4   4   2379.8596   2379.8596    0.0000
    8   0   8       9   1   9   2379.8683   2379.8682    0.0001
    4   4   0       5   5   1   2379.8870   2379.8873   -0.0003
    7   2   6       8   1   7   2379.8906   2379.8908   -0.0002
    6   2   4       7   3   5   2379.9135   2379.9138   -0.0003
    7   1   7       8   0   8   2379.9386   2379.9386    0.0000
    6   1   5       7   2   6   2379.9610   2379.9607    0.0003
    5   4   2       6   3   3   2379.9727   2379.9730   -0.0003
    5   3   3       6   2   4   2379.9873   2379.9875   -0.0002
    3   3   1       4   4   0   2380.0090   2380.0104
    6   0   6       7   1   7   2380.0090   2380.0082
                            Blend   2380.0091   -0.0001
```

| | | | | | | | | |
|---|---|---|---|---|---|---|---|---|
| 5 | 2 | 4 | 6 | 1 | 5 | 2380.0297 | 2380.0298 | −0.0001 |
| 4 | 2 | 2 | 5 | 3 | 3 | 2380.0389 | 2380.0394 | −0.0004 |
| 5 | 1 | 5 | 6 | 0 | 6 | 2380.0767 | 2380.0768 | −0.0001 |
| 4 | 1 | 3 | 5 | 2 | 4 | 2380.0969 | 2380.0969 | 0.0000 |
| 4 | 0 | 4 | 5 | 1 | 5 | 2380.1443 | 2380.1444 | −0.0001 |
| 2 | 2 | 0 | 3 | 3 | 1 | 2380.1539 | 2380.1541 | −0.0002 |
| 3 | 2 | 2 | 4 | 1 | 3 | 2380.1693 | 2380.1694 | −0.0001 |
| 8 | 2 | 6 | 8 | 3 | 5 | 2380.2056 | 2380.2055 | 0.0001 |
| 3 | 1 | 3 | 4 | 0 | 4 | 2380.2110 | 2380.2112 | −0.0002 |
| 2 | 1 | 1 | 3 | 2 | 2 | 2380.2183 | 2380.2184 | |
| 6 | 0 | 6 | 6 | 1 | 5 | 2380.2183 | 2380.2180 | |
| | | | | | Blend | 2380.2183 | 2380.2183 | 0.0000 |
| 1 | 1 | 1 | 2 | 2 | 0 | 2380.2478 | 2380.2481 | −0.0003 |
| 7 | 3 | 5 | 7 | 4 | 4 | 2380.2564 | 2380.2560 | 0.0004 |
| 6 | 1 | 5 | 6 | 2 | 4 | 2380.2663 | 2380.2661 | 0.0002 |
| 5 | 1 | 5 | 5 | 2 | 4 | 2380.2701 | 2380.2697 | |
| 8 | 3 | 5 | 8 | 4 | 4 | 2380.2701 | 2380.2702 | |
| | | | | | Blend | 2380.2701 | 2380.2699 | 0.0001 |
| 2 | 0 | 2 | 3 | 1 | 3 | 2380.2758 | 2380.2759 | −0.0001 |
| 7 | 4 | 4 | 7 | 5 | 3 | 2380.2878 | 2380.2874 | 0.0004 |
| 6 | 5 | 1 | 6 | 6 | 0 | 2380.2947 | 2380.2945 | 0.0002 |
| 5 | 2 | 4 | 5 | 3 | 3 | 2380.3129 | 2380.3127 | 0.0002 |
| 5 | 4 | 2 | 5 | 5 | 1 | 2380.3198 | 2380.3193 | 0.0005 |
| 4 | 0 | 4 | 4 | 1 | 3 | 2380.3228 | 2380.3228 | 0.0001 |
| 6 | 2 | 4 | 6 | 3 | 3 | 2380.3294 | 2380.3293 | 0.0002 |
| 5 | 3 | 3 | 5 | 4 | 2 | 2380.3371 | 2380.3368 | 0.0003 |
| 1 | 1 | 1 | 2 | 0 | 2 | 2380.3464 | 2380.3463 | 0.0001 |
| 3 | 1 | 3 | 3 | 2 | 2 | 2380.3661 | 2380.3658 | 0.0003 |
| 6 | 4 | 2 | 6 | 5 | 1 | 2380.3781 | 2380.3779 | 0.0003 |
| 4 | 1 | 3 | 4 | 2 | 2 | 2380.3828 | 2380.3825 | |
| 3 | 2 | 2 | 3 | 3 | 1 | 2380.3828 | 2380.3830 | |
| | | | | | Blend | 2380.3828 | 2380.3827 | 0.0001 |
| 6 | 3 | 3 | 6 | 4 | 2 | 2380.3922 | 2380.3916 | |
| 4 | 3 | 1 | 4 | 4 | 0 | 2380.3922 | 2380.3923 | |
| | | | | | Blend | 2380.3922 | 2380.3919 | 0.0003 |
| 0 | 0 | 0 | 1 | 1 | 1 | 2380.4002 | 2380.4001 | 0.0001 |
| 4 | 2 | 2 | 4 | 3 | 1 | 2380.4244 | 2380.4243 | 0.0001 |
| 2 | 0 | 2 | 2 | 1 | 1 | 2380.4288 | 2380.4286 | 0.0001 |
| 2 | 1 | 1 | 2 | 2 | 0 | 2380.4501 | 2380.4500 | 0.0001 |
| 2 | 2 | 0 | 2 | 1 | 1 | 2380.5227 | 2380.5227 | 0.0000 |
| 4 | 3 | 1 | 4 | 2 | 2 | 2380.5423 | 2380.5421 | 0.0001 |
| 2 | 1 | 1 | 2 | 0 | 2 | 2380.5483 | 2380.5482 | |
| 4 | 4 | 0 | 4 | 3 | 1 | 2380.5483 | 2380.5480 | |
| | | | | | Blend | 2380.5481 | 2380.5481 | 0.0002 |
| 6 | 5 | 1 | 6 | 4 | 2 | 2380.5554 | 2380.5551 | 0.0003 |
| 8 | 6 | 2 | 8 | 5 | 3 | 2380.5652 | 2380.5661 | |
| 6 | 4 | 2 | 6 | 3 | 3 | 2380.5652 | 2380.5650 | |
| | | | | | Blend | 2380.5652 | 2380.5653 | −0.0001 |
| 4 | 2 | 2 | 4 | 1 | 3 | 2380.5829 | 2380.5829 | 0.0000 |
| 8 | 5 | 3 | 8 | 4 | 4 | 2380.5910 | 2380.5910 | |
| 9 | 8 | 2 | 9 | 7 | 3 | 2380.5910 | 2380.5906 | |
| | | | | | Blend | 2380.5909 | 2380.5909 | 0.0001 |
| 7 | 6 | 2 | 7 | 5 | 3 | 2380.5949 | 2380.5941 | |
| 5 | 4 | 2 | 5 | 3 | 3 | 2380.5949 | 2380.5949 | |
| | | | | | Blend | 2380.5946 | 2380.5946 | 0.0003 |
| 3 | 2 | 2 | 3 | 1 | 3 | 2380.5988 | 2380.5989 | −0.0002 |
| 9 | 7 | 3 | 9 | 6 | 4 | 2380.6092 | 2380.6088 | 0.0005 |
| 6 | 3 | 3 | 6 | 2 | 4 | 2380.6159 | 2380.6157 | 0.0002 |
| 10 | 6 | 4 | 10 | 5 | 5 | 2380.6183 | 2380.6183 | |
| 7 | 5 | 3 | 7 | 4 | 4 | 2380.6183 | 2380.6187 | |
| | | | | | Blend | 2380.6186 | 2380.6186 | −0.0003 |
| 4 | 1 | 3 | 4 | 0 | 4 | 2380.6370 | 2380.6369 | 0.0001 |
| 9 | 6 | 4 | 9 | 5 | 5 | 2380.6407 | 2380.6407 | 0.0000 |
| 8 | 4 | 4 | 8 | 3 | 5 | 2380.6446 | 2380.6450 | -0.0004 |
| 7 | 4 | 4 | 7 | 3 | 5 | 2380.6595 | 2380.6596 | 0.0000 |
| 6 | 2 | 4 | 6 | 1 | 5 | 2380.6685 | 2380.6684 | 0.0001 |
| 5 | 2 | 4 | 5 | 1 | 5 | 2380.6778 | 2380.6778 | -0.0001 |
| 3 | 1 | 3 | 2 | 0 | 2 | 2380.6955 | 2380.6956 | -0.0001 |
| 4 | 0 | 4 | 3 | 1 | 3 | 2380.7520 | 2380.7523 | |
| 3 | 2 | 2 | 2 | 1 | 1 | 2380.7520 | 2380.7517 | |
| | | | Blend | | | | 2380.7521 | -0.0001 |
| 4 | 1 | 3 | 3 | 2 | 2 | 2380.7913 | 2380.7916 | -0.0003 |
| 3 | 3 | 1 | 2 | 2 | 0 | 2380.8005 | 2380.8008 | -0.0003 |
| 5 | 1 | 5 | 4 | 0 | 4 | 2380.8099 | 2380.8098 | 0.0001 |
| 5 | 2 | 4 | 4 | 1 | 3 | 2380.8562 | 2380.8562 | 0.0000 |
| 6 | 0 | 6 | 5 | 1 | 5 | 2380.8662 | 2380.8660 | 0.0002 |
| 6 | 1 | 5 | 5 | 2 | 4 | 2380.9102 | 2380.9099 | 0.0003 |
| 5 | 3 | 3 | 4 | 2 | 2 | 2380.9171 | 2380.9169 | 0.0002 |
| 7 | 1 | 7 | 6 | 0 | 6 | 2380.9209 | 2380.9213 | |
| 4 | 4 | 0 | 3 | 3 | 1 | 2380.9209 | 2380.9202 | |
| | | | Blend | | | | 2380.9208 | 0.0001 |
| 6 | 2 | 4 | 5 | 3 | 3 | 2380.9509 | 2380.9512 | |
| 4 | 3 | 1 | 3 | 2 | 2 | 2380.9509 | 2380.9512 | |
| | | | Blend | | | | 2380.9512 | -0.0003 |
| 7 | 2 | 6 | 6 | 1 | 5 | 2380.9647 | 2380.9650 | -0.0004 |
| 8 | 0 | 8 | 7 | 1 | 7 | 2380.9758 | 2380.9757 | 0.0001 |
| 5 | 4 | 2 | 4 | 3 | 1 | 2380.9794 | 2380.9799 | -0.0005 |
| 8 | 1 | 7 | 7 | 2 | 6 | 2381.0187 | 2381.0188 | -0.0001 |
| 8 | 2 | 6 | 7 | 3 | 5 | 2381.0630 | 2381.0634 | -0.0004 |
| 9 | 2 | 8 | 8 | 1 | 7 | 2381.0715 | 2381.0718 | -0.0002 |
| 7 | 4 | 4 | 6 | 3 | 3 | 2381.0750 | 2381.0750 | 0.0000 |
| 8 | 3 | 5 | 7 | 4 | 4 | 2381.1075 | 2381.1071 | |
| 6 | 6 | 0 | 5 | 5 | 1 | 2381.1075 | 2381.1081 | |
| | | | Blend | | | | 2381.1078 | -0.0003 |
| 9 | 3 | 7 | 8 | 2 | 6 | 2381.1158 | 2381.1160 | -0.0002 |
| 6 | 5 | 1 | 5 | 4 | 2 | 2381.1291 | 2381.1286 | 0.0005 |
| 7 | 5 | 3 | 6 | 4 | 2 | 2381.1505 | 2381.1507 | -0.0002 |
| 7 | 6 | 2 | 6 | 5 | 1 | 2381.1956 | 2381.1946 | |
| 7 | 7 | 1 | 6 | 6 | 0 | 2381.1956 | 2381.1961 | |
| | | | Blend | | | | 2381.1955 | 0.0001 |
| 8 | 8 | 0 | 7 | 7 | 1 | 2381.2818 | 2381.2819 | -0.0001 |
| 9 | 9 | 1 | 8 | 8 | 0 | 2381.3621 | 2381.3623 | -0.0003 |
| 9 | 7 | 3 | 8 | 6 | 2 | 2381.3746 | 2381.3742 | 0.0004 |
| 9 | 8 | 2 | 8 | 7 | 1 | 2381.3809 | 2381.3803 | 0.0006 |
| 10 | 1 | 9 | 11 | 2 | 10 | 2379.6752 | 2379.6754 | -0.0002 |
| 9 | 1 | 9 | 10 | 0 | 10 | 2379.7970 | 2379.7968 | 0.0002 |
| 5 | 5 | 1 | 4 | 4 | 0 | 2381.0103 | 2381.0101 | 0.0003 |
| 9 | 1 | 9 | 8 | 0 | 8 | 2381.0292 | 2381.0291 | 0.0001 |
| 10 | 0 | 10 | 9 | 1 | 9 | 2381.0818 | 2381.0815 | 0.0003 |
| 10 | 1 | 9 | 9 | 2 | 8 | 2381.1238 | 2381.1238 | 0.0001 |
| 11 | 1 | 11 | 10 | 0 | 10 | 2381.1337 | 2381.1331 | 0.0006 |
| 11 | 2 | 10 | 10 | 1 | 9 | 2381.1745 | 2381.1748 | -0.0004 |

\*\*\*\*\*\*\*\*\*\*\*\*\*\*\*\*\*\*\*\*\*\*\*\*\*\*\*\*\*\*\*\*\*\*\*\*\*\*\*\*\*\*\*\*\*\*\*\*\*\*\*\*\*\*\*\*\*\*\*\*

Table A-10. Observed and calculated line positions for $CO_2$-$Kr_2$ fundamental band (in $cm^{-1}$).

```
*****************************************************************************
              J'  Ka'  Kc'   J"  Ka"  Kc"     Obs        Calc        O-C
*****************************************************************************
CO2-Kr2-84-84  9   4    5    10   5    5    2346.8152   2346.8128
CO2-Kr2-83-84  9   4    5    10   5    5    2346.8152   2346.8114
CO2-Kr2-83-84 10   1    9    11   2    9    2346.8152   2346.8150
CO2-Kr2-84-84  6   6    1     7   7    1    2346.8152   2346.8154
CO2-Kr2-84-86  9   4    5    10   5    5    2346.8152   2346.8156
CO2-Kr2-82-84  6   6    0     7   7    0    2346.8152   2346.8138
CO2-Kr2-82-84  6   6    1     7   7    1    2346.8152   2346.8138
CO2-Kr2-83-84  6   6    0     7   7    0    2346.8152   2346.8146
CO2-Kr2-83-84  6   6    1     7   7    1    2346.8152   2346.8146
CO2-Kr2-84-86  6   6    0     7   7    0    2346.8152   2346.8169
CO2-Kr2-84-86  6   6    1     7   7    1    2346.8152   2346.8169
                                    Blend              2346.8147    0.0004
CO2-Kr2-84-84  7   4    3     8   5    3    2346.8851   2346.8869
CO2-Kr2-84-84  9   1    8    10   2    8    2346.8851   2346.8853
                                    Blend              2346.8863   -0.0011
CO2-Kr2-84-84  5   5    0     6   6    0    2346.9119   2346.9106
CO2-Kr2-84-84  9   2    7    10   3    7    2346.9119   2346.9141
CO2-Kr2-84-84  8   2    7     9   3    7    2346.9119   2346.9140
CO2-Kr2-84-84  8   1    8     9   2    8    2346.9119   2346.9126
CO2-Kr2-84-86  5   5    0     6   6    0    2346.9119   2346.9119
CO2-Kr2-84-86  5   5    1     6   6    1    2346.9119   2346.9119
                                    Blend              2346.9124   -0.0006
CO2-Kr2-84-84  5   4    1     6   5    1    2346.9654   2346.9657
CO2-Kr2-84-86  5   4    1     6   5    1    2346.9654   2346.9672
CO2-Kr2-84-86  5   4    2     6   5    2    2346.9654   2346.9672
                                    Blend              2346.9663   -0.0009
CO2-Kr2-84-84  4   4    1     5   5    1    2347.0045   2347.0057
CO2-Kr2-84-86  4   4    0     5   5    0    2347.0045   2347.0068
CO2-Kr2-84-86  4   4    1     5   5    1    2347.0045   2347.0068
                                    Blend              2347.0061   -0.0016
CO2-Kr2-84-84  3   3    0     4   4    0    2347.1002   2347.1007
CO2-Kr2-84-86  3   3    0     4   4    0    2347.1002   2347.1016
CO2-Kr2-84-86  3   3    1     4   4    1    2347.1002   2347.1015
                                    Blend              2347.1010   -0.0008
CO2-Kr2-84-84  4   2    3     5   3    3    2347.1105   2347.1105
CO2-Kr2-84-84  5   0    5     6   1    5    2347.1105   2347.1118
CO2-Kr2-84-86  4   2    3     5   3    3    2347.1105   2347.1122
                                    Blend              2347.1112   -0.0008
CO2-Kr2-84-84  2   2    1     3   3    1    2347.1950   2347.1951
CO2-Kr2-84-86  2   2    0     3   3    0    2347.1950   2347.1967
CO2-Kr2-84-86  2   2    1     3   3    1    2347.1950   2347.1958
                                    Blend              2347.1955   -0.0005
CO2-Kr2-84-84  6   3    4     6   4    2    2347.2607   2347.2602
CO2-Kr2-84-84  4   3    2     4   4    0    2347.2607   2347.2608
CO2-Kr2-84-86  6   3    4     6   4    2    2347.2607   2347.2595
CO2-Kr2-84-86  5   3    3     5   4    1    2347.2607   2347.2602
                                    Blend              2347.2602    0.0004
CO2-Kr2-84-84  1   1    0     2   2    0    2347.2921   2347.2925
CO2-Kr2-84-86  1   1    0     2   2    0    2347.2921   2347.2930
                                    Blend              2347.2926   -0.0005
CO2-Kr2-84-84  1   0    1     2   1    1    2347.3360   2347.3361
CO2-Kr2-84-84  5   2    3     5   3    3    2347.3360   2347.3358
CO2-Kr2-84-86  1   0    1     2   1    1    2347.3360   2347.3369
                                    Blend              2347.3361   -0.0001
CO2-Kr2-84-84  2   1    2     2   2    0    2347.3594   2347.3592
CO2-Kr2-84-86  2   1    2     2   2    0    2347.3594   2347.3591
CO2-Kr2-84-84  6   4    3     7   3    5    2347.3594   2347.3611
```

| | | | | | | | | | |
|---|---|---|---|---|---|---|---|---|---|
| | | | | | | Blend | 2347.3596 | -0.0002 | |
| CO2-Kr2-84-84 | 7 | 2 | 5 | 7 | 3 | 5 | 2347.3657 | 2347.3666 | |
| CO2-Kr2-84-86 | 7 | 2 | 5 | 7 | 3 | 5 | 2347.3657 | 2347.3650 | |
| | | | | | | Blend | 2347.3663 | -0.0005 | |
| CO2-Kr2-84-86 | 8 | 2 | 6 | 8 | 3 | 6 | 2347.3827 | 2347.3828 | |
| CO2-Kr2-84-86 | 0 | 0 | 0 | 1 | 1 | 0 | 2347.3827 | 2347.3823 | |
| CO2-Kr2-82-84 | 0 | 0 | 0 | 1 | 1 | 0 | 2347.3827 | 2347.3816 | |
| CO2-Kr2-83-84 | 0 | 0 | 0 | 1 | 1 | 0 | 2347.3827 | 2347.3818 | |
| | | | | | | Blend | 2347.3823 | 0.0004 | |
| CO2-Kr2-84-84 | 3 | 1 | 2 | 3 | 2 | 2 | 2347.3875 | 2347.3876 | |
| CO2-Kr2-84-86 | 3 | 1 | 2 | 3 | 2 | 2 | 2347.3875 | 2347.3870 | |
| | | | | | | Blend | 2347.3874 | 0.0001 | |
| CO2-Kr2-84-84 | 5 | 1 | 4 | 5 | 2 | 4 | 2347.4098 | 2347.4103 | |
| CO2-Kr2-84-86 | 5 | 1 | 4 | 5 | 2 | 4 | 2347.4098 | 2347.4095 | |
| | | | | | | Blend | 2347.4101 | -0.0003 | |
| CO2-Kr2-84-84 | 1 | 0 | 1 | 1 | 1 | 1 | 2347.4285 | 2347.4278 | |
| CO2-Kr2-84-84 | 11 | 2 | 9 | 11 | 3 | 9 | 2347.4285 | 2347.4299 | |
| CO2-Kr2-84-86 | 1 | 0 | 1 | 1 | 1 | 1 | 2347.4285 | 2347.4277 | |
| CO2-Kr2-84-86 | 11 | 2 | 9 | 11 | 3 | 9 | 2347.4285 | 2347.4286 | |
| | | | | | | Blend | 2347.4286 | -0.0001 | |
| CO2-Kr2-84-84 | 3 | 0 | 3 | 3 | 1 | 3 | 2347.4392 | 2347.4389 | |
| CO2-Kr2-84-86 | 3 | 0 | 3 | 3 | 1 | 3 | 2347.4392 | 2347.4386 | |
| | | | | | | Blend | 2347.4388 | 0.0004 | |
| CO2-Kr2-84-84 | 9 | 1 | 8 | 9 | 2 | 8 | 2347.4446 | 2347.4450 | |
| CO2-Kr2-84-86 | 4 | 0 | 4 | 4 | 1 | 4 | 2347.4446 | 2347.4438 | |
| CO2-Kr2-82-84 | 4 | 0 | 4 | 4 | 1 | 4 | 2347.4446 | 2347.4442 | |
| CO2-Kr2-83-84 | 4 | 0 | 4 | 4 | 1 | 4 | 2347.4446 | 2347.4441 | |
| CO2-Kr2-84-86 | 9 | 1 | 8 | 9 | 2 | 8 | 2347.4446 | 2347.4446 | |
| CO2-Kr2-84-84 | 13 | 2 | 11 | 13 | 3 | 11 | 2347.4446 | 2347.4434 | |
| | | | | | | Blend | 2347.4444 | 0.0003 | |
| CO2-Kr2-84-84 | 7 | 0 | 7 | 7 | 1 | 7 | 2347.4517 | 2347.4507 | |
| CO2-Kr2-84-84 | 6 | 1 | 6 | 6 | 0 | 6 | 2347.4517 | 2347.4541 | |
| CO2-Kr2-84-84 | 8 | 1 | 8 | 8 | 0 | 8 | 2347.4517 | 2347.4522 | |
| CO2-Kr2-84-84 | 9 | 0 | 9 | 9 | 1 | 9 | 2347.4517 | 2347.4514 | |
| CO2-Kr2-84-84 | 10 | 1 | 10 | 10 | 0 | 10 | 2347.4517 | 2347.4516 | |
| CO2-Kr2-84-84 | 11 | 0 | 11 | 11 | 1 | 11 | 2347.4517 | 2347.4514 | |
| CO2-Kr2-84-84 | 12 | 1 | 12 | 12 | 0 | 12 | 2347.4517 | 2347.4513 | |
| CO2-Kr2-84-84 | 13 | 0 | 13 | 13 | 1 | 13 | 2347.4517 | 2347.4512 | |
| CO2-Kr2-84-84 | 12 | 2 | 11 | 12 | 1 | 11 | 2347.4517 | 2347.4516 | |
| CO2-Kr2-84-84 | 14 | 1 | 14 | 14 | 0 | 14 | 2347.4517 | 2347.4510 | |
| | | | | | | Blend | 2347.4518 | -0.0001 | |
| CO2-Kr2-84-84 | 4 | 1 | 4 | 4 | 0 | 4 | 2347.4605 | 2347.4600 | |
| CO2-Kr2-84-86 | 4 | 1 | 4 | 4 | 0 | 4 | 2347.4605 | 2347.4603 | |
| | | | | | | Blend | 2347.4601 | 0.0004 | |
| CO2-Kr2-84-84 | 6 | 2 | 5 | 6 | 1 | 5 | 2347.4820 | 2347.4811 | |
| CO2-Kr2-84-86 | 6 | 2 | 5 | 6 | 1 | 5 | 2347.4820 | 2347.4819 | |
| | | | | | | Blend | 2347.4813 | 0.0007 | |
| CO2-Kr2-84-84 | 4 | 2 | 3 | 4 | 1 | 3 | 2347.5067 | 2347.5054 | |
| CO2-Kr2-84-86 | 4 | 2 | 3 | 4 | 1 | 3 | 2347.5067 | 2347.5061 | |
| | | | | | | Blend | 2347.5056 | 0.0012 | |
| CO2-Kr2-84-84 | 1 | 1 | 0 | 0 | 0 | 0 | 2347.5228 | 2347.5223 | |
| CO2-Kr2-84-84 | 12 | 4 | 9 | 12 | 3 | 9 | 2347.5228 | 2347.5211 | |
| CO2-Kr2-84-86 | 1 | 1 | 0 | 0 | 0 | 0 | 2347.5228 | 2347.5220 | |
| | | | | | | Blend | 2347.5219 | 0.0009 | |
| CO2-Kr2-84-84 | 5 | 3 | 2 | 5 | 2 | 4 | 2347.5999 | 2347.6001 | |
| CO2-Kr2-84-86 | 5 | 3 | 2 | 5 | 2 | 4 | 2347.5999 | 2347.6001 | |
| | | | | | | Blend | 2347.6001 | -0.0001 | |
| CO2-Kr2-84-84 | 8 | 4 | 5 | 8 | 3 | 5 | 2347.6114 | 2347.6096 | |
| CO2-Kr2-84-86 | 2 | 2 | 0 | 1 | 1 | 0 | 2347.6114 | 2347.6110 | |
| CO2-Kr2-82-84 | 2 | 2 | 0 | 1 | 1 | 0 | 2347.6114 | 2347.6120 | |
| CO2-Kr2-83-84 | 2 | 2 | 0 | 1 | 1 | 0 | 2347.6114 | 2347.6118 | |
| CO2-Kr2-84-84 | 5 | 2 | 3 | 5 | 1 | 5 | 2347.6114 | 2347.6135 | |
| CO2-Kr2-84-86 | 8 | 4 | 5 | 8 | 3 | 5 | 2347.6114 | 2347.6116 | |
| CO2-Kr2-84-86 | 6 | 3 | 3 | 6 | 2 | 5 | 2347.6114 | 2347.6128 | |
| | | | | Blend | | | | 2347.6113 | 0.0001 |
| CO2-Kr2-84-84 | 3 | 1 | 2 | 2 | 0 | 2 | 2347.6169 | 2347.6174 | |
| CO2-Kr2-84-84 | 2 | 2 | 1 | 1 | 1 | 1 | 2347.6169 | 2347.6167 | |
| CO2-Kr2-84-86 | 3 | 1 | 2 | 2 | 0 | 2 | 2347.6169 | 2347.6160 | |
| | | | | Blend | | | | 2347.6169 | -0.0001 |
| CO2-Kr2-84-84 | 6 | 4 | 3 | 6 | 3 | 3 | 2347.6331 | 2347.6333 | |
| CO2-Kr2-84-84 | 7 | 3 | 4 | 7 | 2 | 6 | 2347.6331 | 2347.6351 | |
| CO2-Kr2-84-86 | 6 | 4 | 3 | 6 | 3 | 3 | 2347.6331 | 2347.6344 | |
| | | | | Blend | | | | 2347.6341 | -0.0009 |
| CO2-Kr2-84-84 | 7 | 5 | 2 | 7 | 4 | 4 | 2347.6941 | 2347.6938 | |
| CO2-Kr2-84-84 | 6 | 5 | 2 | 6 | 4 | 2 | 2347.6941 | 2347.6945 | |
| CO2-Kr2-84-84 | 9 | 5 | 4 | 9 | 4 | 6 | 2347.6941 | 2347.6923 | |
| CO2-Kr2-84-84 | 5 | 5 | 0 | 5 | 4 | 2 | 2347.6941 | 2347.6958 | |
| CO2-Kr2-84-86 | 7 | 5 | 3 | 7 | 4 | 3 | 2347.6941 | 2347.6933 | |
| CO2-Kr2-84-86 | 7 | 5 | 2 | 7 | 4 | 4 | 2347.6941 | 2347.6948 | |
| CO2-Kr2-84-86 | 8 | 5 | 3 | 8 | 4 | 5 | 2347.6941 | 2347.6937 | |
| CO2-Kr2-84-86 | 6 | 5 | 2 | 6 | 4 | 2 | 2347.6941 | 2347.6955 | |
| CO2-Kr2-84-86 | 6 | 5 | 1 | 6 | 4 | 3 | 2347.6941 | 2347.6959 | |
| CO2-Kr2-84-86 | 9 | 5 | 4 | 9 | 4 | 6 | 2347.6941 | 2347.6932 | |
| | | | | Blend | | | | 2347.6941 | -0.0001 |
| CO2-Kr2-84-84 | 3 | 3 | 0 | 2 | 2 | 0 | 2347.7084 | 2347.7075 | |
| CO2-Kr2-84-86 | 3 | 3 | 0 | 2 | 2 | 0 | 2347.7084 | 2347.7069 | |
| CO2-Kr2-84-86 | 3 | 3 | 1 | 2 | 2 | 1 | 2347.7084 | 2347.7078 | |
| | | | | Blend | | | | 2347.7075 | 0.0010 |
| CO2-Kr2-84-84 | 5 | 2 | 3 | 4 | 1 | 3 | 2347.7301 | 2347.7307 | |
| CO2-Kr2-84-84 | 5 | 1 | 4 | 4 | 0 | 4 | 2347.7301 | 2347.7294 | |
| CO2-Kr2-84-86 | 5 | 2 | 3 | 4 | 1 | 3 | 2347.7301 | 2347.7286 | |
| | | | | Blend | | | | 2347.7299 | 0.0001 |
| CO2-Kr2-84-84 | 5 | 3 | 2 | 4 | 2 | 2 | 2347.7812 | 2347.7816 | |
| CO2-Kr2-84-86 | 5 | 3 | 2 | 4 | 2 | 2 | 2347.7812 | 2347.7802 | |
| | | | | Blend | | | | 2347.7813 | -0.0001 |
| CO2-Kr2-84-84 | 4 | 4 | 1 | 3 | 3 | 1 | 2347.8024 | 2347.8022 | |
| CO2-Kr2-84-86 | 4 | 4 | 0 | 3 | 3 | 0 | 2347.8024 | 2347.8012 | |
| CO2-Kr2-84-86 | 4 | 4 | 1 | 3 | 3 | 1 | 2347.8024 | 2347.8013 | |
| | | | | Blend | | | | 2347.8019 | 0.0005 |
| CO2-Kr2-84-84 | 5 | 4 | 1 | 4 | 3 | 1 | 2347.8425 | 2347.8416 | |
| CO2-Kr2-84-86 | 5 | 4 | 1 | 4 | 3 | 1 | 2347.8425 | 2347.8403 | |
| CO2-Kr2-84-86 | 5 | 4 | 2 | 4 | 3 | 2 | 2347.8425 | 2347.8410 | |
| | | | | Blend | | | | 2347.8412 | 0.0013 |
| CO2-Kr2-84-84 | 6 | 4 | 3 | 5 | 3 | 3 | 2347.8828 | 2347.8827 | |
| CO2-Kr2-84-86 | 6 | 4 | 3 | 5 | 3 | 3 | 2347.8828 | 2347.8809 | |
| CO2-Kr2-84-86 | 7 | 3 | 5 | 6 | 2 | 5 | 2347.8828 | 2347.8825 | |
| | | | | Blend | | | | 2347.8823 | 0.0005 |
| CO2-Kr2-84-84 | 5 | 5 | 0 | 4 | 4 | 0 | 2347.8959 | 2347.8962 | |
| CO2-Kr2-84-86 | 5 | 5 | 0 | 4 | 4 | 0 | 2347.8959 | 2347.8951 | |
| CO2-Kr2-84-86 | 5 | 5 | 1 | 4 | 4 | 1 | 2347.8959 | 2347.8952 | |
| | | | | Blend | | | | 2347.8958 | 0.0001 |
| CO2-Kr2-84-84 | 7 | 4 | 3 | 6 | 3 | 3 | 2347.9170 | 2347.9168 | |
| CO2-Kr2-84-86 | 7 | 4 | 3 | 6 | 3 | 3 | 2347.9170 | 2347.9149 | |
| | | | | Blend | | | | 2347.9164 | 0.0006 |
| CO2-Kr2-84-84 | 6 | 5 | 2 | 5 | 4 | 2 | 2347.9363 | 2347.9361 | |
| CO2-Kr2-84-84 | 9 | 3 | 6 | 8 | 2 | 6 | 2347.9363 | 2347.9369 | |
| CO2-Kr2-84-84 | 8 | 3 | 6 | 7 | 2 | 6 | 2347.9363 | 2347.9355 | |
| CO2-Kr2-84-86 | 6 | 5 | 1 | 5 | 4 | 1 | 2347.9363 | 2347.9345 | |
| CO2-Kr2-84-86 | 6 | 5 | 2 | 5 | 4 | 2 | 2347.9363 | 2347.9346 | |
| | | | | Blend | | | | 2347.9359 | 0.0004 |
| CO2-Kr2-84-84 | 8 | 5 | 4 | 7 | 4 | 4 | 2348.0156 | 2348.0157 | |
| CO2-Kr2-84-86 | 8 | 5 | 4 | 7 | 4 | 4 | 2348.0156 | 2348.0133 | |
| | | | | Blend | | | | 2348.0151 | 0.0004 |
| CO2-Kr2-84-84 | 7 | 6 | 1 | 6 | 5 | 1 | 2348.0299 | 2348.0300 | -0.0001 |
| CO2-Kr2-84-86 | 7 | 6 | 2 | 6 | 5 | 2 | 2348.0299 | 2348.0283 | 0.0016 |
| CO2-Kr2-84-86 | 7 | 6 | 1 | 6 | 5 | 1 | 2348.0299 | 2348.0283 | |

```
                                                      Blend          2348.0293      0.0005
CO2-Kr2-84-84    11    3    8    10    2    8    2348.0436    2348.0443
CO2-Kr2-84-84    10    3    8     9    2    8    2348.0436    2348.0435
                                                      Blend          2348.0439     -0.0003
CO2-Kr2-84-84     7    7    0     6    6    0    2348.0837    2348.0840
CO2-Kr2-84-84    11    2    9    10    1    9    2348.0837    2348.0833
CO2-Kr2-84-86     7    7    1     6    6    1    2348.0837    2348.0825
CO2-Kr2-84-86     7    7    0     6    6    0    2348.0837    2348.0825
                                                      Blend          2348.0834      0.0003
*********************************************************************************
```

Table A-11. Observed and calculated line positions for CO$_2$-Kr$_2$ combination band (in cm$^{-1}$).

```
*********************************************************************************
                J'   Ka'  Kc'   J"   Ka"  Kc"    Obs          Calc          O-C
*********************************************************************************
CO2-Kr2-84-84     9    7    3    10    8    2    2380.3686    2380.3677
CO2-Kr2-84-84    11    6    6    12    7    5    2380.3686    2380.3685
CO2-Kr2-84-86     9    7    3    10    8    2    2380.3686    2380.3681
CO2-Kr2-84-86     9    7    2    10    8    3    2380.3686    2380.3681
                                                      Blend          2380.3680      0.0006
CO2-Kr2-84-84     8    7    1     9    8    2    2380.4088    2380.4084
CO2-Kr2-84-84    10    6    4    11    7    5    2380.4088    2380.4088
CO2-Kr2-84-86     8    7    2     9    8    1    2380.4088    2380.4085
CO2-Kr2-84-86     8    7    1     9    8    2    2380.4088    2380.4085
                                                      Blend          2380.4086      0.0002
CO2-Kr2-84-84    12    5    7    13    6    8    2380.4220    2380.4220
CO2-Kr2-84-86    12    5    7    13    6    8    2380.4220    2380.4227
                                                      Blend          2380.4222     -0.0001
CO2-Kr2-84-84     7    7    1     8    8    0    2380.4495    2380.4492
CO2-Kr2-84-84     9    6    4    10    7    3    2380.4495    2380.4490
CO2-Kr2-84-86     7    7    1     8    8    0    2380.4495    2380.4489
CO2-Kr2-84-86     7    7    0     8    8    1    2380.4495    2380.4489
                                                      Blend          2380.4491      0.0004
CO2-Kr2-84-84     8    6    2     9    7    3    2380.4899    2380.4896
CO2-Kr2-84-84    10    5    5    11    6    6    2380.4899    2380.4896
CO2-Kr2-84-86     8    6    3     9    7    2    2380.4899    2380.4897
CO2-Kr2-84-86     8    6    2     9    7    3    2380.4899    2380.4897
                                                      Blend          2380.4896      0.0003
CO2-Kr2-84-84     9    4    6    10    5    5    2380.5961    2380.5955
CO2-Kr2-84-86     9    4    6    10    5    5    2380.5961    2380.5968
                                                      Blend          2380.5958      0.0003
CO2-Kr2-84-84     7    5    3     8    6    2    2380.6073    2380.6073
CO2-Kr2-84-86     7    5    3     8    6    2    2380.6073    2380.6072
CO2-Kr2-84-86     7    5    2     8    6    3    2380.6073    2380.6073
                                                      Blend          2380.6073      0.0000
CO2-Kr2-84-86     8    4    5     9    5    4    2380.6387    2380.6392
CO2-Kr2-82-84     8    4    5     9    5    4    2380.6387    2380.6378
CO2-Kr2-83-84     8    4    5     9    5    4    2380.6387    2380.6382
                                                      Blend          2380.6385      0.0002
CO2-Kr2-84-84     9    3    7    10    4    6    2380.6263    2380.6264     -0.0001
CO2-Kr2-84-84     6    5    1     7    6    2    2380.6473    2380.6478
CO2-Kr2-84-84     8    4    4     9    5    5    2380.6473    2380.6453
CO2-Kr2-84-86     6    5    2     7    6    1    2380.6473    2380.6473
CO2-Kr2-84-86     6    5    1     7    6    2    2380.6473    2380.6474
                                                      Blend          2380.6470      0.0003
CO2-Kr2-84-84     7    4    4     8    5    3    2380.6810    2380.6800
CO2-Kr2-84-86     7    4    4     8    5    3    2380.6810    2380.6802
CO2-Kr2-84-86     7    4    3     8    5    4    2380.6810    2380.6823
                                                      Blend          2380.6805      0.0006
CO2-Kr2-84-84     5    5    1     6    6    0    2380.6883    2380.6884     -0.0001
```

| | | | | | | | | | |
|---|---|---|---|---|---|---|---|---|---|
| CO2-Kr2-84-84 | 7 | 3 | 5 | 8 | 4 | 4 | 2380.7398 | 2380.7390 | |
| CO2-Kr2-84-84 | 8 | 3 | 5 | 9 | 4 | 6 | 2380.7398 | 2380.7404 | |
| CO2-Kr2-84-86 | 7 | 3 | 5 | 8 | 4 | 4 | 2380.7398 | 2380.7399 | |
| | | | | | | Blend | 2380.7397 | 0.0001 |
| CO2-Kr2-84-84 | 5 | 3 | 3 | 6 | 4 | 2 | 2380.8294 | 2380.8287 | 0.0007 |
| CO2-Kr2-84-84 | 11 | 1 | 11 | 12 | 0 | 12 | 2380.8440 | 2380.8427 | |
| CO2-Kr2-84-86 | 11 | 0 | 11 | 12 | 1 | 12 | 2380.8440 | 2380.8443 | |
| CO2-Kr2-84-86 | 11 | 1 | 11 | 12 | 0 | 12 | 2380.8440 | 2380.8443 | |
| | | | | | | Blend | 2380.8433 | 0.0007 |
| CO2-Kr2-84-84 | 4 | 3 | 1 | 5 | 4 | 2 | 2380.8707 | 2380.8712 | -0.0005 |
| CO2-Kr2-84-84 | 10 | 0 | 10 | 11 | 1 | 11 | 2380.8802 | 2380.8798 | |
| CO2-Kr2-84-86 | 10 | 0 | 10 | 11 | 1 | 11 | 2380.8802 | 2380.8810 | |
| CO2-Kr2-84-86 | 10 | 1 | 10 | 11 | 0 | 11 | 2380.8802 | 2380.8812 | |
| | | | | | | Blend | 2380.8803 | -0.0001 |
| CO2-Kr2-84-84 | 9 | 2 | 8 | 10 | 1 | 9 | 2380.8900 | 2380.8894 | |
| CO2-Kr2-84-84 | 6 | 2 | 4 | 7 | 3 | 5 | 2380.8900 | 2380.8900 | |
| CO2-Kr2-84-86 | 9 | 2 | 8 | 10 | 1 | 9 | 2380.8900 | 2380.8909 | |
| | | | | | | Blend | 2380.8899 | 0.0001 |
| CO2-Kr2-84-84 | 9 | 1 | 9 | 10 | 0 | 10 | 2380.9164 | 2380.9168 | |
| CO2-Kr2-84-84 | 8 | 1 | 7 | 9 | 2 | 8 | 2380.9164 | 2380.9155 | |
| CO2-Kr2-84-86 | 9 | 0 | 9 | 10 | 1 | 10 | 2380.9164 | 2380.9175 | |
| CO2-Kr2-84-86 | 9 | 1 | 9 | 10 | 0 | 10 | 2380.9164 | 2380.9177 | |
| | | | | | | Blend | 2380.9166 | -0.0002 |
| CO2-Kr2-84-84 | 8 | 5 | 3 | 8 | 6 | 2 | 2380.9294 | 2380.9291 | |
| CO2-Kr2-84-84 | 9 | 5 | 5 | 9 | 6 | 4 | 2380.9294 | 2380.9297 | |
| CO2-Kr2-84-84 | 7 | 5 | 3 | 7 | 6 | 2 | 2380.9294 | 2380.9282 | |
| CO2-Kr2-84-86 | 8 | 2 | 7 | 9 | 1 | 8 | 2380.9294 | 2380.9312 | |
| CO2-Kr2-82-84 | 8 | 2 | 7 | 9 | 1 | 8 | 2380.9294 | 2380.9284 | |
| CO2-Kr2-83-84 | 8 | 2 | 7 | 9 | 1 | 8 | 2380.9294 | 2380.9291 | |
| CO2-Kr2-84-84 | 6 | 5 | 1 | 6 | 6 | 0 | 2380.9294 | 2380.9281 | |
| CO2-Kr2-84-86 | 10 | 5 | 5 | 10 | 6 | 4 | 2380.9294 | 2380.9302 | |
| | | | | | | Blend | 2380.9293 | 0.0001 |
| CO2-Kr2-84-84 | 4 | 2 | 2 | 5 | 3 | 3 | 2380.9467 | 2380.9454 | |
| CO2-Kr2-84-86 | 7 | 1 | 6 | 8 | 2 | 7 | 2380.9467 | 2380.9465 | |
| | | | | | | Blend | 2380.9457 | 0.0011 |
| CO2-Kr2-84-84 | 8 | 0 | 8 | 9 | 1 | 9 | 2380.9532 | 2380.9529 | |
| CO2-Kr2-84-86 | 8 | 0 | 8 | 9 | 1 | 9 | 2380.9532 | 2380.9535 | |
| CO2-Kr2-84-86 | 8 | 1 | 8 | 9 | 0 | 9 | 2380.9532 | 2380.9541 | |
| | | | | | | Blend | 2380.9532 | 0.0000 |
| CO2-Kr2-84-84 | 6 | 1 | 5 | 7 | 2 | 6 | 2380.9734 | 2380.9749 | |
| CO2-Kr2-84-84 | 3 | 2 | 2 | 4 | 3 | 1 | 2380.9734 | 2380.9727 | |
| CO2-Kr2-84-84 | 7 | 2 | 6 | 8 | 1 | 7 | 2380.9734 | 2380.9724 | |
| CO2-Kr2-84-86 | 6 | 1 | 5 | 7 | 2 | 6 | 2380.9734 | 2380.9744 | |
| CO2-Kr2-84-86 | 7 | 2 | 6 | 8 | 1 | 7 | 2380.9734 | 2380.9737 | |
| | | | | | | Blend | 2380.9735 | -0.0001 |
| CO2-Kr2-84-84 | 7 | 1 | 7 | 8 | 0 | 8 | 2380.9898 | 2380.9900 | |
| CO2-Kr2-84-86 | 7 | 1 | 7 | 8 | 0 | 8 | 2380.9898 | 2380.9903 | |
| | | | | | | Blend | 2380.9901 | -0.0003 |
| CO2-Kr2-84-84 | 7 | 4 | 4 | 7 | 5 | 3 | 2381.0027 | 2381.0023 | |
| CO2-Kr2-84-84 | 9 | 4 | 6 | 9 | 5 | 5 | 2381.0027 | 2381.0025 | |
| CO2-Kr2-84-84 | 6 | 4 | 2 | 6 | 5 | 1 | 2381.0027 | 2381.0024 | |
| CO2-Kr2-84-84 | 3 | 1 | 3 | 4 | 2 | 2 | 2381.0027 | 2381.0035 | |
| CO2-Kr2-84-84 | 5 | 4 | 2 | 5 | 5 | 1 | 2381.0027 | 2381.0016 | |
| CO2-Kr2-82-84 | 5 | 1 | 4 | 6 | 2 | 5 | 2381.0027 | 2381.0023 | |
| CO2-Kr2-83-84 | 5 | 1 | 4 | 6 | 2 | 5 | 2381.0027 | 2381.0019 | |
| | | | | | | Blend | 2381.0024 | 0.0003 |
| CO2-Kr2-84-84 | 2 | 2 | 0 | 3 | 3 | 1 | 2381.0167 | 2381.0165 | 0.0001 |
| CO2-Kr2-84-86 | 6 | 0 | 6 | 7 | 1 | 7 | 2381.0243 | 2381.0239 | 0.0004 |
| CO2-Kr2-84-86 | 6 | 1 | 6 | 7 | 0 | 7 | 2381.0265 | 2381.0266 | -0.0001 |
| CO2-Kr2-84-84 | 9 | 1 | 9 | 9 | 2 | 8 | 2381.0398 | 2381.0391 | |
| CO2-Kr2-84-86 | 10 | 3 | 8 | 10 | 4 | 7 | 2381.0398 | 2381.0395 | |
| | | | | | | Blend | 2381.0392 | 0.0006 |
| CO2-Kr2-84-84 | 5 | 1 | 5 | 6 | 0 | 6 | 2381.0640 | 2381.0636 | |
| CO2-Kr2-84-84 | 9 | 2 | 8 | 9 | 3 | 7 | 2381.0640 | 2381.0630 | |
| | | | | | | Blend | | 2381.0634 | 0.0006 |
| CO2-Kr2-84-84 | 6 | 3 | 3 | 6 | 4 | 2 | 2381.0800 | 2381.0792 | |
| CO2-Kr2-84-84 | 8 | 0 | 8 | 8 | 1 | 7 | 2381.0800 | 2381.0791 | |
| | | | | | | Blend | | 2381.0791 | 0.0008 |
| CO2-Kr2-84-84 | 4 | 0 | 4 | 5 | 1 | 5 | 2381.0897 | 2381.0906 | −0.0009 |
| CO2-Kr2-84-84 | 7 | 2 | 6 | 7 | 3 | 5 | 2381.1006 | 2381.1015 | |
| CO2-Kr2-84-84 | 8 | 3 | 5 | 8 | 4 | 4 | 2381.1006 | 2381.1001 | |
| CO2-Kr2-84-86 | 4 | 1 | 4 | 5 | 0 | 5 | 2381.1006 | 2381.1011 | |
| CO2-Kr2-83-84 | 4 | 1 | 4 | 5 | 0 | 5 | 2381.1006 | 2381.1017 | |
| | | | | | | Blend | | 2381.1010 | −0.0005 |
| CO2-Kr2-84-84 | 10 | 3 | 7 | 10 | 4 | 6 | 2381.1220 | 2381.1221 | |
| CO2-Kr2-83-84 | 3 | 0 | 3 | 4 | 1 | 4 | 2381.1220 | 2381.1223 | |
| | | | | | | Blend | | 2381.1221 | −0.0001 |
| CO2-Kr2-84-84 | 3 | 1 | 3 | 4 | 0 | 4 | 2381.1418 | 2381.1410 | |
| CO2-Kr2-84-84 | 8 | 1 | 7 | 8 | 2 | 6 | 2381.1418 | 2381.1426 | |
| CO2-Kr2-84-84 | 5 | 1 | 5 | 5 | 2 | 4 | 2381.1418 | 2381.1423 | |
| | | | | | | Blend | | 2381.1418 | 0.0000 |
| CO2-Kr2-84-84 | 6 | 1 | 5 | 6 | 2 | 4 | 2381.1927 | 2381.1926 | 0.0002 |
| CO2-Kr2-84-84 | 4 | 1 | 3 | 4 | 2 | 2 | 2381.2085 | 2381.2096 | |
| CO2-Kr2-84-84 | 4 | 0 | 4 | 4 | 1 | 3 | 2381.2085 | 2381.2078 | |
| | | | | | | Blend | | 2381.2088 | −0.0003 |
| CO2-Kr2-84-84 | 0 | 0 | 0 | 1 | 1 | 1 | 2381.2170 | 2381.2167 | 0.0004 |
| CO2-Kr2-84-84 | 1 | 1 | 1 | 2 | 0 | 2 | 2381.2263 | 2381.2246 | |
| CO2-Kr2-84-86 | 3 | 0 | 3 | 3 | 1 | 2 | 2381.2263 | 2381.2266 | |
| | | | | | | Blend | | 2381.2256 | 0.0007 |
| CO2-Kr2-84-84 | 2 | 0 | 2 | 2 | 1 | 1 | 2381.2420 | 2381.2421 | −0.0001 |
| CO2-Kr2-84-84 | 2 | 1 | 1 | 2 | 0 | 2 | 2381.3166 | 2381.3163 | 0.0002 |
| CO2-Kr2-84-84 | 4 | 2 | 2 | 4 | 1 | 3 | 2381.3404 | 2381.3403 | |
| CO2-Kr2-84-84 | 1 | 1 | 1 | 0 | 0 | 0 | 2381.3404 | 2381.3424 | |
| CO2-Kr2-84-86 | 5 | 2 | 3 | 5 | 1 | 4 | 2381.3404 | 2381.3414 | |
| | | | | | | Blend | | 2381.3410 | −0.0006 |
| CO2-Kr2-84-84 | 6 | 2 | 4 | 6 | 1 | 5 | 2381.3538 | 2381.3536 | 0.0002 |
| CO2-Kr2-84-84 | 8 | 3 | 5 | 8 | 2 | 6 | 2381.3672 | 2381.3667 | |
| CO2-Kr2-84-86 | 7 | 2 | 5 | 7 | 1 | 6 | 2381.3672 | 2381.3669 | |
| | | | | | | Blend | | 2381.3668 | 0.0004 |
| CO2-Kr2-84-84 | 6 | 3 | 3 | 6 | 2 | 4 | 2381.3721 | 2381.3720 | |
| CO2-Kr2-84-84 | 3 | 2 | 2 | 3 | 1 | 3 | 2381.3721 | 2381.3735 | |
| | | | | | | Blend | | 2381.3725 | −0.0004 |
| CO2-Kr2-84-84 | 10 | 3 | 7 | 10 | 2 | 8 | 2381.3893 | 2381.3885 | |
| CO2-Kr2-84-84 | 4 | 3 | 1 | 4 | 2 | 2 | 2381.3893 | 2381.3888 | |
| CO2-Kr2-84-84 | 10 | 4 | 6 | 10 | 3 | 7 | 2381.3893 | 2381.3896 | |
| CO2-Kr2-84-84 | 12 | 4 | 8 | 12 | 3 | 9 | 2381.3893 | 2381.3906 | |
| CO2-Kr2-84-86 | 8 | 2 | 6 | 8 | 1 | 7 | 2381.3893 | 2381.3894 | |
| CO2-Kr2-84-86 | 10 | 4 | 6 | 10 | 3 | 7 | 2381.3893 | 2381.3886 | |
| | | | | | | Blend | | 2381.3892 | 0.0001 |
| CO2-Kr2-84-84 | 3 | 1 | 3 | 2 | 0 | 2 | 2381.4053 | 2381.4052 | |
| CO2-Kr2-84-84 | 5 | 3 | 3 | 5 | 2 | 4 | 2381.4053 | 2381.4064 | |
| | | | | | | Blend | | 2381.4057 | −0.0004 |
| CO2-Kr2-84-84 | 7 | 3 | 5 | 7 | 2 | 6 | 2381.4229 | 2381.4235 | |
| CO2-Kr2-84-84 | 14 | 4 | 10 | 14 | 3 | 11 | 2381.4229 | 2381.4228 | |
| | | | | | | Blend | | 2381.4234 | −0.0005 |
| CO2-Kr2-84-84 | 6 | 4 | 2 | 6 | 3 | 3 | 2381.4276 | 2381.4290 | −0.0015 |
| CO2-Kr2-84-84 | 7 | 4 | 4 | 7 | 3 | 5 | 2381.4369 | 2381.4378 | |
| CO2-Kr2-84-84 | 2 | 2 | 0 | 1 | 1 | 1 | 2381.4369 | 2381.4382 | |
| CO2-Kr2-84-84 | 5 | 4 | 2 | 5 | 3 | 3 | 2381.4369 | 2381.4372 | |
| CO2-Kr2-84-84 | 12 | 3 | 9 | 12 | 2 | 10 | 2381.4369 | 2381.4372 | |
| CO2-Kr2-84-84 | 4 | 4 | 0 | 4 | 3 | 1 | 2381.4369 | 2381.4372 | |
| CO2-Kr2-82-84 | 4 | 1 | 4 | 3 | 0 | 3 | 2381.4369 | 2381.4367 | |
| CO2-Kr2-84-86 | 8 | 4 | 5 | 8 | 3 | 6 | 2381.4369 | 2381.4387 | |
| | | | | | | Blend | | 2381.4376 | −0.0007 |
| CO2-Kr2-84-84 | 6 | 0 | 6 | 5 | 1 | 5 | 2381.4843 | 2381.4838 | |
| CO2-Kr2-84-86 | 6 | 1 | 6 | 5 | 0 | 5 | 2381.4843 | 2381.4860 | |
| | | | | | | Blend | | 2381.4843 | 0.0000 |
| CO2-Kr2-84-84 | 7 | 1 | 7 | 6 | 0 | 6 | 2381.5192 | 2381.5192 | |
| CO2-Kr2-84-84 | 3 | 3 | 1 | 2 | 2 | 0 | 2381.5192 | 2381.5176 | |
| CO2-Kr2-84-84 | 5 | 2 | 4 | 4 | 1 | 3 | 2381.5192 | 2381.5191 | |
| | | | | | | Blend | | 2381.5187 | 0.0005 |
| CO2-Kr2-84-84 | 8 | 0 | 8 | 7 | 1 | 7 | 2381.5477 | 2381.5475 | 0.0003 |
| CO2-Kr2-84-84 | 8 | 1 | 7 | 7 | 2 | 6 | 2381.5601 | 2381.5605 | |
| CO2-Kr2-84-84 | 4 | 3 | 1 | 3 | 2 | 2 | 2381.5601 | 2381.5606 | |
| CO2-Kr2-84-86 | 7 | 2 | 6 | 6 | 1 | 5 | 2381.5601 | 2381.5613 | |
| | | | | | | Blend | | 2381.5606 | −0.0005 |
| CO2-Kr2-84-84 | 9 | 1 | 9 | 8 | 0 | 8 | 2381.5791 | 2381.5787 | 0.0004 |
| CO2-Kr2-84-84 | 11 | 1 | 11 | 10 | 0 | 10 | 2381.6379 | 2381.6379 | |
| CO2-Kr2-84-84 | 5 | 4 | 2 | 4 | 3 | 1 | 2381.6379 | 2381.6371 | |
| | | | | | | Blend | | 2381.6375 | 0.0003 |
| CO2-Kr2-84-84 | 6 | 3 | 3 | 5 | 2 | 3 | 2381.6569 | 2381.6568 | |
| CO2-Kr2-84-86 | 11 | 2 | 10 | 10 | 1 | 9 | 2381.6569 | 2381.6587 | |
| CO2-Kr2-84-86 | 8 | 3 | 6 | 7 | 2 | 5 | 2381.6569 | 2381.6581 | |
| | | | | | | Blend | | 2381.6575 | −0.0006 |
| CO2-Kr2-84-84 | 6 | 4 | 2 | 5 | 3 | 3 | 2381.6783 | 2381.6784 | |
| CO2-Kr2-84-84 | 9 | 3 | 7 | 8 | 2 | 6 | 2381.6783 | 2381.6797 | |
| CO2-Kr2-82-84 | 5 | 5 | 1 | 4 | 4 | 0 | 2381.6783 | 2381.6773 | |
| CO2-Kr2-82-84 | 5 | 5 | 0 | 4 | 4 | 1 | 2381.6783 | 2381.6773 | |
| | | | | | | Blend | | 2381.6785 | −0.0002 |
| CO2-Kr2-84-84 | 12 | 1 | 11 | 11 | 2 | 10 | 2381.6894 | 2381.6894 | |
| CO2-Kr2-84-86 | 13 | 0 | 13 | 12 | 1 | 12 | 2381.6894 | 2381.6901 | |
| CO2-Kr2-84-86 | 13 | 1 | 13 | 12 | 0 | 12 | 2381.6894 | 2381.6902 | |
| | | | | | | Blend | | 2381.6897 | −0.0003 |
| CO2-Kr2-84-84 | 13 | 1 | 13 | 12 | 0 | 12 | 2381.6961 | 2381.6960 | 0.0000 |
| CO2-Kr2-84-84 | 6 | 5 | 1 | 5 | 4 | 2 | 2381.7127 | 2381.7134 | |
| CO2-Kr2-84-84 | 11 | 3 | 9 | 10 | 2 | 8 | 2381.7127 | 2381.7134 | |
| | | | | | | Blend | | 2381.7134 | −0.0006 |
| CO2-Kr2-84-84 | 6 | 6 | 0 | 5 | 5 | 1 | 2381.7455 | 2381.7457 | |
| CO2-Kr2-84-84 | 14 | 1 | 13 | 13 | 2 | 12 | 2381.7455 | 2381.7464 | |
| | | | | | | Blend | | 2381.7459 | −0.0004 |
| CO2-Kr2-84-84 | 8 | 4 | 4 | 7 | 3 | 5 | 2381.7666 | 2381.7663 | |
| CO2-Kr2-84-84 | 9 | 4 | 6 | 8 | 3 | 5 | 2381.7666 | 2381.7669 | |
| CO2-Kr2-84-84 | 14 | 2 | 12 | 13 | 3 | 11 | 2381.7666 | 2381.7680 | |
| CO2-Kr2-84-84 | 14 | 3 | 11 | 13 | 4 | 10 | 2381.7666 | 2381.7674 | |
| | | | | | | Blend | | 2381.7669 | −0.0003 |
| CO2-Kr2-84-84 | 7 | 6 | 2 | 6 | 5 | 1 | 2381.7843 | 2381.7849 | −0.0005 |
| CO2-Kr2-84-84 | 5 | 4 | 2 | 6 | 5 | 1 | 2380.7616 | 2380.7612 | |
| CO2-Kr2-84-86 | 5 | 4 | 2 | 6 | 5 | 1 | 2380.7616 | 2380.7605 | |
| CO2-Kr2-84-86 | 5 | 4 | 1 | 6 | 5 | 2 | 2380.7616 | 2380.7606 | |
| | | | | | | Blend | | 2380.7610 | 0.0006 |
| CO2-Kr2-84-84 | 13 | 1 | 13 | 14 | 0 | 14 | 2380.7683 | 2380.7674 | |
| CO2-Kr2-84-86 | 13 | 0 | 13 | 14 | 1 | 14 | 2380.7683 | 2380.7697 | |
| CO2-Kr2-84-86 | 13 | 1 | 13 | 14 | 0 | 14 | 2380.7683 | 2380.7697 | |
| | | | | | | Blend | | 2380.7683 | 0.0000 |
| CO2-Kr2-84-86 | 6 | 3 | 4 | 7 | 4 | 3 | 2380.7854 | 2380.7857 | |
| CO2-Kr2-82-84 | 6 | 3 | 4 | 7 | 4 | 3 | 2380.7854 | 2380.7853 | |
| CO2-Kr2-83-84 | 6 | 3 | 4 | 7 | 4 | 3 | 2380.7854 | 2380.7854 | |
| | | | | | | Blend | | 2380.7855 | −0.0001 |
| CO2-Kr2-84-84 | 6 | 3 | 3 | 7 | 4 | 4 | 2380.7982 | 2380.7975 | |
| CO2-Kr2-84-84 | 10 | 2 | 8 | 11 | 3 | 9 | 2380.7982 | 2380.7966 | |
| CO2-Kr2-84-86 | 6 | 3 | 3 | 7 | 4 | 4 | 2380.7982 | 2380.7970 | |
| | | | | | | Blend | | 2380.7971 | 0.0011 |
| CO2-Kr2-84-84 | 11 | 2 | 10 | 12 | 1 | 11 | 2380.8109 | 2380.8115 | |
| CO2-Kr2-84-86 | 11 | 1 | 10 | 12 | 2 | 11 | 2380.8109 | 2380.8116 | |
| | | | | | | Blend | | 2380.8115 | −0.0007 |
| CO2-Kr2-84-86 | 5 | 0 | 5 | 6 | 1 | 6 | 2381.0583 | 2381.0575 | |
| CO2-Kr2-82-84 | 5 | 0 | 5 | 6 | 1 | 6 | 2381.0583 | 2381.0583 | |
| CO2-Kr2-83-84 | 5 | 0 | 5 | 6 | 1 | 6 | 2381.0583 | 2381.0582 | |
| CO2-Kr2-84-84 | 12 | 4 | 8 | 12 | 5 | 7 | 2381.0583 | 2381.0593 | |

```
CO2-Kr2-84-86    8    3    6     8    4    5    2381.0583    2381.0591
CO2-Kr2-82-84    3    1    2     4    2    3    2381.0583    2381.0572
CO2-Kr2-83-84    3    1    2     4    2    3    2381.0583    2381.0566
                         Blend                   2381.0581       0.0002
CO2-Kr2-84-84   12    3    9    12    4    8    2381.1173    2381.1170
CO2-Kr2-84-86    1    1    0     2    2    1    2381.1173    2381.1182
CO2-Kr2-84-86    6    1    6     6    2    5    2381.1173    2381.1190
CO2-Kr2-84-86   10    3    7    10    4    6    2381.1173    2381.1184
CO2-Kr2-82-84    6    2    5     6    3    4    2381.1173    2381.1170
CO2-Kr2-83-84    6    2    5     6    3    4    2381.1173    2381.1159
CO2-Kr2-84-86    4    2    3     5    1    4    2381.1173    2381.1172
CO2-Kr2-82-84    9    3    6     9    4    5    2381.1173    2381.1167
CO2-Kr2-84-86   12    3    9    12    4    8    2381.1173    2381.1156
CO2-Kr2-82-84    4    2    3     5    1    4    2381.1173    2381.1161
                         Blend                   2381.1173       0.0000
CO2-Kr2-84-84    6    2    4     6    3    3    2381.1628    2381.1622
CO2-Kr2-84-86    7    2    5     7    3    4    2381.1628    2381.1644
CO2-Kr2-84-86    8    2    6     8    3    5    2381.1628    2381.1637
                         Blend                   2381.1629      -0.0001
CO2-Kr2-84-84    8    2    6     8    3    5    2381.1666    2381.1660
CO2-Kr2-84-84    3    2    2     4    1    3    2381.1666    2381.1677
                         Blend                   2381.1664       0.0003
CO2-Kr2-84-84    4    1    3     4    0    4    2381.3464    2381.3471
CO2-Kr2-84-84    2    2    0     2    1    1    2381.3464    2381.3464
CO2-Kr2-82-84    5    2    3     5    1    4    2381.3464    2381.3461
CO2-Kr2-83-84    5    2    3     5    1    4    2381.3464    2381.3450
                         Blend                   2381.3465      -0.0001
CO2-Kr2-84-84    8    4    4     8    3    5    2381.4091    2381.4097
CO2-Kr2-84-86    4    0    4     3    1    3    2381.4091    2381.4098
CO2-Kr2-84-86    6    3    4     6    2    5    2381.4091    2381.4112
CO2-Kr2-84-86    8    4    4     8    3    5    2381.4091    2381.4091
CO2-Kr2-82-84    3    1    3     2    0    2    2381.4091    2381.4082
                         Blend                   2381.4097      -0.0006
CO2-Kr2-84-86    9    0    9     8    1    8    2381.5732    2381.5732
CO2-Kr2-84-86    9    1    9     8    0    8    2381.5732    2381.5739
                         Blend                   2381.5735      -0.0004
CO2-Kr2-84-84    7    5    3     6    4    2    2381.7523    2381.7522
CO2-Kr2-84-84   15    1   15    14    0   14    2381.7523    2381.7529
CO2-Kr2-84-84   13    3   11    12    2   10    2381.7523    2381.7534
                         Blend                   2381.7526      -0.0003
CO2-Kr2-84-84    8    5    3     9    6    4    2380.5668    2380.5672
CO2-Kr2-84-86    8    5    4     9    6    3    2380.5668    2380.5672
CO2-Kr2-84-86    8    5    3     9    6    4    2380.5668    2380.5676
                         Blend                   2380.5673      -0.0005
CO2-Kr2-84-84    6    6    0     7    7    1    2380.5709    2380.5709
CO2-Kr2-84-86    6    6    1     7    7    0    2380.5709    2380.5703
CO2-Kr2-84-86    6    6    0     7    7    1    2380.5709    2380.5703
                         Blend                   2380.5706       0.0003
CO2-Kr2-82-84    6    1    6     5    0    5    2381.4945    2381.4939
CO2-Kr2-84-84   10    6    4    10    5    5    2381.4945    2381.4951
CO2-Kr2-84-84   11    6    6    11    5    7    2381.4945    2381.4944
CO2-Kr2-84-84   13    4   10    13    3   11    2381.4945    2381.4945
CO2-Kr2-82-84    4    2    3     3    1    2    2381.4945    2381.4963
CO2-Kr2-83-84    4    2    3     3    1    2    2381.4945    2381.4946
CO2-Kr2-84-84   14    3   11    14    2   12    2381.4945    2381.4950
CO2-Kr2-84-86   10    6    4    10    5    5    2381.4945    2381.4945
CO2-Kr2-84-86   10    6    5    10    5    6    2381.4945    2381.4963
                         Blend                   2381.4948      -0.0003
CO2-Kr2-84-84    4    4    0     3    3    1    2381.5975    2381.5982      -0.0007
********************************************************************
```

Table A-12. Observed and calculated line positions for CO$_2$-Xe$_2$ fundamental band (in cm$^{-1}$).

```
**********************************************************************
                    J'  Ka' Kc'   J"   Ka"  Kc"    Obs       Calc       O-C
**********************************************************************
CO2-Xe2-132-132      7   7   0     8    8    0   2345.7034  2345.7034
CO2-Xe2-129-132      7   7   0     8    8    0   2345.7034  2345.7023
CO2-Xe2-129-132      7   7   1     8    8    1   2345.7034  2345.7023
CO2-Xe2-129-129      7   7   0     8    8    0   2345.7034  2345.7013
CO2-Xe2-131-132      7   7   0     8    8    0   2345.7034  2345.7030
CO2-Xe2-131-132      7   7   1     8    8    1   2345.7034  2345.7030
CO2-Xe2-129-131      7   7   0     8    8    0   2345.7034  2345.7020
CO2-Xe2-129-131      7   7   1     8    8    1   2345.7034  2345.7020
CO2-Xe2-131-131      7   7   1     8    8    1   2345.7034  2345.7041
CO2-Xe2-132-134      7   7   0     8    8    0   2345.7034  2345.7040
                                              Blend  2345.7026   0.0009
CO2-Xe2-132-132      9   6   3    10    7    3   2345.7169  2345.7177
CO2-Xe2-129-132      9   6   3    10    7    3   2345.7169  2345.7161
CO2-Xe2-129-132      9   6   4    10    7    4   2345.7169  2345.7161
CO2-Xe2-131-132      9   6   3    10    7    3   2345.7169  2345.7172
CO2-Xe2-131-132      9   6   4    10    7    4   2345.7169  2345.7172
CO2-Xe2-129-131      9   6   3    10    7    3   2345.7169  2345.7155
CO2-Xe2-129-131      9   6   4    10    7    4   2345.7169  2345.7155
CO2-Xe2-129-129     14   4  11    15    5   11   2345.7169  2345.7180
CO2-Xe2-132-132     16   3  14    17    4   14   2345.7169  2345.7161
CO2-Xe2-131-131      9   6   4    10    7    4   2345.7169  2345.7179
                                              Blend  2345.7166   0.0002
CO2-Xe2-132-132      6   6   1     7    7    1   2345.7863  2345.7863
CO2-Xe2-129-132      6   6   0     7    7    0   2345.7863  2345.7854
CO2-Xe2-129-132      6   6   1     7    7    1   2345.7863  2345.7854
CO2-Xe2-129-129      6   6   1     7    7    1   2345.7863  2345.7845
CO2-Xe2-131-132      6   6   0     7    7    0   2345.7863  2345.7860
CO2-Xe2-131-132      6   6   1     7    7    1   2345.7863  2345.7860
CO2-Xe2-129-131      6   6   0     7    7    0   2345.7863  2345.7851
CO2-Xe2-129-131      6   6   1     7    7    1   2345.7863  2345.7851
CO2-Xe2-129-129     15   3  12    16    4   12   2345.7863  2345.7875
CO2-Xe2-129-132     14   2  13    15    3   13   2345.7863  2345.7865
                                              Blend  2345.7856   0.0007
CO2-Xe2-132-132      8   5   4     9    6    4   2345.7994  2345.8006
CO2-Xe2-129-132      8   5   3     9    6    3   2345.7994  2345.7991
CO2-Xe2-129-132      8   5   4     9    6    4   2345.7994  2345.7991
CO2-Xe2-131-132      8   5   3     9    6    3   2345.7994  2345.8001
CO2-Xe2-131-132      8   5   4     9    6    4   2345.7994  2345.8001
CO2-Xe2-129-131      8   5   3     9    6    3   2345.7994  2345.7986
CO2-Xe2-129-131      8   5   4     9    6    4   2345.7994  2345.7986
CO2-Xe2-131-131      8   5   3     9    6    3   2345.7994  2345.8007
CO2-Xe2-129-134      8   5   3     9    6    3   2345.7994  2345.8000
CO2-Xe2-129-134      8   5   4     9    6    4   2345.7994  2345.8000
                                              Blend  2345.7996  -0.0002
CO2-Xe2-129-132      6   5   1     7    6    1   2345.8438  2345.8452
CO2-Xe2-129-132      6   5   2     7    6    2   2345.8438  2345.8452
CO2-Xe2-129-129      6   5   2     7    6    2   2345.8438  2345.8442
CO2-Xe2-131-132      6   5   1     7    6    1   2345.8438  2345.8459
CO2-Xe2-131-132      6   5   2     7    6    2   2345.8438  2345.8459
CO2-Xe2-129-131      6   5   1     7    6    1   2345.8438  2345.8449
CO2-Xe2-129-131      6   5   2     7    6    2   2345.8438  2345.8449
CO2-Xe2-129-131     11   3   9    12    4    9   2345.8438  2345.8454
                                              Blend  2345.8452  -0.0014
CO2-Xe2-132-132     13   2  11    14    3   11   2345.8884  2345.8875
CO2-Xe2-129-132     11   1  11    12    2   11   2345.8884  2345.8883
CO2-Xe2-129-131     11   1  11    12    2   11   2345.8884  2345.8870
```

```
CO2-Xe2-132-134   13   2   11    14   3   11   2345.8884   2345.8898
CO2-Xe2-132-134   11   2   10    12   3   10   2345.8884   2345.8885
CO2-Xe2-131-134   13   2   11    14   3   11   2345.8884   2345.8887
CO2-Xe2-131-134   11   2   10    12   3   10   2345.8884   2345.8874
                                        Blend        2345.8880     0.0004
CO2-Xe2-132-132    6   4    3     7   5    3   2345.9053   2345.9062
CO2-Xe2-129-132    6   4    2     7   5    2   2345.9053   2345.9051
CO2-Xe2-129-132    6   4    3     7   5    3   2345.9053   2345.9051
CO2-Xe2-131-132    6   4    2     7   5    2   2345.9053   2345.9059
CO2-Xe2-131-132    6   4    3     7   5    3   2345.9053   2345.9059
CO2-Xe2-129-131    6   4    2     7   5    2   2345.9053   2345.9047
CO2-Xe2-129-131    6   4    3     7   5    3   2345.9053   2345.9047
CO2-Xe2-129-131   12   1   11    13   2   11   2345.9053   2345.9072
CO2-Xe2-129-132   11   0   11    12   1   11   2345.9053   2345.9056
                                        Blend        2345.9055    -0.0002
CO2-Xe2-132-132    5   4    1     6   5    1   2345.9279   2345.9291
CO2-Xe2-129-132    5   4    1     6   5    1   2345.9279   2345.9282
CO2-Xe2-129-132    5   4    2     6   5    2   2345.9279   2345.9282
CO2-Xe2-129-129    5   4    1     6   5    1   2345.9279   2345.9273
CO2-Xe2-131-132    5   4    1     6   5    1   2345.9279   2345.9288
CO2-Xe2-131-132    5   4    2     6   5    2   2345.9279   2345.9288
CO2-Xe2-129-131    5   4    1     6   5    1   2345.9279   2345.9279
CO2-Xe2-129-131    5   4    2     6   5    2   2345.9279   2345.9279
CO2-Xe2-129-129   11   2    9    12   3    9   2345.9279   2345.9294
CO2-Xe2-132-132   10   1   10    11   2   10   2345.9279   2345.9277
                                        Blend        2345.9283    -0.0005
CO2-Xe2-132-132   11   1   10    12   2   10   2345.9486   2345.9500
CO2-Xe2-131-132   11   1   10    12   2   10   2345.9486   2345.9486
CO2-Xe2-129-132   10   0   10    11   1   10   2345.9486   2345.9467
CO2-Xe2-131-132   10   0   10    11   1   10   2345.9486   2345.9496
CO2-Xe2-129-134   11   1   10    12   2   10   2345.9486   2345.9485
CO2-Xe2-132-136    9   2    8    10   3    8   2345.9486   2345.9471
CO2-Xe2-132-132   11   6    5    11   7    5   2345.9486   2345.9477
CO2-Xe2-132-132   10   6    5    10   7    3   2345.9486   2345.9474
CO2-Xe2-129-132   11   6    5    11   7    5   2345.9486   2345.9485
                                        Blend        2345.9485     0.0001
CO2-Xe2-132-132    4   4    1     5   5    1   2345.9513   2345.9520
CO2-Xe2-129-132    4   4    0     5   5    0   2345.9513   2345.9513
CO2-Xe2-129-132    4   4    1     5   5    1   2345.9513   2345.9513
CO2-Xe2-129-129    4   4    1     5   5    1   2345.9513   2345.9506
CO2-Xe2-131-132    4   4    0     5   5    0   2345.9513   2345.9518
CO2-Xe2-131-132    4   4    1     5   5    1   2345.9513   2345.9518
CO2-Xe2-129-131    4   4    0     5   5    0   2345.9513   2345.9511
CO2-Xe2-129-131    4   4    1     5   5    1   2345.9513   2345.9511
CO2-Xe2-129-131   10   2    8    11   3    8   2345.9513   2345.9523
                                        Blend        2345.9514    -0.0002
CO2-Xe2-129-129    9   1    8    10   2    8   2346.0076   2346.0088
CO2-Xe2-132-132   10   5    6    10   6    4   2346.0076   2346.0078
CO2-Xe2-129-132   10   5    5    10   6    5   2346.0076   2346.0085
CO2-Xe2-129-132   10   5    6    10   6    4   2346.0076   2346.0085
CO2-Xe2-132-132    9   5    4     9   6    4   2346.0076   2346.0074
CO2-Xe2-132-132   11   5    6    11   6    6   2346.0076   2346.0083
CO2-Xe2-129-132    9   5    4     9   6    4   2346.0076   2346.0081
CO2-Xe2-129-132    9   5    5     9   6    3   2346.0076   2346.0081
                                        Blend        2346.0083    -0.0007
CO2-Xe2-132-132    4   3    2     5   4    2   2346.0109   2346.0118
CO2-Xe2-129-132    4   3    1     5   4    1   2346.0109   2346.0111
CO2-Xe2-129-132    4   3    2     5   4    2   2346.0109   2346.0111
CO2-Xe2-129-129    4   3    2     5   4    2   2346.0109   2346.0103
CO2-Xe2-129-132    9   1    8    10   2    8   2346.0109   2346.0119
CO2-Xe2-132-132    7   2    5     8   3    5   2346.0109   2346.0120
CO2-Xe2-129-132    7   2    5     8   3    5   2346.0109   2346.0106
CO2-Xe2-131-132    4   3    1     5   4    1   2346.0109   2346.0116
```

```
CO2-Xe2-131-132    4   3   2      5   4   2  2346.0109  2346.0116
CO2-Xe2-129-131    4   3   1      5   4   1  2346.0109  2346.0108
                                     Blend         2346.0113  -0.0004
CO2-Xe2-132-132    3   3   0      4   4   0  2346.0347  2346.0348
CO2-Xe2-129-132    3   3   0      4   4   0  2346.0347  2346.0342
CO2-Xe2-129-132    3   3   1      4   4   1  2346.0347  2346.0342
CO2-Xe2-129-129    3   3   0      4   4   0  2346.0347  2346.0337
CO2-Xe2-131-132    3   3   0      4   4   0  2346.0347  2346.0346
CO2-Xe2-131-132    3   3   1      4   4   1  2346.0347  2346.0346
CO2-Xe2-129-131    3   3   0      4   4   0  2346.0347  2346.0340
CO2-Xe2-129-131    3   3   1      4   4   1  2346.0347  2346.0340
CO2-Xe2-131-131    3   3   1      4   4   1  2346.0347  2346.0351
CO2-Xe2-132-134    3   3   0      4   4   0  2346.0347  2346.0351
                                     Blend         2346.0343   0.0003
CO2-Xe2-129-132    5   2   4      6   3   4  2346.0460  2346.0457
CO2-Xe2-131-132    5   2   4      6   3   4  2346.0460  2346.0465
CO2-Xe2-129-131    5   2   4      6   3   4  2346.0460  2346.0453
CO2-Xe2-132-134    8   1   7      9   2   7  2346.0460  2346.0452
CO2-Xe2-131-131    5   2   4      6   3   4  2346.0460  2346.0466
CO2-Xe2-132-136    8   1   7      9   2   7  2346.0460  2346.0467
CO2-Xe2-129-134    5   2   4      6   3   4  2346.0460  2346.0465
CO2-Xe2-131-136    8   1   7      9   2   7  2346.0460  2346.0459
CO2-Xe2-131-134    5   2   4      6   3   4  2346.0460  2346.0473
                                     Blend         2346.0460  -0.0001
CO2-Xe2-132-132    5   2   3      6   3   3  2346.0506  2346.0517
CO2-Xe2-129-132    5   2   3      6   3   3  2346.0506  2346.0507
CO2-Xe2-129-129    5   2   3      6   3   3  2346.0506  2346.0496
CO2-Xe2-131-132    5   2   3      6   3   3  2346.0506  2346.0514
CO2-Xe2-129-131    5   2   3      6   3   3  2346.0506  2346.0503
CO2-Xe2-132-134    5   2   3      6   3   3  2346.0506  2346.0524
CO2-Xe2-129-134    5   2   3      6   3   3  2346.0506  2346.0513
CO2-Xe2-131-134    5   2   3      6   3   3  2346.0506  2346.0520
                                     Blend         2346.0509  -0.0004
CO2-Xe2-132-132    7   1   6      8   2   6  2346.0687  2346.0701
CO2-Xe2-129-132    7   1   6      8   2   6  2346.0687  2346.0680
CO2-Xe2-129-129    4   2   3      5   3   3  2346.0687  2346.0689
CO2-Xe2-131-132    7   1   6      8   2   6  2346.0687  2346.0694
CO2-Xe2-129-131    7   1   6      8   2   6  2346.0687  2346.0674
CO2-Xe2-132-132    7   0   7      8   1   7  2346.0687  2346.0692
CO2-Xe2-129-131    4   2   3      5   3   3  2346.0687  2346.0695
CO2-Xe2-131-132    7   0   7      8   1   7  2346.0687  2346.0683
CO2-Xe2-132-132    9   4   5      9   5   5  2346.0687  2346.0681
CO2-Xe2-129-132    9   4   5      9   5   5  2346.0687  2346.0687
                                     Blend         2346.0688  -0.0001
CO2-Xe2-129-132    6   1   5      7   2   5  2346.0941  2346.0929
CO2-Xe2-132-132    3   2   1      4   3   1  2346.0941  2346.0950
CO2-Xe2-129-132    3   2   1      4   3   1  2346.0941  2346.0943
CO2-Xe2-129-132    3   2   2      4   3   2  2346.0941  2346.0936
CO2-Xe2-129-129    3   2   1      4   3   1  2346.0941  2346.0937
CO2-Xe2-131-132    6   1   5      7   2   5  2346.0941  2346.0940
CO2-Xe2-131-132    3   2   1      4   3   1  2346.0941  2346.0948
CO2-Xe2-131-132    3   2   2      4   3   2  2346.0941  2346.0940
CO2-Xe2-129-131    3   2   1      4   3   1  2346.0941  2346.0941
CO2-Xe2-129-131    3   2   2      4   3   2  2346.0941  2346.0933
                                     Blend         2346.0939   0.0002
CO2-Xe2-132-132    5   1   4      6   2   4  2346.1171  2346.1173
CO2-Xe2-129-132    5   1   4      6   2   4  2346.1171  2346.1161
CO2-Xe2-132-132    2   2   1      3   3   1  2346.1171  2346.1174
CO2-Xe2-129-132    2   2   0      3   3   0  2346.1171  2346.1171
CO2-Xe2-129-132    2   2   1      3   3   1  2346.1171  2346.1170
CO2-Xe2-129-129    2   2   1      3   3   1  2346.1171  2346.1165
CO2-Xe2-131-132    5   1   4      6   2   4  2346.1171  2346.1169
CO2-Xe2-131-132    2   2   0      3   3   0  2346.1171  2346.1174
```

```
CO2-Xe2-131-132    2   2   1      3   3   1  2346.1171  2346.1172
                                     Blend   2346.1170   0.0002
CO2-Xe2-132-132    8   3   6      8   4   4  2346.1271  2346.1271
CO2-Xe2-129-132    8   3   6      8   4   4  2346.1271  2346.1275
CO2-Xe2-129-129    8   3   6      8   4   4  2346.1271  2346.1279
CO2-Xe2-129-132    9   3   7      9   4   5  2346.1271  2346.1273
CO2-Xe2-132-132    7   3   4      7   4   4  2346.1271  2346.1283
CO2-Xe2-129-132    7   3   5      7   4   3  2346.1271  2346.1274
CO2-Xe2-132-132   10   3   8     10   4   6  2346.1271  2346.1263
CO2-Xe2-129-132   10   3   8     10   4   6  2346.1271  2346.1266
CO2-Xe2-129-129   10   3   8     10   4   6  2346.1271  2346.1269
CO2-Xe2-132-132    6   3   4      6   4   2  2346.1271  2346.1268
                                     Blend   2346.1272  -0.0002
CO2-Xe2-132-132    3   1   2      4   2   2  2346.1589  2346.1597
CO2-Xe2-129-132    3   1   2      4   2   2  2346.1589  2346.1589
CO2-Xe2-129-129    3   1   2      4   2   2  2346.1589  2346.1582
CO2-Xe2-131-132    3   1   2      4   2   2  2346.1589  2346.1594
CO2-Xe2-129-131    3   1   2      4   2   2  2346.1589  2346.1587
CO2-Xe2-132-132   10   2   9     10   3   7  2346.1589  2346.1593
CO2-Xe2-129-132   10   2   9     10   3   7  2346.1589  2346.1584
CO2-Xe2-132-132   13   3  10     13   4  10  2346.1589  2346.1588
CO2-Xe2-131-132   10   2   9     10   3   7  2346.1589  2346.1590
                                     Blend   2346.1589  -0.0001
CO2-Xe2-129-132    8   1   7      8   2   7  2346.2873  2346.2871
CO2-Xe2-131-132    8   1   7      8   2   7  2346.2873  2346.2866
CO2-Xe2-129-131    8   1   7      8   2   7  2346.2873  2346.2874
CO2-Xe2-129-129   15   2  13     15   3  13  2346.2873  2346.2860
CO2-Xe2-131-131    8   1   7      8   2   7  2346.2873  2346.2871
CO2-Xe2-132-134    8   1   7      8   2   7  2346.2873  2346.2858
CO2-Xe2-129-134    8   1   7      8   2   7  2346.2873  2346.2866
CO2-Xe2-131-134    8   1   7      8   2   7  2346.2873  2346.2861
CO2-Xe2-131-136    8   1   7      8   2   7  2346.2873  2346.2856
                                     Blend   2346.2867   0.0006
CO2-Xe2-132-132    9   1   8      9   2   8  2346.2941  2346.2945
CO2-Xe2-129-132    9   1   8      9   2   8  2346.2941  2346.2953
CO2-Xe2-131-132    9   1   8      9   2   8  2346.2941  2346.2948
CO2-Xe2-129-131    9   1   8      9   2   8  2346.2941  2346.2956
CO2-Xe2-132-134    9   1   8      9   2   8  2346.2941  2346.2939
CO2-Xe2-129-134    9   1   8      9   2   8  2346.2941  2346.2948
CO2-Xe2-129-132   16   2  14     16   3  14  2346.2941  2346.2949
CO2-Xe2-132-136    9   1   8      9   2   8  2346.2941  2346.2934
CO2-Xe2-131-134    9   1   8      9   2   8  2346.2941  2346.2942
                                     Blend   2346.2948  -0.0006
CO2-Xe2-129-132    4   0   4      4   1   4  2346.3164  2346.3164
CO2-Xe2-131-132    4   0   4      4   1   4  2346.3164  2346.3163
CO2-Xe2-129-131    4   0   4      4   1   4  2346.3164  2346.3165
CO2-Xe2-129-132   12   1  11     12   2  11  2346.3164  2346.3167
CO2-Xe2-131-132   12   1  11     12   2  11  2346.3164  2346.3162
CO2-Xe2-129-131   12   1  11     12   2  11  2346.3164  2346.3170
CO2-Xe2-131-131    4   0   4      4   1   4  2346.3164  2346.3164
CO2-Xe2-132-134    4   0   4      4   1   4  2346.3164  2346.3161
CO2-Xe2-129-134    4   0   4      4   1   4  2346.3164  2346.3163
CO2-Xe2-132-136    4   0   4      4   1   4  2346.3164  2346.3159
                                     Blend   2346.3164   0.0000
CO2-Xe2-132-134    7   0   7      7   1   7  2346.3273  2346.3268
CO2-Xe2-132-132    7   0   7      7   1   7  2346.3273  2346.3269
CO2-Xe2-129-132    7   0   7      7   1   7  2346.3273  2346.3272
CO2-Xe2-129-129    7   0   7      7   1   7  2346.3273  2346.3274
CO2-Xe2-131-132    7   0   7      7   1   7  2346.3273  2346.3270
CO2-Xe2-129-131    7   0   7      7   1   7  2346.3273  2346.3272
CO2-Xe2-132-132   15   1  14     15   2  14  2346.3273  2346.3284
CO2-Xe2-129-134    7   0   7      7   1   7  2346.3273  2346.3270
                                     Blend   2346.3272   0.0001
```

| | | | | | | | | |
|---|---|---|---|---|---|---|---|---|
| CO2-Xe2-129-132 | 11 | 1 | 11 | 11 | 0 | 11 | 2346.3356 | 2346.3370 |
| CO2-Xe2-129-132 | 12 | 1 | 12 | 12 | 0 | 12 | 2346.3356 | 2346.3363 |
| CO2-Xe2-129-132 | 12 | 0 | 12 | 12 | 1 | 12 | 2346.3356 | 2346.3342 |
| CO2-Xe2-129-129 | 12 | 1 | 12 | 12 | 0 | 12 | 2346.3356 | 2346.3363 |
| CO2-Xe2-132-132 | 13 | 0 | 13 | 13 | 1 | 13 | 2346.3356 | 2346.3345 |
| CO2-Xe2-129-132 | 13 | 1 | 13 | 13 | 0 | 13 | 2346.3356 | 2346.3359 |
| CO2-Xe2-129-132 | 13 | 0 | 13 | 13 | 1 | 13 | 2346.3356 | 2346.3346 |
| CO2-Xe2-129-129 | 13 | 0 | 13 | 13 | 1 | 13 | 2346.3356 | 2346.3346 |
| CO2-Xe2-131-132 | 11 | 1 | 11 | 11 | 0 | 11 | 2346.3356 | 2346.3371 |
| CO2-Xe2-132-132 | 12 | 1 | 12 | 12 | 0 | 12 | 2346.3356 | 2346.3364 |
| | | | | | | Blend | 2346.3357 | -0.0001 |
| CO2-Xe2-132-132 | 6 | 1 | 6 | 6 | 0 | 6 | 2346.3468 | 2346.3471 |
| CO2-Xe2-129-132 | 6 | 1 | 6 | 6 | 0 | 6 | 2346.3468 | 2346.3469 |
| CO2-Xe2-129-129 | 6 | 1 | 6 | 6 | 0 | 6 | 2346.3468 | 2346.3466 |
| CO2-Xe2-131-132 | 6 | 1 | 6 | 6 | 0 | 6 | 2346.3468 | 2346.3470 |
| CO2-Xe2-129-131 | 6 | 1 | 6 | 6 | 0 | 6 | 2346.3468 | 2346.3468 |
| CO2-Xe2-132-132 | 14 | 2 | 13 | 14 | 1 | 13 | 2346.3468 | 2346.3457 |
| CO2-Xe2-132-134 | 6 | 1 | 6 | 6 | 0 | 6 | 2346.3468 | 2346.3473 |
| CO2-Xe2-131-132 | 14 | 2 | 13 | 14 | 1 | 13 | 2346.3468 | 2346.3455 |
| CO2-Xe2-129-134 | 6 | 1 | 6 | 6 | 0 | 6 | 2346.3468 | 2346.3470 |
| | | | | | | Blend | 2346.3467 | 0.0001 |
| CO2-Xe2-129-132 | 5 | 1 | 5 | 5 | 0 | 5 | 2346.3504 | 2346.3505 |
| CO2-Xe2-131-132 | 5 | 1 | 5 | 5 | 0 | 5 | 2346.3504 | 2346.3507 |
| CO2-Xe2-129-131 | 5 | 1 | 5 | 5 | 0 | 5 | 2346.3504 | 2346.3505 |
| CO2-Xe2-131-132 | 13 | 2 | 12 | 13 | 1 | 12 | 2346.3504 | 2346.3497 |
| CO2-Xe2-131-131 | 5 | 1 | 5 | 5 | 0 | 5 | 2346.3504 | 2346.3505 |
| CO2-Xe2-132-134 | 5 | 1 | 5 | 5 | 0 | 5 | 2346.3504 | 2346.3510 |
| CO2-Xe2-129-134 | 5 | 1 | 5 | 5 | 0 | 5 | 2346.3504 | 2346.3507 |
| CO2-Xe2-132-136 | 5 | 1 | 5 | 5 | 0 | 5 | 2346.3504 | 2346.3511 |
| CO2-Xe2-131-134 | 5 | 1 | 5 | 5 | 0 | 5 | 2346.3504 | 2346.3509 |
| CO2-Xe2-131-136 | 5 | 1 | 5 | 5 | 0 | 5 | 2346.3504 | 2346.3510 |
| | | | | | | Blend | 2346.3506 | -0.0002 |
| CO2-Xe2-132-132 | 4 | 1 | 4 | 4 | 0 | 4 | 2346.3549 | 2346.3547 |
| CO2-Xe2-129-132 | 4 | 1 | 4 | 4 | 0 | 4 | 2346.3549 | 2346.3545 |
| CO2-Xe2-129-129 | 4 | 1 | 4 | 4 | 0 | 4 | 2346.3549 | 2346.3543 |
| CO2-Xe2-131-132 | 4 | 1 | 4 | 4 | 0 | 4 | 2346.3549 | 2346.3546 |
| CO2-Xe2-129-131 | 4 | 1 | 4 | 4 | 0 | 4 | 2346.3549 | 2346.3544 |
| CO2-Xe2-132-132 | 12 | 2 | 11 | 12 | 1 | 11 | 2346.3549 | 2346.3551 |
| CO2-Xe2-129-132 | 12 | 2 | 11 | 12 | 1 | 11 | 2346.3549 | 2346.3544 |
| CO2-Xe2-129-129 | 12 | 2 | 11 | 12 | 1 | 11 | 2346.3549 | 2346.3536 |
| CO2-Xe2-131-132 | 12 | 2 | 11 | 12 | 1 | 11 | 2346.3549 | 2346.3549 |
| CO2-Xe2-129-131 | 12 | 2 | 11 | 12 | 1 | 11 | 2346.3549 | 2346.3541 |
| | | | | | | Blend | 2346.3545 | 0.0005 |
| CO2-Xe2-129-132 | 11 | 2 | 10 | 11 | 1 | 10 | 2346.3612 | 2346.3606 |
| CO2-Xe2-131-132 | 11 | 2 | 10 | 11 | 1 | 10 | 2346.3612 | 2346.3612 |
| CO2-Xe2-129-131 | 11 | 2 | 10 | 11 | 1 | 10 | 2346.3612 | 2346.3604 |
| CO2-Xe2-132-132 | 2 | 1 | 2 | 2 | 0 | 2 | 2346.3612 | 2346.3617 |
| CO2-Xe2-129-132 | 2 | 1 | 2 | 2 | 0 | 2 | 2346.3612 | 2346.3616 |
| CO2-Xe2-129-129 | 2 | 1 | 2 | 2 | 0 | 2 | 2346.3612 | 2346.3615 |
| CO2-Xe2-131-132 | 2 | 1 | 2 | 2 | 0 | 2 | 2346.3612 | 2346.3617 |
| CO2-Xe2-129-131 | 2 | 1 | 2 | 2 | 0 | 2 | 2346.3612 | 2346.3615 |
| CO2-Xe2-131-131 | 11 | 2 | 10 | 11 | 1 | 10 | 2346.3612 | 2346.3607 |
| CO2-Xe2-132-134 | 11 | 2 | 10 | 11 | 1 | 10 | 2346.3612 | 2346.3620 |
| | | | | | | Blend | 2346.3612 | 0.0000 |
| CO2-Xe2-132-132 | 6 | 2 | 5 | 6 | 1 | 5 | 2346.3996 | 2346.4002 |
| CO2-Xe2-129-132 | 6 | 2 | 5 | 6 | 1 | 5 | 2346.3996 | 2346.3996 |
| CO2-Xe2-129-129 | 6 | 2 | 5 | 6 | 1 | 5 | 2346.3996 | 2346.3990 |
| CO2-Xe2-131-132 | 6 | 2 | 5 | 6 | 1 | 5 | 2346.3996 | 2346.4000 |
| CO2-Xe2-129-131 | 6 | 2 | 5 | 6 | 1 | 5 | 2346.3996 | 2346.3994 |
| CO2-Xe2-132-132 | 14 | 3 | 12 | 14 | 2 | 12 | 2346.3996 | 2346.4010 |
| CO2-Xe2-129-132 | 14 | 3 | 12 | 14 | 2 | 12 | 2346.3996 | 2346.3991 |
| CO2-Xe2-131-132 | 14 | 3 | 12 | 14 | 2 | 12 | 2346.3996 | 2346.4003 |
| CO2-Xe2-132-134 | 6 | 2 | 5 | 6 | 1 | 5 | 2346.3996 | 2346.4006 |

```
CO2-Xe2-129-134    6   2   5     6   1   5  2346.3996   2346.4000
                                          Blend         2346.3998  -0.0002
CO2-Xe2-129-132    5   2   4     5   1   4  2346.4066   2346.4065
CO2-Xe2-131-132    5   2   4     5   1   4  2346.4066   2346.4068
CO2-Xe2-129-131    5   2   4     5   1   4  2346.4066   2346.4063
CO2-Xe2-131-131    5   2   4     5   1   4  2346.4066   2346.4063
CO2-Xe2-132-134    5   2   4     5   1   4  2346.4066   2346.4073
CO2-Xe2-129-134    5   2   4     5   1   4  2346.4066   2346.4068
CO2-Xe2-132-136    5   2   4     5   1   4  2346.4066   2346.4076
CO2-Xe2-131-134    5   2   4     5   1   4  2346.4066   2346.4071
CO2-Xe2-131-136    5   2   4     5   1   4  2346.4066   2346.4074
                                          Blend         2346.4067  -0.0001
CO2-Xe2-132-132    5   2   3     4   1   3  2346.5333   2346.5320
CO2-Xe2-129-132    5   2   3     4   1   3  2346.5333   2346.5330
CO2-Xe2-129-129    5   2   3     4   1   3  2346.5333   2346.5340
CO2-Xe2-129-132    6   1   5     5   0   5  2346.5333   2346.5334
CO2-Xe2-131-132    5   2   3     4   1   3  2346.5333   2346.5324
CO2-Xe2-129-131    5   2   3     4   1   3  2346.5333   2346.5334
CO2-Xe2-131-132    6   1   5     5   0   5  2346.5333   2346.5321
CO2-Xe2-129-131    6   1   5     5   0   5  2346.5333   2346.5341
CO2-Xe2-132-132   10   4   7    10   3   7  2346.5333   2346.5343
CO2-Xe2-129-132   10   4   7    10   3   7  2346.5333   2346.5335
                                          Blend         2346.5332   0.0001
CO2-Xe2-132-132    9   4   5     9   3   7  2346.5434   2346.5436
CO2-Xe2-129-132    8   4   4     8   3   6  2346.5434   2346.5430
CO2-Xe2-129-132    9   4   5     9   3   7  2346.5434   2346.5433
CO2-Xe2-132-132    7   4   3     7   3   5  2346.5434   2346.5434
CO2-Xe2-129-129    9   4   5     9   3   7  2346.5434   2346.5429
CO2-Xe2-129-132    7   4   3     7   3   5  2346.5434   2346.5430
CO2-Xe2-129-129    7   4   3     7   3   5  2346.5434   2346.5426
CO2-Xe2-129-132   10   4   6    10   3   8  2346.5434   2346.5441
CO2-Xe2-132-132    6   4   3     6   3   3  2346.5434   2346.5430
CO2-Xe2-129-132    6   4   3     6   3   3  2346.5434   2346.5426
                                          Blend         2346.5432   0.0003
CO2-Xe2-132-132    3   3   0     2   2   0  2346.5535   2346.5530
CO2-Xe2-129-132    3   3   0     2   2   0  2346.5535   2346.5535
CO2-Xe2-129-132    3   3   1     2   2   1  2346.5535   2346.5536
CO2-Xe2-129-129    3   3   0     2   2   0  2346.5535   2346.5539
CO2-Xe2-131-132    6   2   4     5   1   4  2346.5535   2346.5541
CO2-Xe2-131-132    3   3   0     2   2   0  2346.5535   2346.5532
CO2-Xe2-131-132    3   3   1     2   2   1  2346.5535   2346.5533
CO2-Xe2-129-131    3   3   0     2   2   0  2346.5535   2346.5536
CO2-Xe2-129-131    3   3   1     2   2   1  2346.5535   2346.5538
                                          Blend         2346.5536   0.0000
CO2-Xe2-132-132    7   4   3     6   3   3  2346.7061   2346.7039
CO2-Xe2-129-132    7   4   3     6   3   3  2346.7061   2346.7051
CO2-Xe2-129-132    7   4   4     6   3   4  2346.7061   2346.7056
CO2-Xe2-129-129    7   4   3     6   3   3  2346.7061   2346.7064
CO2-Xe2-131-132    7   4   3     6   3   3  2346.7061   2346.7043
CO2-Xe2-131-132    7   4   4     6   3   4  2346.7061   2346.7048
CO2-Xe2-129-131    7   4   3     6   3   3  2346.7061   2346.7055
CO2-Xe2-129-131    7   4   4     6   3   4  2346.7061   2346.7060
CO2-Xe2-131-131    7   4   4     6   3   4  2346.7061   2346.7045
                                          Blend         2346.7052   0.0009
CO2-Xe2-132-132    5   5   0     4   4   0  2346.7194   2346.7178
CO2-Xe2-129-132    5   5   0     4   4   0  2346.7194   2346.7185
CO2-Xe2-129-132    5   5   1     4   4   1  2346.7194   2346.7185
CO2-Xe2-129-129    5   5   0     4   4   0  2346.7194   2346.7192
CO2-Xe2-131-132    5   5   0     4   4   0  2346.7194   2346.7181
CO2-Xe2-131-132    5   5   1     4   4   1  2346.7194   2346.7181
CO2-Xe2-129-132   11   3   8    10   2   8  2346.7194   2346.7184
CO2-Xe2-129-131    5   5   0     4   4   0  2346.7194   2346.7187
CO2-Xe2-129-131    5   5   1     4   4   1  2346.7194   2346.7187
```

```
CO2-Xe2-129-129   11   3    8     10   2    8  2346.7194  2346.7206
                                          Blend   2346.7187     0.0007
CO2-Xe2-132-132    8   4    5      7   3    5  2346.7281  2346.7274
CO2-Xe2-129-132    8   4    4      7   3    4  2346.7281  2346.7277
CO2-Xe2-129-132    8   4    5      7   3    5  2346.7281  2346.7289
CO2-Xe2-132-132   10   3    8      9   2    8  2346.7281  2346.7270
CO2-Xe2-131-132    8   4    4      7   3    4  2346.7281  2346.7267
CO2-Xe2-131-132    8   4    5      7   3    5  2346.7281  2346.7279
CO2-Xe2-129-131    8   4    4      7   3    4  2346.7281  2346.7282
CO2-Xe2-129-131    8   4    5      7   3    5  2346.7281  2346.7294
CO2-Xe2-131-132   10   3    8      9   2    8  2346.7281  2346.7278
CO2-Xe2-129-131   12   2   10     11   1   10  2346.7281  2346.7272
                                          Blend   2346.7278     0.0003
CO2-Xe2-132-132    6   5    2      5   4    2  2346.7412  2346.7408
CO2-Xe2-129-132    6   5    1      5   4    1  2346.7412  2346.7417
CO2-Xe2-129-132    6   5    2      5   4    2  2346.7412  2346.7417
CO2-Xe2-129-129    6   5    2      5   4    2  2346.7412  2346.7426
CO2-Xe2-131-132    6   5    1      5   4    1  2346.7412  2346.7411
CO2-Xe2-131-132    6   5    2      5   4    2  2346.7412  2346.7411
CO2-Xe2-129-131    6   5    1      5   4    1  2346.7412  2346.7420
CO2-Xe2-129-131    6   5    2      5   4    2  2346.7412  2346.7420
CO2-Xe2-131-131    6   5    1      5   4    1  2346.7412  2346.7405
CO2-Xe2-132-134    6   5    1      5   4    1  2346.7412  2346.7402
                                          Blend   2346.7415    -0.0003
CO2-Xe2-132-132   12   2   11     11   1   11  2346.7802  2346.7784
CO2-Xe2-131-132   12   2   11     11   1   11  2346.7802  2346.7797
CO2-Xe2-132-134   14   3   11     13   2   11  2346.7802  2346.7821
CO2-Xe2-132-136   14   3   11     13   2   11  2346.7802  2346.7798
CO2-Xe2-132-136   12   3   10     11   2   10  2346.7802  2346.7804
CO2-Xe2-131-136   14   3   11     13   2   11  2346.7802  2346.7810
CO2-Xe2-131-136   12   3   10     11   2   10  2346.7802  2346.7815
CO2-Xe2-132-132   11   8    3     11   7    5  2346.7802  2346.7809
CO2-Xe2-132-132   12   8    5     12   7    5  2346.7802  2346.7807
CO2-Xe2-129-132   11   8    3     11   7    5  2346.7802  2346.7800
                                          Blend   2346.7802     0.0000
CO2-Xe2-132-132    7   6    1      6   5    1  2346.8242  2346.8231
CO2-Xe2-129-132    7   6    2      6   5    2  2346.8242  2346.8241
CO2-Xe2-129-132    7   6    1      6   5    1  2346.8242  2346.8241
CO2-Xe2-129-129    7   6    1      6   5    1  2346.8242  2346.8251
CO2-Xe2-131-132    7   6    2      6   5    2  2346.8242  2346.8234
CO2-Xe2-131-132    7   6    1      6   5    1  2346.8242  2346.8234
CO2-Xe2-129-131    7   6    2      6   5    2  2346.8242  2346.8244
CO2-Xe2-129-131    7   6    1      6   5    1  2346.8242  2346.8244
CO2-Xe2-129-132   12   4    9     11   3    9  2346.8242  2346.8245
CO2-Xe2-131-132   12   4    9     11   3    9  2346.8242  2346.8227
                                          Blend   2346.8240     0.0003
CO2-Xe2-132-132    8   6    3      7   5    3  2346.8471  2346.8460
CO2-Xe2-129-132    8   6    3      7   5    3  2346.8471  2346.8473
CO2-Xe2-129-132    8   6    2      7   5    2  2346.8471  2346.8473
CO2-Xe2-131-132    8   6    3      7   5    3  2346.8471  2346.8464
CO2-Xe2-131-132    8   6    2      7   5    2  2346.8471  2346.8464
CO2-Xe2-129-131    8   6    3      7   5    3  2346.8471  2346.8477
CO2-Xe2-129-131    8   6    2      7   5    2  2346.8471  2346.8477
CO2-Xe2-129-132   14   4   10     13   3   10  2346.8471  2346.8464
CO2-Xe2-131-132   13   4   10     12   3   10  2346.8471  2346.8477
CO2-Xe2-132-132   14   3   12     13   2   12  2346.8471  2346.8470
                                          Blend   2346.8470     0.0002
CO2-Xe2-132-132    8   7    2      7   6    2  2346.9059  2346.9053
CO2-Xe2-129-132    8   7    2      7   6    2  2346.9059  2346.9065
CO2-Xe2-129-132    8   7    1      7   6    1  2346.9059  2346.9065
CO2-Xe2-131-132    8   7    2      7   6    2  2346.9059  2346.9057
CO2-Xe2-131-132    8   7    1      7   6    1  2346.9059  2346.9057
CO2-Xe2-129-131    8   7    2      7   6    2  2346.9059  2346.9068
```

```
CO2-Xe2-129-131    8   7   1      7   6   1  2346.9059  2346.9068
CO2-Xe2-131-131    8   7   1      7   6   1  2346.9059  2346.9048
CO2-Xe2-132-134    8   7   2      7   6   2  2346.9059  2346.9046
CO2-Xe2-132-134    8   7   1      7   6   1  2346.9059  2346.9046
                                     Blend     2346.9059    0.0000
CO2-Xe2-132-132    9   7   2      8   6   2  2346.9289  2346.9282
CO2-Xe2-129-132    9   7   3      8   6   3  2346.9289  2346.9296
CO2-Xe2-129-132    9   7   2      8   6   2  2346.9289  2346.9296
CO2-Xe2-131-132    9   7   3      8   6   3  2346.9289  2346.9287
CO2-Xe2-131-132    9   7   2      8   6   2  2346.9289  2346.9287
CO2-Xe2-129-131    9   7   3      8   6   3  2346.9289  2346.9301
CO2-Xe2-129-131    9   7   2      8   6   2  2346.9289  2346.9301
CO2-Xe2-129-129   14   5  10     13   4  10  2346.9289  2346.9292
CO2-Xe2-129-132   16   4  13     15   3  13  2346.9289  2346.9308
CO2-Xe2-131-131    9   7   3      8   6   3  2346.9289  2346.9279
                                     Blend     2346.9293   -0.0004
CO2-Xe2-132-132    9   8   1      8   7   1  2346.9877  2346.9875
CO2-Xe2-129-132    9   8   2      8   7   2  2346.9877  2346.9888
CO2-Xe2-129-132    9   8   1      8   7   1  2346.9877  2346.9888
CO2-Xe2-131-132    9   8   2      8   7   2  2346.9877  2346.9879
CO2-Xe2-131-132    9   8   1      8   7   1  2346.9877  2346.9879
CO2-Xe2-129-131    9   8   2      8   7   2  2346.9877  2346.9892
CO2-Xe2-129-131    9   8   1      8   7   1  2346.9877  2346.9892
CO2-Xe2-129-129   14   6   9     13   5   9  2346.9877  2346.9882
CO2-Xe2-131-131    9   8   2      8   7   2  2346.9877  2346.9869
CO2-Xe2-132-134    9   8   2      8   7   2  2346.9877  2346.9867
                                     Blend     2346.9882   -0.0005
```

Table A-13. Observed and calculated line positions for $CO_2$-$Xe_2$ combination band (in $cm^{-1}$).

```
****************************************************************
                J'  Ka' Kc'  J"  Ka" Kc"   Obs        Calc        O-C
****************************************************************
CO2-Xe2-132-132   10   4   6     11   5   7  2380.5062  2380.5053
CO2-Xe2-129-132   10   4   6     11   5   7  2380.5062  2380.5055
CO2-Xe2-129-132   10   4   7     11   5   6  2380.5062  2380.5048
CO2-Xe2-129-129   10   4   6     11   5   7  2380.5062  2380.5058
CO2-Xe2-131-132   10   4   6     11   5   7  2380.5062  2380.5054
CO2-Xe2-131-132   10   4   7     11   5   6  2380.5062  2380.5046
CO2-Xe2-129-131   10   4   6     11   5   7  2380.5062  2380.5056
CO2-Xe2-129-131   10   4   7     11   5   6  2380.5062  2380.5048
                                     Blend     2380.5052    0.0009
CO2-Xe2-129-132    7   5   3      8   6   2  2380.5106  2380.5100
CO2-Xe2-129-132    7   5   2      8   6   3  2380.5106  2380.5100
CO2-Xe2-129-129    7   5   3      8   6   2  2380.5106  2380.5110
CO2-Xe2-131-132    7   5   3      8   6   2  2380.5106  2380.5093
CO2-Xe2-131-132    7   5   2      8   6   3  2380.5106  2380.5093
CO2-Xe2-129-131    7   5   3      8   6   2  2380.5106  2380.5103
CO2-Xe2-129-131    7   5   2      8   6   3  2380.5106  2380.5103
CO2-Xe2-129-132   12   3  10     13   4   9  2380.5106  2380.5106
CO2-Xe2-131-131    7   5   2      8   6   3  2380.5106  2380.5107
CO2-Xe2-129-134    7   5   3      8   6   2  2380.5106  2380.5093
                                     Blend     2380.5101    0.0005
CO2-Xe2-132-132    9   4   6     10   5   5  2380.5288  2380.5277
CO2-Xe2-129-132    9   4   5     10   5   6  2380.5288  2380.5285
CO2-Xe2-129-132    9   4   6     10   5   5  2380.5288  2380.5282
CO2-Xe2-129-129    9   4   6     10   5   5  2380.5288  2380.5286
CO2-Xe2-131-132    9   4   5     10   5   6  2380.5288  2380.5282
CO2-Xe2-131-132    9   4   6     10   5   5  2380.5288  2380.5279
CO2-Xe2-129-131    9   4   5     10   5   6  2380.5288  2380.5287
```

```
CO2-Xe2-129-131     9   4   6   10   5   5  2380.5288 2380.5284
                                       Blend     2380.5283   0.0005
CO2-Xe2-132-132    11   3   9   12   4   8  2380.5382 2380.5392
CO2-Xe2-129-132    11   3   9   12   4   8  2380.5382 2380.5386
CO2-Xe2-129-129    11   3   9   12   4   8  2380.5382 2380.5380
CO2-Xe2-131-132    11   3   9   12   4   8  2380.5382 2380.5390
CO2-Xe2-129-131    11   3   9   12   4   8  2380.5382 2380.5385
CO2-Xe2-132-134    11   3   9   12   4   8  2380.5382 2380.5395
CO2-Xe2-129-134    11   3   9   12   4   8  2380.5382 2380.5390
CO2-Xe2-132-136    11   3   9   12   4   8  2380.5382 2380.5398
CO2-Xe2-131-134    11   3   9   12   4   8  2380.5382 2380.5393
                                       Blend     2380.5388  -0.0006
CO2-Xe2-132-132     8   4   4    9   5   5  2380.5521 2380.5511
CO2-Xe2-129-132     8   4   4    9   5   5  2380.5521 2380.5518
CO2-Xe2-129-132     8   4   5    9   5   4  2380.5521 2380.5516
CO2-Xe2-129-129     8   4   4    9   5   5  2380.5521 2380.5524
CO2-Xe2-131-132     8   4   4    9   5   5  2380.5521 2380.5513
CO2-Xe2-131-132     8   4   5    9   5   4  2380.5521 2380.5512
CO2-Xe2-129-131     8   4   4    9   5   5  2380.5521 2380.5520
CO2-Xe2-129-131     8   4   5    9   5   4  2380.5521 2380.5519
                                       Blend     2380.5517   0.0005
CO2-Xe2-132-132     7   4   4    8   5   3  2380.5749 2380.5741
CO2-Xe2-129-132     7   4   3    8   5   4  2380.5749 2380.5751
CO2-Xe2-129-132     7   4   4    8   5   3  2380.5749 2380.5751
CO2-Xe2-129-129     7   4   4    8   5   3  2380.5749 2380.5760
CO2-Xe2-131-132     7   4   3    8   5   4  2380.5749 2380.5745
CO2-Xe2-131-132     7   4   4    8   5   3  2380.5749 2380.5744
CO2-Xe2-129-131     7   4   3    8   5   4  2380.5749 2380.5754
CO2-Xe2-129-131     7   4   4    8   5   3  2380.5749 2380.5754
                                       Blend     2380.5750  -0.0001
CO2-Xe2-132-132     9   3   7   10   4   6  2380.5890 2380.5900
CO2-Xe2-129-132     9   3   7   10   4   6  2380.5890 2380.5902
CO2-Xe2-129-129     9   3   7   10   4   6  2380.5890 2380.5904
CO2-Xe2-131-132     9   3   7   10   4   6  2380.5890 2380.5901
CO2-Xe2-129-131     9   3   7   10   4   6  2380.5890 2380.5903
CO2-Xe2-132-134     9   3   7   10   4   6  2380.5890 2380.5898
CO2-Xe2-129-134     9   3   7   10   4   6  2380.5890 2380.5901
CO2-Xe2-131-134     9   3   7   10   4   6  2380.5890 2380.5899
CO2-Xe2-131-136     9   3   7   10   4   6  2380.5890 2380.5897
                                       Blend     2380.5901  -0.0011
CO2-Xe2-132-132     6   4   2    7   5   3  2380.5975 2380.5973
CO2-Xe2-129-132     6   4   2    7   5   3  2380.5975 2380.5985
CO2-Xe2-129-132     6   4   3    7   5   2  2380.5975 2380.5985
CO2-Xe2-131-132     6   4   2    7   5   3  2380.5975 2380.5977
CO2-Xe2-131-132     6   4   3    7   5   2  2380.5975 2380.5977
CO2-Xe2-129-131     6   4   2    7   5   3  2380.5975 2380.5989
CO2-Xe2-129-131     6   4   3    7   5   2  2380.5975 2380.5989
CO2-Xe2-129-132     9   3   6   10   4   7  2380.5975 2380.5971
CO2-Xe2-131-132     9   3   6   10   4   7  2380.5975 2380.5967
CO2-Xe2-129-131     9   3   6   10   4   7  2380.5975 2380.5973
                                       Blend     2380.5979  -0.0005
CO2-Xe2-132-132     5   4   2    6   5   1  2380.6208 2380.6205
CO2-Xe2-129-132     5   4   1    6   5   2  2380.6208 2380.6219
CO2-Xe2-129-132     5   4   2    6   5   1  2380.6208 2380.6219
CO2-Xe2-131-132     5   4   1    6   5   2  2380.6208 2380.6210
CO2-Xe2-131-132     5   4   2    6   5   1  2380.6208 2380.6210
CO2-Xe2-132-132    18   0  18   19   1  19  2380.6208 2380.6213
CO2-Xe2-129-132    18   0  18   19   1  19  2380.6208 2380.6198
CO2-Xe2-129-132    18   1  18   19   0  19  2380.6208 2380.6199
CO2-Xe2-131-132    18   0  18   19   1  19  2380.6208 2380.6208
                                       Blend     2380.6210  -0.0002
CO2-Xe2-132-132     9   2   8   10   3   7  2380.6295 2380.6305
CO2-Xe2-132-132    17   2  16   18   1  17  2380.6295 2380.6301
```

```
CO2-Xe2-129-132    9   2   8   10   3   7   2380.6295 2380.6297
CO2-Xe2-129-132   17   2  16   18   1  17   2380.6295 2380.6284
CO2-Xe2-129-129    9   2   8   10   3   7   2380.6295 2380.6289
CO2-Xe2-131-132    9   2   8   10   3   7   2380.6295 2380.6302
CO2-Xe2-131-132   17   2  16   18   1  17   2380.6295 2380.6296
CO2-Xe2-129-131    9   2   8   10   3   7   2380.6295 2380.6295
CO2-Xe2-129-131   17   2  16   18   1  17   2380.6295 2380.6278
                                     Blend      2380.6294    0.0001
CO2-Xe2-132-132    6   3   3    7   4   4   2380.6620 2380.6618
CO2-Xe2-129-132    6   3   3    7   4   4   2380.6620 2380.6629
CO2-Xe2-129-132    6   3   4    7   4   3   2380.6620 2380.6622
CO2-Xe2-131-132    6   3   3    7   4   4   2380.6620 2380.6621
CO2-Xe2-131-132    6   3   4    7   4   3   2380.6620 2380.6615
CO2-Xe2-129-131    6   3   3    7   4   4   2380.6620 2380.6632
CO2-Xe2-129-131    6   3   4    7   4   3   2380.6620 2380.6626
CO2-Xe2-129-132    8   2   7    9   3   6   2380.6620 2380.6623
CO2-Xe2-131-132    8   2   7    9   3   6   2380.6620 2380.6624
CO2-Xe2-129-131    8   2   7    9   3   6   2380.6620 2380.6623
                                     Blend      2380.6623   -0.0003
CO2-Xe2-132-132   16   0  16   17   1  17   2380.6656 2380.6661
CO2-Xe2-129-132   16   0  16   17   1  17   2380.6656 2380.6650
CO2-Xe2-129-132   16   1  16   17   0  17   2380.6656 2380.6652
CO2-Xe2-131-132   16   0  16   17   1  17   2380.6656 2380.6658
CO2-Xe2-131-132   16   1  16   17   0  17   2380.6656 2380.6660
CO2-Xe2-129-131   16   0  16   17   1  17   2380.6656 2380.6646
CO2-Xe2-129-131   16   1  16   17   0  17   2380.6656 2380.6648
CO2-Xe2-129-132   15   1  14   16   2  15   2380.6656 2380.6658
                                     Blend      2380.6654    0.0002
CO2-Xe2-132-132   15   1  15   16   0  16   2380.6877 2380.6887
CO2-Xe2-129-132   15   0  15   16   1  16   2380.6877 2380.6874
CO2-Xe2-129-132   15   1  15   16   0  16   2380.6877 2380.6878
CO2-Xe2-129-132    5   3   2    6   4   3   2380.6877 2380.6859
CO2-Xe2-129-132    5   3   3    6   4   2   2380.6877 2380.6857
CO2-Xe2-129-129   15   1  15   16   0  16   2380.6877 2380.6868
CO2-Xe2-129-129    5   3   3    6   4   2   2380.6877 2380.6870
CO2-Xe2-131-132   15   0  15   16   1  16   2380.6877 2380.6880
CO2-Xe2-131-132   15   1  15   16   0  16   2380.6877 2380.6884
CO2-Xe2-129-131   15   0  15   16   1  16   2380.6877 2380.6871
                                     Blend      2380.6873    0.0005
CO2-Xe2-132-132   13   1  13   14   0  14   2380.7321 2380.7332
CO2-Xe2-129-132   13   0  13   14   1  14   2380.7321 2380.7316
CO2-Xe2-129-132   13   1  13   14   0  14   2380.7321 2380.7326
CO2-Xe2-129-129   13   1  13   14   0  14   2380.7321 2380.7320
CO2-Xe2-129-132    3   3   0    4   4   1   2380.7321 2380.7325
CO2-Xe2-129-132    3   3   1    4   4   0   2380.7321 2380.7325
CO2-Xe2-131-132   13   0  13   14   1  14   2380.7321 2380.7319
CO2-Xe2-131-132   13   1  13   14   0  14   2380.7321 2380.7330
CO2-Xe2-129-131   13   0  13   14   1  14   2380.7321 2380.7315
CO2-Xe2-129-131   13   1  13   14   0  14   2380.7321 2380.7324
                                     Blend      2380.7323   -0.0002
CO2-Xe2-132-132   12   0  12   13   1  13   2380.7540 2380.7537
CO2-Xe2-129-132   12   0  12   13   1  13   2380.7540 2380.7534
CO2-Xe2-129-132   12   1  12   13   0  13   2380.7540 2380.7550
CO2-Xe2-129-129   12   0  12   13   1  13   2380.7540 2380.7531
CO2-Xe2-131-132   12   0  12   13   1  13   2380.7540 2380.7536
CO2-Xe2-131-132   12   1  12   13   0  13   2380.7540 2380.7553
CO2-Xe2-129-131   12   0  12   13   1  13   2380.7540 2380.7533
CO2-Xe2-129-131   12   1  12   13   0  13   2380.7540 2380.7549
CO2-Xe2-132-132   10   1   9   11   2  10   2380.7540 2380.7535
CO2-Xe2-129-132   10   1   9   11   2  10   2380.7540 2380.7542
                                     Blend      2380.7540    0.0000
CO2-Xe2-132-132   10   0  10   11   1  11   2380.7958 2380.7958
CO2-Xe2-129-132   10   0  10   11   1  11   2380.7958 2380.7960
```

| | | | | | | | | |
|---|---|---|---|---|---|---|---|---|
| CO2-Xe2-129-129 | 10 | 0 | 10 | 11 | 1 | 11 | 2380.7958 | 2380.7962 |
| CO2-Xe2-131-132 | 10 | 0 | 10 | 11 | 1 | 11 | 2380.7958 | 2380.7959 |
| CO2-Xe2-129-131 | 10 | 0 | 10 | 11 | 1 | 11 | 2380.7958 | 2380.7961 |
| CO2-Xe2-129-132 | 3 | 2 | 1 | 4 | 3 | 2 | 2380.7958 | 2380.7956 |
| CO2-Xe2-129-132 | 3 | 2 | 2 | 4 | 3 | 1 | 2380.7958 | 2380.7947 |
| CO2-Xe2-129-129 | 3 | 2 | 2 | 4 | 3 | 1 | 2380.7958 | 2380.7963 |
| CO2-Xe2-131-132 | 3 | 2 | 1 | 4 | 3 | 2 | 2380.7958 | 2380.7945 |
| CO2-Xe2-129-131 | 3 | 2 | 1 | 4 | 3 | 2 | 2380.7958 | 2380.7962 |
| | | | | | Blend | | 2380.7958 | 0.0000 |
| CO2-Xe2-129-132 | 10 | 1 | 10 | 11 | 0 | 11 | 2380.7996 | 2380.8000 |
| CO2-Xe2-131-132 | 10 | 1 | 10 | 11 | 0 | 11 | 2380.7996 | 2380.8000 |
| CO2-Xe2-129-131 | 10 | 1 | 10 | 11 | 0 | 11 | 2380.7996 | 2380.8000 |
| CO2-Xe2-129-132 | 7 | 1 | 6 | 8 | 2 | 7 | 2380.7996 | 2380.7982 |
| CO2-Xe2-129-131 | 7 | 1 | 6 | 8 | 2 | 7 | 2380.7996 | 2380.7986 |
| CO2-Xe2-131-131 | 10 | 1 | 10 | 11 | 0 | 11 | 2380.7996 | 2380.8000 |
| CO2-Xe2-132-134 | 10 | 1 | 10 | 11 | 0 | 11 | 2380.7996 | 2380.8000 |
| CO2-Xe2-129-134 | 10 | 1 | 10 | 11 | 0 | 11 | 2380.7996 | 2380.8000 |
| CO2-Xe2-132-136 | 10 | 1 | 10 | 11 | 0 | 11 | 2380.7996 | 2380.8000 |
| CO2-Xe2-131-134 | 10 | 1 | 10 | 11 | 0 | 11 | 2380.7996 | 2380.8000 |
| | | | | | Blend | | 2380.7997 | -0.0002 |
| CO2-Xe2-132-132 | 9 | 1 | 9 | 10 | 0 | 10 | 2380.8223 | 2380.8227 |
| CO2-Xe2-129-132 | 9 | 1 | 9 | 10 | 0 | 10 | 2380.8223 | 2380.8228 |
| CO2-Xe2-129-129 | 9 | 1 | 9 | 10 | 0 | 10 | 2380.8223 | 2380.8229 |
| CO2-Xe2-131-132 | 9 | 1 | 9 | 10 | 0 | 10 | 2380.8223 | 2380.8227 |
| CO2-Xe2-129-131 | 9 | 1 | 9 | 10 | 0 | 10 | 2380.8223 | 2380.8229 |
| CO2-Xe2-129-129 | 2 | 2 | 0 | 3 | 3 | 1 | 2380.8223 | 2380.8204 |
| CO2-Xe2-132-132 | 8 | 3 | 5 | 8 | 4 | 4 | 2380.8223 | 2380.8245 |
| CO2-Xe2-129-132 | 9 | 3 | 7 | 9 | 4 | 6 | 2380.8223 | 2380.8238 |
| CO2-Xe2-129-132 | 10 | 3 | 8 | 10 | 4 | 7 | 2380.8223 | 2380.8228 |
| CO2-Xe2-132-132 | 7 | 3 | 5 | 7 | 4 | 4 | 2380.8223 | 2380.8218 |
| | | | | | Blend | | 2380.8227 | -0.0004 |
| CO2-Xe2-129-132 | 8 | 1 | 8 | 9 | 0 | 9 | 2380.8456 | 2380.8460 |
| CO2-Xe2-131-132 | 8 | 1 | 8 | 9 | 0 | 9 | 2380.8456 | 2380.8458 |
| CO2-Xe2-129-131 | 8 | 1 | 8 | 9 | 0 | 9 | 2380.8456 | 2380.8461 |
| CO2-Xe2-132-132 | 4 | 1 | 3 | 5 | 2 | 4 | 2380.8456 | 2380.8447 |
| CO2-Xe2-129-132 | 4 | 1 | 3 | 5 | 2 | 4 | 2380.8456 | 2380.8463 |
| CO2-Xe2-129-132 | 12 | 3 | 9 | 12 | 4 | 8 | 2380.8456 | 2380.8446 |
| CO2-Xe2-131-131 | 8 | 1 | 8 | 9 | 0 | 9 | 2380.8456 | 2380.8458 |
| CO2-Xe2-132-134 | 8 | 1 | 8 | 9 | 0 | 9 | 2380.8456 | 2380.8454 |
| CO2-Xe2-129-134 | 8 | 1 | 8 | 9 | 0 | 9 | 2380.8456 | 2380.8458 |
| CO2-Xe2-131-132 | 4 | 1 | 3 | 5 | 2 | 4 | 2380.8456 | 2380.8452 |
| | | | | | Blend | | 2380.8457 | -0.0001 |
| CO2-Xe2-129-132 | 7 | 0 | 7 | 8 | 1 | 8 | 2380.8562 | 2380.8562 |
| CO2-Xe2-131-132 | 7 | 0 | 7 | 8 | 1 | 8 | 2380.8562 | 2380.8556 |
| CO2-Xe2-129-131 | 7 | 0 | 7 | 8 | 1 | 8 | 2380.8562 | 2380.8565 |
| CO2-Xe2-132-132 | 11 | 2 | 10 | 11 | 3 | 9 | 2380.8562 | 2380.8563 |
| CO2-Xe2-131-131 | 7 | 0 | 7 | 8 | 1 | 8 | 2380.8562 | 2380.8560 |
| CO2-Xe2-132-134 | 7 | 0 | 7 | 8 | 1 | 8 | 2380.8562 | 2380.8547 |
| CO2-Xe2-129-134 | 7 | 0 | 7 | 8 | 1 | 8 | 2380.8562 | 2380.8556 |
| CO2-Xe2-131-132 | 11 | 2 | 10 | 11 | 3 | 9 | 2380.8562 | 2380.8568 |
| CO2-Xe2-132-132 | 14 | 3 | 11 | 14 | 4 | 10 | 2380.8562 | 2380.8563 |
| CO2-Xe2-131-134 | 7 | 0 | 7 | 8 | 1 | 8 | 2380.8562 | 2380.8550 |
| | | | | | Blend | | 2380.8560 | 0.0003 |
| CO2-Xe2-132-132 | 7 | 1 | 7 | 8 | 0 | 8 | 2380.8696 | 2380.8692 |
| CO2-Xe2-129-132 | 7 | 1 | 7 | 8 | 0 | 8 | 2380.8696 | 2380.8697 |
| CO2-Xe2-129-129 | 7 | 1 | 7 | 8 | 0 | 8 | 2380.8696 | 2380.8701 |
| CO2-Xe2-131-132 | 7 | 1 | 7 | 8 | 0 | 8 | 2380.8696 | 2380.8693 |
| CO2-Xe2-129-131 | 7 | 1 | 7 | 8 | 0 | 8 | 2380.8696 | 2380.8698 |
| CO2-Xe2-132-132 | 9 | 2 | 8 | 9 | 3 | 7 | 2380.8696 | 2380.8701 |
| CO2-Xe2-131-132 | 9 | 2 | 8 | 9 | 3 | 7 | 2380.8696 | 2380.8708 |
| CO2-Xe2-129-132 | 8 | 2 | 7 | 9 | 1 | 8 | 2380.8696 | 2380.8698 |
| CO2-Xe2-129-134 | 7 | 1 | 7 | 8 | 0 | 8 | 2380.8696 | 2380.8694 |
| CO2-Xe2-129-132 | 10 | 1 | 10 | 10 | 2 | 9 | 2380.8696 | 2380.8705 |

```
                                                    Blend          2380.8698  -0.0002
CO2-Xe2-129-132    6   1   6    7   0   7  2380.8939 2380.8939
CO2-Xe2-129-132    5   0   5    6   1   6  2380.8939 2380.8931
CO2-Xe2-131-132    6   1   6    7   0   7  2380.8939 2380.8935
CO2-Xe2-129-131    6   1   6    7   0   7  2380.8939 2380.8942
CO2-Xe2-131-132    5   0   5    6   1   6  2380.8939 2380.8922
CO2-Xe2-129-131    5   0   5    6   1   6  2380.8939 2380.8935
CO2-Xe2-129-132    6   2   4    6   3   3  2380.8939 2380.8944
CO2-Xe2-131-132    6   2   4    6   3   3  2380.8939 2380.8925
CO2-Xe2-129-131    5   2   3    5   3   2  2380.8939 2380.8922
CO2-Xe2-131-131    6   1   6    7   0   7  2380.8939 2380.8937
                                                    Blend          2380.8934   0.0005
CO2-Xe2-132-132    8   2   6    8   3   5  2380.9006 2380.9011
CO2-Xe2-129-132    7   2   5    7   3   4  2380.9006 2380.8988
CO2-Xe2-129-131    7   2   5    7   3   4  2380.9006 2380.8998
CO2-Xe2-129-132   13   1  12   13   2  11  2380.9006 2380.9013
CO2-Xe2-132-132   10   0  10   10   1   9  2380.9006 2380.9003
CO2-Xe2-129-132   10   0  10   10   1   9  2380.9006 2380.9006
CO2-Xe2-129-129   10   0  10   10   1   9  2380.9006 2380.9009
CO2-Xe2-132-132    1   1   1    2   2   0  2380.9006 2380.8995
CO2-Xe2-132-132    7   2   6    8   1   7  2380.9006 2380.9011
CO2-Xe2-129-132    1   1   1    2   2   0  2380.9006 2380.9016
                                                    Blend          2380.9004   0.0002
CO2-Xe2-132-132    4   0   4    5   1   5  2380.9102 2380.9095
CO2-Xe2-129-132    9   2   7    9   3   6  2380.9102 2380.9099
CO2-Xe2-131-132    4   0   4    5   1   5  2380.9102 2380.9100
CO2-Xe2-129-132    7   1   7    7   2   6  2380.9102 2380.9094
CO2-Xe2-129-132   15   2  13   15   3  12  2380.9102 2380.9093
CO2-Xe2-129-131    7   1   7    7   2   6  2380.9102 2380.9100
CO2-Xe2-131-132   15   2  13   15   3  12  2380.9102 2380.9085
CO2-Xe2-129-131   15   2  13   15   3  12  2380.9102 2380.9097
CO2-Xe2-129-134    4   0   4    5   1   5  2380.9102 2380.9100
                                                    Blend          2380.9096   0.0006
CO2-Xe2-132-132    5   1   5    6   0   6  2380.9189 2380.9181
CO2-Xe2-129-132    5   1   5    6   0   6  2380.9189 2380.9189
CO2-Xe2-129-129    5   1   5    6   0   6  2380.9189 2380.9197
CO2-Xe2-129-129   10   2   8   10   3   7  2380.9189 2380.9182
CO2-Xe2-129-132   11   2   9   11   3   8  2380.9189 2380.9193
CO2-Xe2-132-132   12   2  10   12   3   9  2380.9189 2380.9186
CO2-Xe2-131-132    5   1   5    6   0   6  2380.9189 2380.9183
CO2-Xe2-129-131    5   1   5    6   0   6  2380.9189 2380.9192
CO2-Xe2-132-132   12   1  11   12   2  10  2380.9189 2380.9190
CO2-Xe2-129-132   12   1  11   12   2  10  2380.9189 2380.9195
                                                    Blend          2380.9189   0.0000
CO2-Xe2-132-132   10   1   9   10   2   8  2380.9453 2380.9450
CO2-Xe2-131-132   10   1   9   10   2   8  2380.9453 2380.9455
CO2-Xe2-129-132    4   1   4    5   0   5  2380.9453 2380.9444
CO2-Xe2-132-132    2   0   2    3   1   3  2380.9453 2380.9458
CO2-Xe2-131-132    4   1   4    5   0   5  2380.9453 2380.9437
CO2-Xe2-129-131    4   1   4    5   0   5  2380.9453 2380.9448
CO2-Xe2-129-129    3   1   3    3   2   2  2380.9453 2380.9433
CO2-Xe2-132-134   10   1   9   10   2   8  2380.9453 2380.9440
CO2-Xe2-129-134   10   1   9   10   2   8  2380.9453 2380.9455
CO2-Xe2-131-134   10   1   9   10   2   8  2380.9453 2380.9445
                                                    Blend          2380.9447   0.0000
CO2-Xe2-132-132    8   2   6    8   1   7  2381.1112 2381.1103
CO2-Xe2-129-132    7   2   5    7   1   6  2381.1112 2381.1108
CO2-Xe2-129-132    6   2   4    6   1   5  2381.1112 2381.1110
CO2-Xe2-131-132    8   2   6    8   1   7  2381.1112 2381.1112
CO2-Xe2-129-131    7   2   5    7   1   6  2381.1112 2381.1116
CO2-Xe2-131-132    6   2   4    6   1   5  2381.1112 2381.1096
CO2-Xe2-129-131    6   2   4    6   1   5  2381.1112 2381.1117
CO2-Xe2-131-132    5   2   3    5   1   4  2381.1112 2381.1115
```

```
CO2-Xe2-129-129    4   0   4    3   1   3   2381.1112 2381.1098
CO2-Xe2-131-131    7   2   5    7   1   6   2381.1112 2381.1098
                                        Blend      2381.1108    0.0004
CO2-Xe2-132-132    8   0   8    7   1   7   2381.1983 2381.1977
CO2-Xe2-129-132    7   1   7    6   0   6   2381.1983 2381.1984
CO2-Xe2-131-132    8   0   8    7   1   7   2381.1983 2381.1991
CO2-Xe2-129-131    7   1   7    6   0   6   2381.1983 2381.1996
CO2-Xe2-129-132   10   3   8   10   2   9   2381.1983 2381.1981
CO2-Xe2-129-131   10   3   8   10   2   9   2381.1983 2381.1990
CO2-Xe2-129-134    8   0   8    7   1   7   2381.1983 2381.1992
CO2-Xe2-132-132   14   4  10   14   3  11   2381.1983 2381.1971
CO2-Xe2-129-132   14   4  10   14   3  11   2381.1983 2381.1980
CO2-Xe2-129-129   14   4  10   14   3  11   2381.1983 2381.1988
                                        Blend      2381.1985   -0.0002
CO2-Xe2-129-131    9   1   9    8   0   8   2381.2336 2381.2342
CO2-Xe2-129-132    9   4   6    9   3   7   2381.2336 2381.2352
CO2-Xe2-129-132   10   4   7   10   3   8   2381.2336 2381.2351
CO2-Xe2-129-132    8   4   5    8   3   6   2381.2336 2381.2355
CO2-Xe2-129-129    8   4   4    8   3   5   2381.2336 2381.2345
CO2-Xe2-129-129   10   1   9    9   2   8   2381.2336 2381.2353
CO2-Xe2-129-132   11   4   8   11   3   9   2381.2336 2381.2355
CO2-Xe2-132-132    7   4   4    7   3   5   2381.2336 2381.2341
CO2-Xe2-129-132    7   4   3    7   3   4   2381.2336 2381.2347
                                        Blend      2381.2348   -0.0013
CO2-Xe2-129-132   12   1  12   11   0  11   2381.2863 2381.2873
CO2-Xe2-129-131   12   0  12   11   1  11   2381.2863 2381.2862
CO2-Xe2-129-132    9   2   8    8   1   7   2381.2863 2381.2873
CO2-Xe2-129-131   12   1  11   11   2  10   2381.2863 2381.2861
CO2-Xe2-131-131   12   1  12   11   0  11   2381.2863 2381.2857
CO2-Xe2-132-132   10   5   5   10   4   6   2381.2863 2381.2853
CO2-Xe2-132-132    9   5   5    9   4   6   2381.2863 2381.2869
CO2-Xe2-129-132   10   5   5   10   4   6   2381.2863 2381.2870
CO2-Xe2-132-132   11   5   7   11   4   8   2381.2863 2381.2848
CO2-Xe2-129-132   11   5   6   11   4   7   2381.2863 2381.2852
                                        Blend      2381.2863    0.0000
CO2-Xe2-132-132    8   0   8    7   1   7   2381.1983 2381.1977
CO2-Xe2-129-132    7   1   7    6   0   6   2381.1983 2381.1984
CO2-Xe2-131-132    8   0   8    7   1   7   2381.1983 2381.1991
CO2-Xe2-129-131    7   1   7    6   0   6   2381.1983 2381.1996
CO2-Xe2-129-132   10   3   8   10   2   9   2381.1983 2381.1981
CO2-Xe2-129-131   10   3   8   10   2   9   2381.1983 2381.1990
CO2-Xe2-129-134    8   0   8    7   1   7   2381.1983 2381.1992
CO2-Xe2-132-132   14   4  10   14   3  11   2381.1983 2381.1971
CO2-Xe2-129-132   14   4  10   14   3  11   2381.1983 2381.1980
CO2-Xe2-129-129   14   4  10   14   3  11   2381.1983 2381.1988
                                        Blend      2381.1985   -0.0002
*****************************************************************
```